\documentclass[a4paper,fleqn,usenatbib]{mnras}

\usepackage{graphicx}
\usepackage{amsmath}
\usepackage{amssymb}
\usepackage{ulem}

\newcommand{\fable}{{\sc fable}}

\title[The FABLE simulations: Group and cluster baryons]{The baryon content of groups and clusters of galaxies in the FABLE simulations}

\author[Henden et al.]{Nicholas A. Henden,$^{1}$\thanks{E-mail: n.henden@ast.cam.ac.uk}
Ewald Puchwein$^{3,1,2}$
and Debora Sijacki$^{1,2}$
\\
$^{1}$Institute of Astronomy, University of Cambridge, Madingley Road, Cambridge, CB3 0HA, UK\\
$^{2}$Kavli Institute for Cosmology, University of Cambridge, Madingley Road, Cambridge CB3 0HA, UK\\
$^{3}$Leibniz-Institut f\"{u}r Astrophysik Potsdam (AIP), An der Sternwarte 16, D-14482 Potsdam, Germany\\
}

\date{Accepted XXX. Received YYY; in original form ZZZ}

\pubyear{2019}

\begin{document}
\label{firstpage}
\pagerange{\pageref{firstpage}--\pageref{lastpage}}
\maketitle

\begin{abstract}
We study the gas and stellar mass content of galaxy groups and clusters in the \fable\ suite of cosmological hydrodynamical simulations, including the evolution of their central brightest cluster galaxies (BCGs), satellite galaxies and intracluster light (ICL). The total gas and stellar mass of \fable\ clusters are in very good agreement with observations and show negligible redshift evolution at fixed halo mass for $M_{500} \gtrsim 3 \times 10^{14} M_{\odot}$ at $z \lesssim 1$, in line with recent findings from Sunyaev-Zel'dovich (SZ)-selected cluster samples. Importantly, the simulations predict significant redshift evolution in these quantities in the low mass ($M_{500} \sim 10^{14} M_{\odot}$) regime, which will be testable with upcoming SZ surveys such as SPT-3G. While the stellar masses of \fable\ BCGs are in reasonable agreement with observations, the total stellar mass in satellite galaxies is lower than observed and the total mass in ICL is somewhat higher. This may be caused by enhanced tidal stripping of satellite galaxies due to their large sizes. BCGs are characterised by moderate stellar mass growth at $z < 1$ coincident with a late-time development of the ICL. The level of BCG mass growth is in good agreement with recent observations, however, we caution that the inferred growth depends sensitively on the mass definition. We further show that in-situ star formation contributes more than half the mass of a BCG over its lifetime, the bulk of which is gained at $z > 1$ where star formation rates are highest. The stellar mass profiles of the BCG+ICL component are similar to observed profiles out to $\sim 100$~kpc at $z \approx 0$ and follow a close to power law shape out to several hundred kpc. We further demonstrate that the inferred size growth of BCGs can be severely biased by the choice of parametric model and the outer radius of the fit.
\end{abstract}


\section{Introduction}\label{sec:intro}
Situated at the nodes of the cosmic web, galaxy cluster formation is characterised by continual accretion of dark matter, gas and stars along connecting filaments. This process is dominated by gravity, which couples equally to all matter types. As such, we expect galaxy clusters to be nearly fair samples of the matter content of the Universe at large \citep{White1991}.
Precise measurements of the baryon content of clusters, including the hot, X-ray emitting gas of the intracluster medium (ICM) and the stars in and out of galaxies, can be used to test this expectation and thus inform our models of cluster formation and evolution.
For example, measurements of the total baryon mass in clusters can provide insight on the key physical processes that act to drive cluster baryon fractions away from the universal value, such as feedback from star formation, active galactic nuclei (AGN) or other sources (e.g. \citealt{Gonzalez2013, Sanderson2013, Eckert2016}).
Several of these processes may leave an imprint on the mass or redshift dependence of the total baryon content, providing a further probe of cluster astrophysics.
For example, observational studies that have measured the baryon fractions as a function of halo mass have shown that feedback has a greater impact on low-mass than high-mass haloes and that star formation efficiency is higher in lower mass systems (e.g. \citealt{Lin2003a, Gonzalez2007, Giodini2009, Zhang2011}).
Furthermore, with the recent rise in the number of high-redshift clusters detected via the Sunyaev-Zel'dovich (SZ) effect, it is possible to investigate how the baryonic contents of clusters evolve with redshift out to $z \sim 1$ (e.g. \citealt{Chiu2016, Chiu2016a, Chiu2018}). Upcoming SZ surveys such as SPT-3G \citep{Benson2014} and Advanced ACTpol \citep{Henderson2016}, combined with X-ray data from observatories such as \textit{Chandra} and \textit{XMM-Newton} and wide-and-deep optical and near infrared data from surveys such as the Dark Energy Survey (DES; \citealt{DES2005}) and the Hyper Suprime-Cam survey \citep{Aihara2018}, will enable precise constraints on the growth and evolution of baryons in clusters over a large portion of their history.

The distribution of cluster baryons encodes additional information on cluster assembly.
For example, the partitioning of stellar mass among satellite galaxies, the brightest cluster galaxy (BCG) and the intracluster light (ICL) can provide clues to the roles played by processes such as star formation, AGN feedback, galaxy mergers and tidal stripping in driving cluster galaxy evolution (see \citealt{Kravtsov2012} for a review).
Most theoretical studies of BCG mass evolution, which have mainly been performed by means of semi-analytic models (SAMs), suggest that BCGs form via rapid star formation at early times, followed by rapid hierarchical mass growth via mergers and accretion of orbiting satellite galaxies at late times (e.g. \citealt{DeLucia2007, Laporte2013, Lee2017}).
However, these studies show substantial quantitative disagreement regarding the rate of BCG mass growth in the latter phase of their evolution. Theoretical predictions of BCG mass growth between $z=1$ and $z=0$ range from a factor of $\sim 3-4$ in earlier studies (e.g. \citealt{Aragon-Salamanca1998, DeLucia2007}) to a factor of $\sim 2$ in some more recent SAMs (e.g. \citealt{Tonini2012, Shankar2015}) and in hydrodynamical simulations (e.g. \citealt{Martizzi2016, Ragone-Figueroa2018}).
Observational works also find different mass growth factors, ranging from close to zero \citep{Whiley2008, Collins2009, Stott2010} to almost a factor of two \citep{Lidman2012, Bellstedt2016, Gozaliasl2018a}.

Linked to the evolution of a BCG is the development of a surrounding low surface brightness envelope generally referred to as ICL.
The mass, spatial distribution, colour and metallicity of the ICL reflects the properties of the galaxies from which these stars were liberated and the mechanisms responsible for unbinding them, which may include violent relaxation during mergers, dwarf galaxy disruption or tidal stripping \citep{Conroy2007, Murante2007, Burke2015, Montes2017, DeMaio2018}.
It is clear from recent observational and theoretical works that the ICL is linked to the formation of BCGs and could provide stringent constraints on models of cluster galaxy evolution (e.g. \citealt{Purcell2007, Puchwein2010, Contini2014, DeMaio2015, Groenewald2017, Morishita2017}).
Unfortunately, the diffuse nature of the ICL means that it is difficult to constrain observationally, with the results thus far disagreeing on the total amount of ICL \citep{Gonzalez2007, Zibetti2007, Gonzalez2013, Burke2015, Montes2017, Jimenez-Teja2018}.
One of the prevailing challenges for such studies is the lack of a distinct boundary between the ICL and the extended surface brightness profile of the BCG. This makes it difficult to compare one observational study of the ICL to another, or to compare with simulations, for which there is also no unambiguous distinction between the BCG and ICL \citep{Conroy2007, Dolag2010, Puchwein2010, Rudick2011, Contini2014, Cooper2015}.

So far, efforts to model baryons in galaxy clusters have typically utilised either SAMs (e.g. \citealt{Conroy2007, Purcell2007, Bower2008, Somerville2008, Guo2011, Contini2014}) or cosmological hydrodynamical simulations (e.g. \citealt{Borgani2004, Kay2004, Sijacki2007, Puchwein2008, Fabjan2010, Martizzi2012, Planelles2013}).
The major advantage of hydrodynamical simulations is their ability to follow self-consistently the complex interactions between physical processes such as gravity, hydrodynamics, gas cooling, heating, star formation, AGN feedback and galaxy--galaxy or galaxy--ICM interactions.
Yet, the computational cost of this approach means that it is only in recent years that cosmological hydrodynamical simulations of representative samples of galaxy clusters have become available (e.g. \citealt{LeBrun2014, Dolag2016, McCarthy2017, MACSIS, Cui2018} for models including some form of AGN feedback).
These simulations have been successful in reproducing the global scaling relations of massive haloes, such as their X-ray properties and total gas and stellar content, however their limited spatial and mass resolution means that typically little attention is paid to the cluster galaxies.

Fortunately, as available computing power increases and numerical codes become more efficient, simulations are beginning to emerge that attempt to reproduce simultaneously the global properties of massive haloes and the properties of cluster galaxy populations.
Two notable examples are the \textsc{c-eagle} and IllustrisTNG projects.
The \textsc{c-eagle} cluster simulations, presented in \cite{CEAGLE} and \cite{Bahe2017}, are a suite of hydrodynamical zoom-in simulations of 30 galaxy clusters ($M_{200} = 10^{14}$--$10^{15.4} M_{\odot}$)\footnote{Spherical-overdensity masses and radii use the critical density of the Universe as a reference point. Hence, $M_{200}$ is the total mass inside a sphere of radius $r_{200}$ within which the average density is 500 times the critical density of the Universe.} performed with the same resolution and galaxy formation physics as the `AGNdT9' \textsc{eagle} simulation \citep{Schaye2015, Crain2015}. The \textsc{c-eagle} clusters show good agreement with observations in terms of their total stellar content, X-ray luminosity, average temperature, and satellite stellar mass functions, but are slightly too gas rich and have somewhat high BCG masses \citep{Bahe2017, CEAGLE}.
The IllustrisTNG project \citep{Marinacci2018, Naiman2018, Nelson2018, Pillepich2018b, Springel2018} consists of a suite of uniformly-sampled cosmological volumes modelled with magneto-hydrodynamics and a set of physical models updated from the Illustris project \citep{Genel2014, Vogelsberger2014, Sijacki2015}.
The IllustrisTNG simulations contain a large number of galaxy groups and clusters, including a handful of objects with $M_{200} \sim 10^{15} M_{\odot}$ \citep{Pillepich2018b}. The IllustrisTNG model reproduces a range of observed properties of galaxies, groups and clusters, including the field galaxy stellar mass function and the stellar-to-halo mass relation, however, some tensions with observations remain with respect to BCG masses and total cluster satellite mass \citep{Pillepich2018b}.

In this paper we study the baryon content of galaxy groups and clusters in the \textit{Feedback Acting on Baryons in Large-scale Environments} (\fable) cosmological hydrodynamical simulations. The simulation suite consists of a uniformly-sampled cosmological volume and a series of 27 zoom-in simulations of galaxy groups and clusters spanning a wide halo mass range ($M_{200} = 10^{13.3}$--$10^{15.5} M_{\odot}$). The \fable\ project shares a similar, though independent, goal to that of IllustrisTNG. Namely, to develop a galaxy formation model that improves upon some of the shortcomings of Illustris and apply the model to objects with a much wider range in halo mass.
Originally presented in \cite{Henden2018} (hereafter Paper I), the \fable\ simulation suite was expanded from 6 to 27 zoom-in simulations in \cite{Henden2019} (Paper II).
We summarise the details of the simulations and describe our analysis methods in Section~\ref{sec:methods}. In Section~\ref{sec:global_baryons} we investigate the total gas mass, total stellar mass and the mass in different stellar components (satellites, BCG and ICL) of \fable\ groups and clusters as a function of halo mass and redshift with comparison to observational constraints.
Then in Section~\ref{sec:BCGs} we study the properties of our simulated BCGs, including the BCG stellar mass as a function of host halo mass and redshift, the stellar mass history of BCG main progenitors, and the rate of in-situ star formation.

Throughout this paper we assume a Planck cosmology \citep{PlanckXII2015} with $\Omega_{\Lambda}=$~0.6911, $\Omega_{\rm M}=$~0.3089, $\Omega_{\rm b}=$~0.0486, $\sigma_8=$~0.8159, $n_s=$~0.9667 and $H_0=67.74$~km~s$^{-1}$~Mpc$^{-1}$.
Unless specified otherwise, power law best fits are performed in log-space using the orthogonal BCES method \citep{Akritas1996}. Intrinsic scatter about the best-fitting relation is estimated via the method described in \cite{Maughan2007} and adopted in Paper II.
Quoted uncertainties on the best-fitting parameters correspond to the 68 per cent confidence interval estimated from bootstrapping with $10^4$ resamples.

\section{Methods}\label{sec:methods}
\subsection{Simulations}
The \fable\ simulations are a suite of cosmological hydrodynamical simulations of galaxies, groups and clusters.
The simulations are performed using the moving-mesh hydrodynamics code \textsc{arepo} \citep{Arepo} together with a set of physical models relevant to galaxy formation. The \fable\ galaxy formation model builds upon the framework developed for the successful Illustris simulation \citep{Vogelsberger2013, Vogelsberger2014, Torrey2014, Genel2014}: it follows a diverse range of astrophysical processes, including primordial and metal-line radiative cooling \citep{Katz1996, Wiersma2009}, the formation of stars and supermassive black holes \citep{Springel2003, Springel2005, Vogelsberger2013}, and stellar evolution and chemical enrichment \citep{Wiersma2009a, Vogelsberger2013}.

\fable\ incorporates revised schemes for AGN and stellar feedback that have been calibrated to reproduce the present-day field galaxy stellar mass function and the gas mass fractions of galaxy groups (see Paper I for details). Briefly, we have updated the Illustris model for stellar feedback \citep{Vogelsberger2013} to allow galactic winds to carry thermal as well as kinetic energy.
In addition, we have modified the Illustris model for AGN feedback \citep{DiMatteo2005, Springel2005, Sijacki2007} by introducing a duty cycle to the high accretion rate quasar-mode.
In Paper~I we show how these relatively minor changes to the Illustris model improve on several of the shortcomings of Illustris, in particular the total gas mass of massive haloes, which were severely underestimated in Illustris \citep{Genel2014}.
In Papers~I and II we show that the \fable\ model reproduces a range of X-ray and SZ properties of groups and clusters, including the X-ray luminosity--halo mass relation, the SZ signal--halo mass relation and the thermodynamic profiles of the ICM.
Some tensions with observations remain, however. In particular, the X-ray luminosity--spectroscopic temperature relation lies on the upper end of the observed scatter. This may be a symptom of X-ray hydrostatic mass bias in the data used to calibrate our model to observed gas fractions. This would manifest in \fable\ clusters that are slightly too gas rich, with correspondingly high X-ray luminosities at fixed halo mass (see discussion in Paper~I).

The calibration of our model was carried out using initial conditions for a uniformly-sampled cosmological volume $40 \, h^{-1}$ co-moving Mpc on a side.
The version of this volume simulated with our preferred model is included in our simulation suite.
The volume contains $512^3$ dark matter particles and an approximately equal number of baryonic resolution elements (gas cells and stars) with masses of $m_{\mathrm{DM}} = 3.4 \times 10^7 h^{-1} M_{\odot}$ and $\overline{m}_{\mathrm{b}} \approx 6.4 \times 10^6 h^{-1} M_{\odot}$, respectively. The gravitational softening length was fixed to $2.393$~$h^{-1}$~kpc in physical coordinates below $z=5$ and fixed in comoving coordinates at higher redshifts.

We have applied our calibrated model to systems of a much wider range in mass than the limited volume of the original Illustris simulation allowed. We have achieved this using the zoom-in technique to simulate haloes drawn from the large volume ($3 \, h^{-1}$ co-moving Gpc on a side) Millennium-XXL simulation \citep{Angulo2012}. Our sample includes 27 zoom-in simulations chosen to span the halo mass range $10^{13} \lesssim M_{500} \lesssim 3 \times 10^{15} M_{\odot}$ at $z=0$ with roughly constant logarithmic spacing. The high-resolution region extends to approximately $5 \, r_{500}$ at $z=0$. Dark matter particles in this region have a mass of $m_{\mathrm{DM}} = 5.5 \times 10^7 \, h^{-1} \, M_{\odot}$. The gravitational softening length was fixed to $2.8125$~$h^{-1}$~kpc in physical coordinates at $z \leq 5$ and fixed in comoving coordinates at higher redshifts.

The softening values stated above directly apply to dark matter, star and black hole particles only. For gas cells, these values provide a lower limit to the softening, which is otherwise set to 2.5 times the cell radius. The cell radius is in this context defined as the radius of a sphere with the same volume as the cell.

\subsection{Halo and galaxy identification}\label{subsec:identification}
Haloes and subhaloes are identified using the friends-of-friends (FoF) and \textsc{subfind} algorithms \citep{Davis1985, Springel2001, Dolag2009}.
We define FoF haloes using a linking length of 0.2 times the mean inter-particle separation. We consider a galaxy group or cluster to be any FoF halo with $M_{500} > 10^{13} M_{\odot}$.
We include all FoF haloes in the zoom-in simulations as long as they are not contaminated by one or more lower resolution particles within $5 \, r_{500}$.

\textsc{subfind} identifies gravitationally self-bound subhaloes within each FoF halo. These consist of a central `main halo', the position of which coincides with the FoF centre (defined as the minimum of the gravitational potential), and any number of satellite subhaloes. All particles in the FoF group that are not part of any satellite subhalo are assigned to the main halo, as long as they are gravitationally bound to it.
We consider a galaxy to be any subhalo or main halo with non-zero stellar content. Hence, \fable\ groups and clusters consist of a central galaxy and multiple satellite galaxies.

\subsection{Definition of stellar components}
Below we describe our operational definitions for the three main stellar components of massive haloes: the central galaxy or BCG, intracluster light, and satellite galaxies.

We take the stellar mass of the BCG and its associated ICL to be the mass of all star particles bound to the main halo.
Note that, with this definition, we are assuming that the BCG is the central galaxy and vice-versa.
Unfortunately, there is no clear boundary between the bright, inner regions of BCGs and their low surface brightness envelopes (ICL), either observationally or in simulations.
We therefore define BCG and ICL masses using fixed apertures of various sizes.
Specifically, the stellar mass of the BCG is calculated within a spherical aperture centred on the main halo, while the ICL mass is the gravitationally bound stellar mass outside this aperture, out to $r_{500}$.
The mass of a satellite galaxy is taken to be the sum of all of its gravitationally-bound star particles. Then, the total stellar mass bound in satellite galaxies is the sum of these masses for all satellites within $r_{500}$ of the halo centre.

\section{Global baryonic properties}\label{sec:global_baryons}
In Sections~\ref{subsec:ICM_mass} and \ref{subsec:stellar_mass} we study how the total mass of gas and stars in \fable\ galaxy groups and clusters depends on halo mass with comparison to observations and other simulation predictions.
In Sections~\ref{subsubsec:ICM_z} and \ref{subsubsec:stellar_z} we compare directly to the results of \cite{Chiu2018} concerning the redshift evolution of the total gas and stellar content of groups and clusters, including predictions for future, lower mass samples.

\begin{figure*}
  \includegraphics[width=0.497\textwidth]{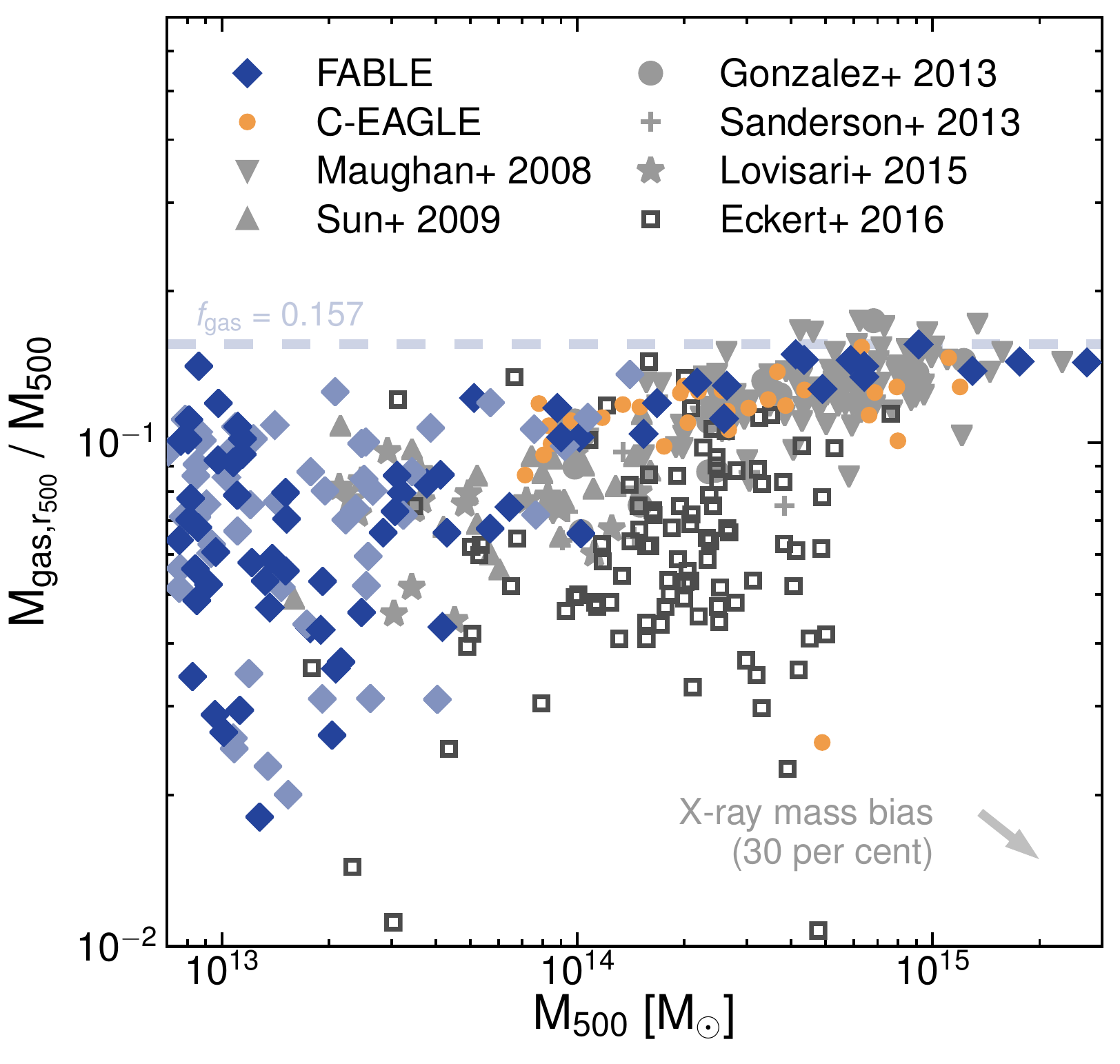}
  \includegraphics[width=0.497\textwidth]{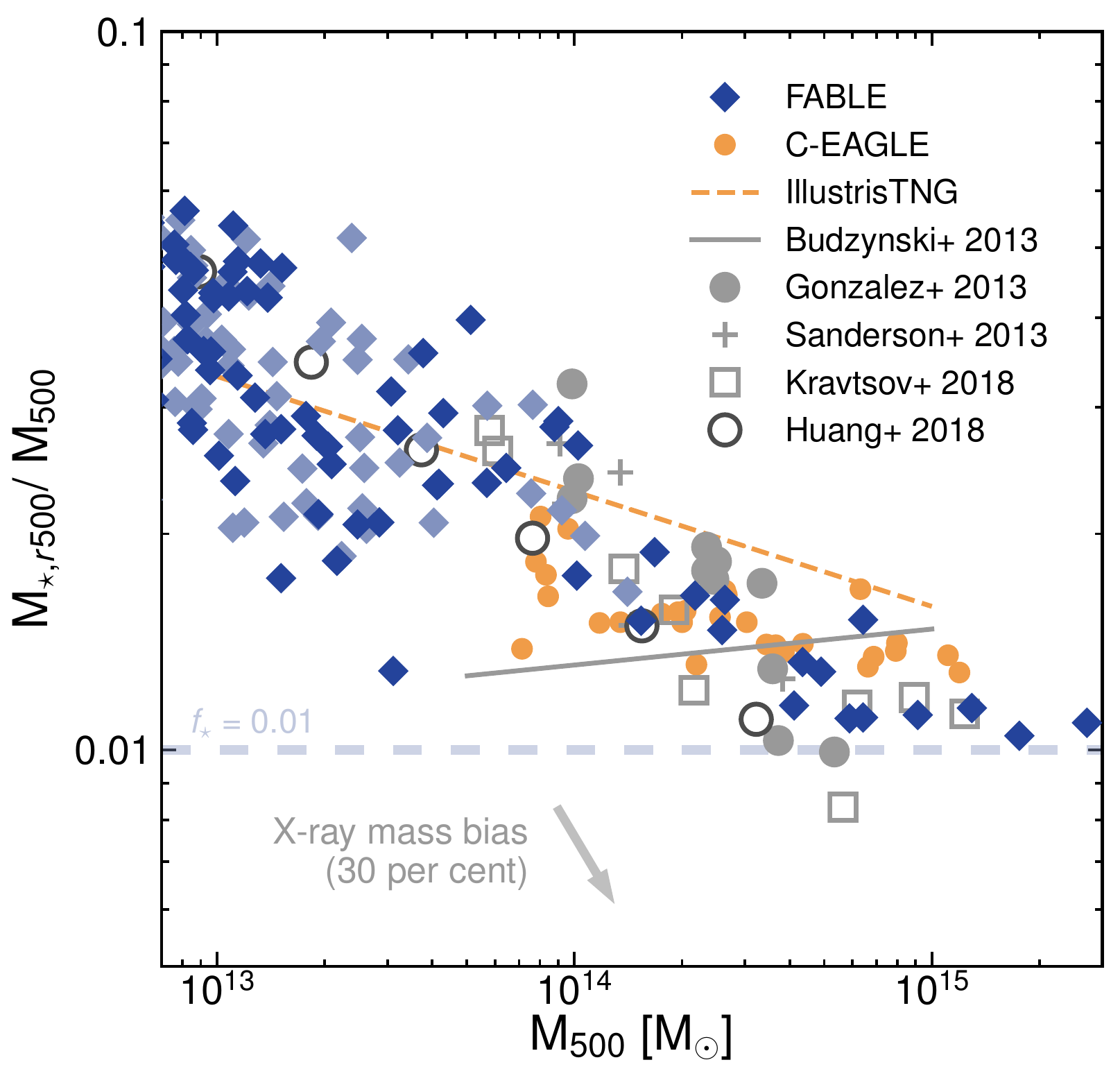}
  \caption{Gas and stellar mass fractions within $r_{500}$ as a function of halo mass at $z=0$ for \fable\ galaxy groups and clusters (blue diamonds) compared to observational data (grey symbols) as well as the \textsc{c-eagle} simulated clusters (orange circles) and the best-fitting relation from the IllustrisTNG simulations (orange dashed line).
    Dark blue diamonds correspond to haloes within our cosmological box and the main halo of each zoom-in simulation. Light blue diamonds indicate secondary haloes in the high-resolution region of the zoom-in simulations.
    Grey open circles in the right-hand panel show the median relation of \protect\cite{Huang2018} and the grey solid line is the best-fitting relation from the stacked analyses of \protect\cite{Budzynski2013}. Other grey symbols indicate individual observed groups and clusters.
    Observational data based on weak lensing mass estimates are shown in dark grey while data based on X-ray hydrostatic masses are shown in light grey. The grey arrow at the bottom of each panel shows how the latter are expected to change when correcting for a potential X-ray hydrostatic mass bias in which X-ray masses underestimate the true mass by $30$ per cent. All of the simulation results shown here use halo masses measured directly from the simulation and should be compared with the less biased weak lensing masses where possible.
    Thick dashed lines correspond to constant gas and stellar mass fractions of $15.7$ per cent and $1$ per cent, respectively, where the former is approximately the cosmic baryon fraction in our fiducial cosmology.
  }
  \label{fig:baryon_mass_vs_halo_mass}
\end{figure*}

\subsection{ICM mass to halo mass relation}\label{subsec:ICM_mass}
In the left-hand panel of Fig.~\ref{fig:baryon_mass_vs_halo_mass} we plot the gas mass to halo mass fraction within $r_{500}$ as a function of halo mass for \fable\ haloes at $z=0$ (blue diamonds).
We compare to $z \approx 0$ observational constraints from \cite{Eckert2016} (open grey squares), which are based on weak lensing mass estimates, and \cite{Maughan2008}, \cite{Sun2009}, \cite{Gonzalez2013}, \cite{Sanderson2013} and \cite{Lovisari2015} (solid grey symbols), which use X-ray hydrostatic masses.
Orange circles correspond to individual clusters from the \textsc{c-eagle} cosmological hydrodynamical zoom-in simulations.
The gas masses of \textsc{c-eagle} clusters are estimated from the mock X-ray pipeline described in \cite{CEAGLE}.  For both \textsc{c-eagle} and \fable\ we use `true' halo masses measured directly from the simulation.
For \fable\ haloes we sum up the total gas mass within $r_{500}$ centred on the gravitational potential minimum of the halo. This definition includes both hot and cold gas, however in fig. 5 of Paper I we show that cold gas makes a negligible contribution to the total gas mass on group and cluster mass scales ($M_{500} \gtrsim 10^{13} M_{\odot}$).

The slope of the simulated relation is in good agreement with the observed relations. These diverge significantly from the self-similar expectation of cluster evolution in the absence of non-gravitational physics, which predicts a constant gas mass fraction at all mass scales. Indeed, whereas on the scale of low-mass clusters and groups ($M_{500} \lesssim 3 \times 10^{14} M_{\odot}$) the ratio of the total gas mass to the halo mass of the cluster is significantly lower than the cosmic baryon fraction ($f_b \equiv \Omega_b / \Omega_m \approx 0.157$; dashed line in the left-hand panel of Fig.~\ref{fig:baryon_mass_vs_halo_mass}), at the high mass end of the relation the slope is consistent with a constant gas mass fraction approaching the cosmic value. This is a clear indication of the impact of non-gravitational processes, such as star formation and AGN feedback, on the total gas content of all but the most massive clusters.

\begin{figure*}
  \includegraphics[width=0.497\textwidth]{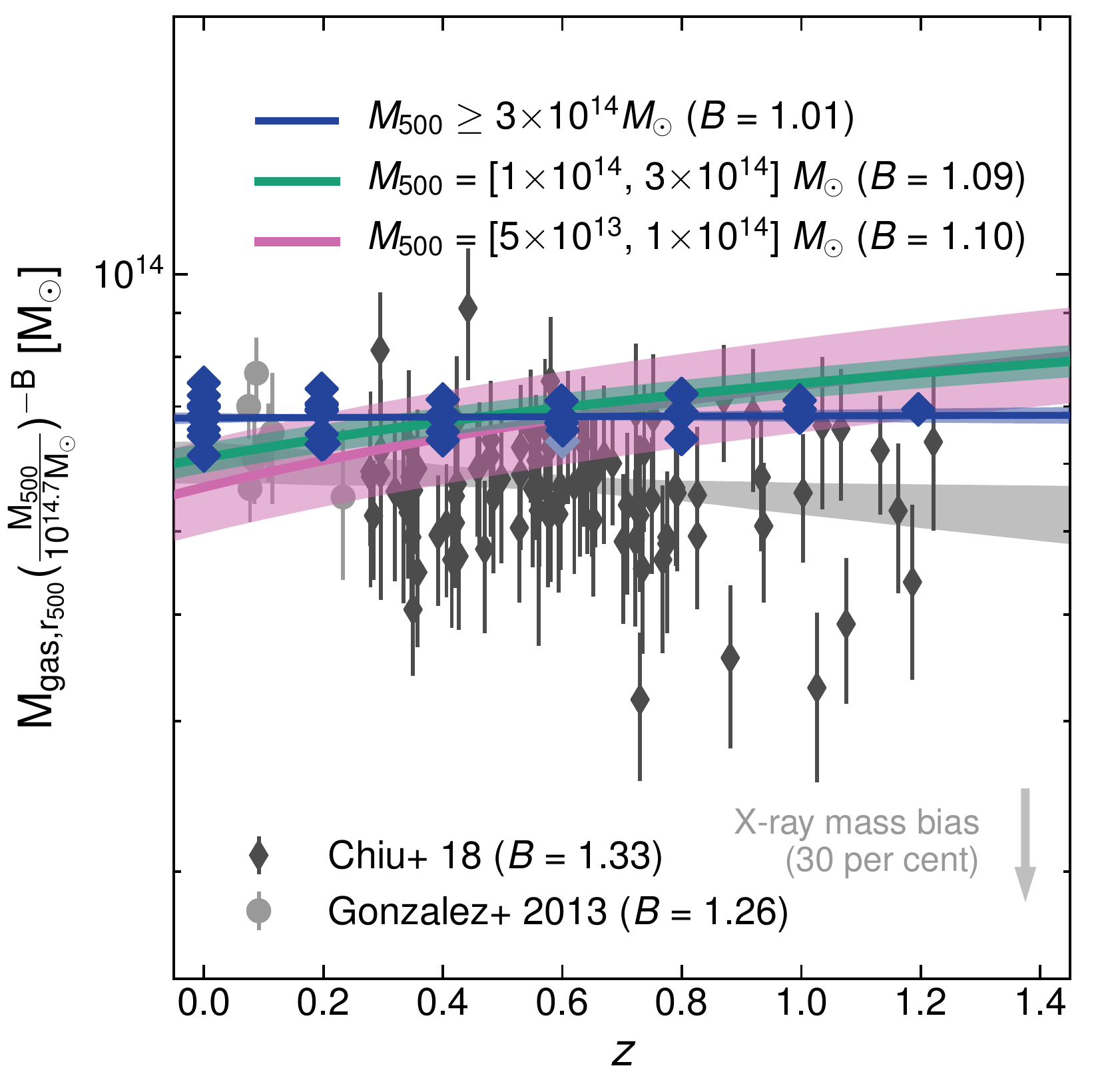}
  \includegraphics[width=0.497\textwidth]{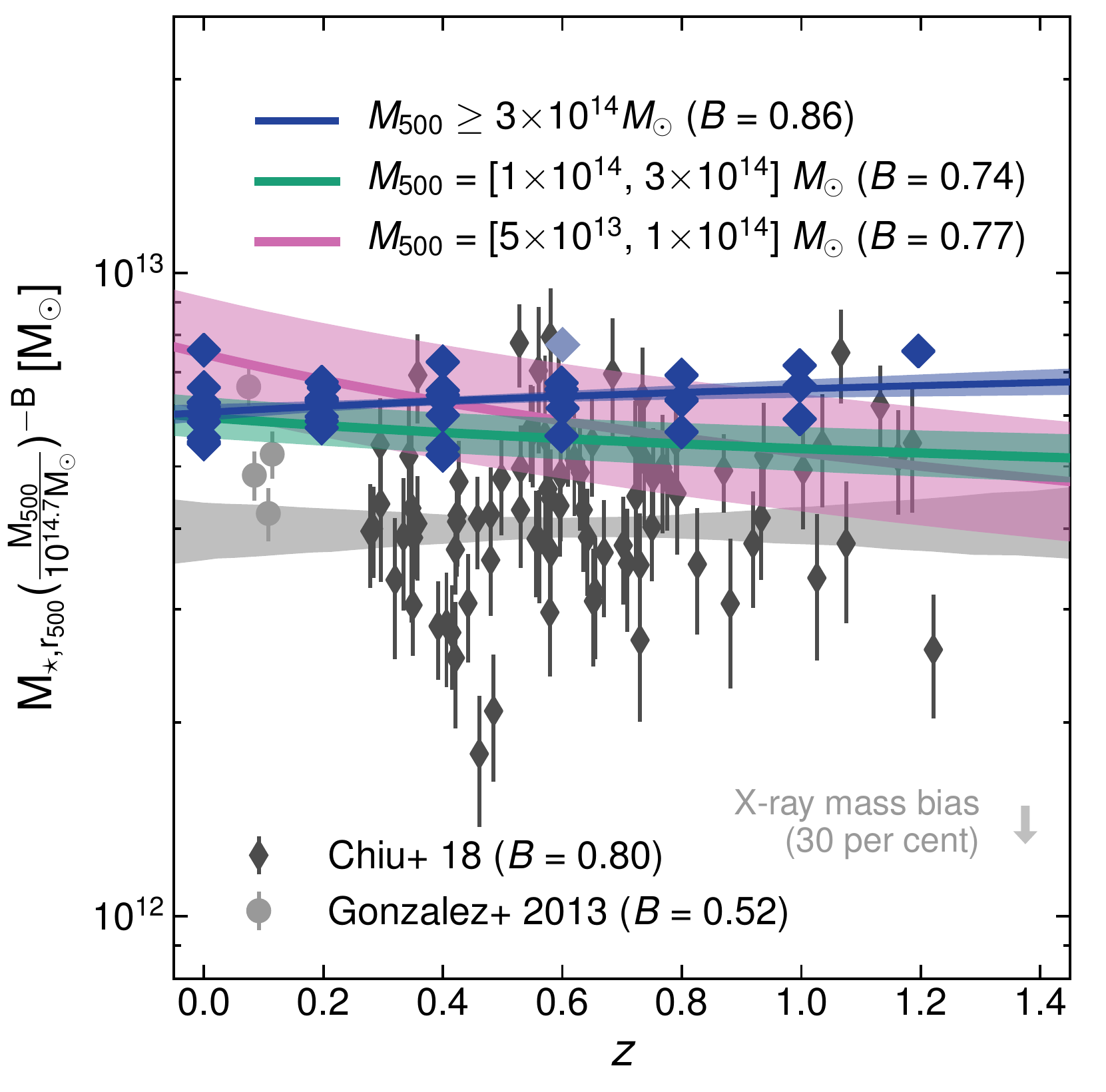}
  \caption{The redshift trend of the total gas mass (left) and total stellar mass (right) within $r_{500}$ with respect to the pivot halo mass of $M_{500} = 4.8 \times 10^{14} M_{\odot}$ in comparison to results from \protect\cite{Chiu2018} for a sample of 91 SPT-selected clusters (grey diamonds). The gas mass and stellar mass are normalised to the pivot halo mass using the best-fitting mass slope of each sample as indicated in the legend.
    Blue diamonds correspond to individual \fable\ clusters with halo mass $M_{500} \geq 3 \times 10^{14} M_{\odot}$ at each redshift in order to roughly match the SZ-selected \protect\cite{Chiu2018} sample for which the minimum halo mass is roughly constant at this value for all redshifts shown. Circles show low-redshift data from \protect\cite{Gonzalez2013} for clusters within the same mass range. Three of the \protect\cite{Gonzalez2013} clusters do not have total stellar mass measurements.
    Grey arrows demonstrate how we would expect the \protect\cite{Gonzalez2013} data points to change if corrected for an X-ray hydrostatic mass bias in which X-ray masses underestimate the true mass by $30$ per cent.
    Solid lines show the best-fitting simulated relation, as calculated from the posterior median of the parameters, for different (redshift-independent) halo mass selections.
    Shaded regions enclose the $16$th to $84$th percentile range of the posterior samples at each redshift. These are larger for lower mass samples largely because the average halo mass is further from the pivot mass.
    Grey shaded regions show the 1-sigma confidence regions of the best-fitting relations derived in \protect\cite{Chiu2018}.
  }
  \label{fig:baryon_mass_vs_z}
\end{figure*}

The simulated relation is similar in normalisation to the observational constraints based on X-ray hydrostatic mass estimates. This is partly by design, since the strength of AGN feedback in the \fable\ model was calibrated to reproduce X-ray measurements of gas mass fractions of galaxy groups with $M_{500} \lesssim 10^{14} M_{\odot}$ (see Paper I for details). Even so, the match to observations at cluster-scale masses ($M_{500} > 10^{14} M_{\odot}$) was not guaranteed by the calibration, as the deeper gravitational potential wells of these systems increase the energy required for AGN feedback to eject gas beyond $r_{500}$.
Conversely, \fable\ haloes have significantly higher gas mass fractions at fixed halo mass compared to data from \cite{Eckert2016}, who estimate halo mass via a weak lensing-calibrated relation between halo mass and X-ray temperature \citep{Lieu2016}. A similar picture is presented by the \textsc{c-eagle} clusters (orange circles in Fig.~\ref{fig:baryon_mass_vs_z}), which have very similar gas mass fractions as the \fable\ systems at fixed halo mass.

The difference between the results of \cite{Eckert2016} and those based on X-ray hydrostatic masses is most readily explained if X-ray masses are biased low compared to weak lensing masses by approximately $28$ per cent \citep{Eckert2016}.
The effect of such a bias is illustrated by the grey arrows in Fig.~\ref{fig:baryon_mass_vs_halo_mass}, which show how the gas and stellar mass fraction measurements based on X-ray hydrostatic mass estimates (light grey symbols) are expected to change when correcting for a mass bias in which the X-ray masses underestimate the true halo mass by $30$ per cent. These include an estimate for the increased gas or stellar mass associated with the increased aperture radius, $r_{500}$. We take this to be the median increase for \fable\ clusters with $M_{500} > 10^{14} M_{\odot}$ at $z=0$, which is 17 and 6 per cent for the gas and stellar mass, respectively.

Many studies favour such a large X-ray mass bias \citep[e.g.,][]{vonderLindenEtAl2014,HoekstraEtAl2015}, thus suggesting that this shift should indeed be taken into account and that consequently \fable\ halos are somewhat too gas rich. This would also explain several discrepancies of our model with respect to observations, such as the rather high X-ray luminosities of \fable\ clusters at fixed temperature (see e.g. fig.~8 in Paper I). By computing X-ray hydrostatic mass estimates from their simulations, \cite{CEAGLE} arrive at a similar result for the {\sc c-eagle} clusters and conclude that they are also too gas rich.

On the other hand, a large X-ray hydrostatic mass bias ($\gtrsim 30$ per cent) implies low baryon fractions in clusters that are also difficult to reconcile with our current models of galaxy cluster formation.
For example, \cite{Eckert2016} constrain the baryon fraction of $10^{14} M_{\odot}$ haloes to be $f_{\rm bar} = 0.067 \pm 0.008$.
This corresponds to a baryon depletion factor of $\mathcal{D} \approx 0.57$, where $\mathcal{D} = 1 - f_{\rm bar}/f_{\rm b}$.
In contrast, simulations typically predict a much smaller depletion factor, on the order of $\mathcal{D} \sim 0.3$ at $10^{14} M_{\odot}$ (e.g. \citealt{Planelles2013, LeBrun2014, McCarthy2017}). Indeed, the baryon depletion factor for \fable\ haloes at $M_{500} \sim 10^{14} M_{\odot}$ is only $\mathcal{D} \sim 0.2$. Models with stronger AGN feedback can yield large depletion factors in agreement with the \cite{Eckert2016} results, however these models struggle to reproduce other cluster observables, such as the thermodynamic profiles of the ICM (see e.g. \citealt{LeBrun2014}).
As discussed in Paper I, a solution to this problem would likely have to come in the form of a more sophisticated modelling of AGN feedback in simulations and/or the inclusion of previously neglected physical processes that, perhaps in combination, are able to efficiently lower baryon fractions in groups and clusters from the universal average without overheating or evacuating gas in the core regions.

\subsection{Stellar mass to halo mass relation}\label{subsec:stellar_mass}
In the right-hand panel of Fig.~\ref{fig:baryon_mass_vs_halo_mass} we show the stellar mass to halo mass fractions of \fable\ groups and clusters within $r_{500}$ as a function of halo mass at $z=0$. We compare to observational data from \cite{Budzynski2013}, \cite{Gonzalez2013}, \cite{Sanderson2013} and \cite{Kravtsov2018}, which are based on X-ray hydrostatic masses, and the median relation from \cite{Huang2018}, which is based on weak lensing mass estimates.
Each of these studies take into account a contribution from the ICL in their stellar mass measurements.
We also compare to the best-fitting relation from IllustrisTNG, which is based on true halo masses measured directly from the simulation \citep{Pillepich2018b}.

Overall, our comparison to observations suggests that most \fable\ clusters and groups have formed a realistic total stellar mass at $z=0$. In massive clusters, approximately $1$ per cent of the halo mass is in the form of stars, while lower mass haloes diverge towards higher stellar mass fractions, in agreement with the majority of the observational data. In the (likely) presence of a significant X-ray mass bias in the data, some of our most massive \fable\ clusters contain, however, still too many stars (see the grey arrow). There is also a difference in the slope compared to the best-fitting relation from \cite{Budzynski2013} that favours an almost constant stellar mass fraction across a wide halo mass range. The origin of this difference is unclear, although it may be related to the fact that \cite{Budzynski2013} find little to no contribution from the ICL on galaxy group scales. This contrasts with our simulation predictions, which show a significant ICL fraction on group scales ($\sim 20$ to $40$ per cent; see Section~\ref{subsec:stellar_mass_components}). A similar, though slightly lower, ICL fraction is predicted by IllustrisTNG ($\sim 10$ to $30$ per cent; \citealt{Pillepich2018b}).

On galaxy group scales ($M_{500} \lesssim 10^{14} M_{\odot}$) the simulations show good agreement with the median relation of \cite{Huang2018}, who measure stellar masses for a large sample of galaxies from the Hyper Suprime-Cam (HSC) survey \citep{Aihara2018}.
On the other hand, the median stellar mass fraction in their highest halo mass bin ($M_{500} = 3.2 \times 10^{14} M_{\odot}$) is slightly lower than the simulation predictions, which are in good agreement with \cite{Gonzalez2013} clusters of similar halo mass.
This may be due to X-ray hydrostatic mass bias, although at a somewhat lower level than that implied by the weak lensing calibrated gas mass fractions of \cite{Eckert2016} discussed in the previous section.
Alternatively, the offset between the two observational studies may be related to the mass-to-light ratios used to convert the measured luminosities into stellar masses.
For example, for a sample of high-redshift ($z \sim 1$) clusters, \cite{VanderBurg2014} show that assuming a fixed mass-to-light ratio for all galaxies (as in \citealt{Gonzalez2013}), rather than deriving it for each galaxy individually based on SED modelling (as in \citealt{Huang2018}), overestimates the total stellar mass in their clusters by at least a factor of two.
If the \cite{Gonzalez2013} stellar masses are indeed biased high -- either due to X-ray mass bias or overestimated mass-to-light ratios -- then the stellar masses of \fable\ clusters are likely overestimated.

The same reasoning also applies to \textsc{c-eagle}, which predicts similar total stellar masses to \fable. Interestingly, the \textsc{c-eagle} galaxy formation model predicts significantly fewer high-mass galaxies ($M_{\star} \gtrsim 10^{11} M_{\odot}$) than \fable\ in the field environment (see the galaxy stellar mass function comparison with \textsc{eagle} in fig.~2 of Paper I).
This suggests that cluster-specific processes responsible for the suppression of star formation may be more effective in \fable\ than in \textsc{c-eagle}.
The prediction from IllustrisTNG is in good agreement with \fable\ and \cite{Huang2018} at the low mass end but lies slightly above the other relations at cluster mass scales ($\gtrsim 10^{14} M_{\odot}$). This may reflect differences in the field galaxy stellar mass function, for which IllustrisTNG predicts a slightly higher abundance of massive galaxies compared to \fable\ and significantly more than \textsc{eagle} \citep{Pillepich2018b}.

The simulations predict a fairly tight relation between stellar mass and halo mass, with a level of intrinsic scatter comparable to, but slightly lower than, the observational constraints.
In particular, a power law fit to \fable\ clusters with $M_{500} > 10^{14} M_{\odot}$ yields a log-normal intrinsic scatter of $0.06^{+0.02}_{-0.01}$ dex in stellar mass at fixed halo mass.
This is similar to the scatter predicted by IllustrisTNG ($0.07$ dex; \citealt{Pillepich2018b}) and consistent with, though somewhat smaller than, the intrinsic scatter of $0.09 \pm 0.05$ dex for the \cite{Kravtsov2018} sample and $0.11 \pm 0.03$ dex for an extended sample including clusters from \cite{Gonzalez2013}.

\subsection{ICM and stellar mass redshift evolution}\label{subsec:ICM_stellar_mass_z}

In the following sections we compare the total gas and stellar mass content of \fable\ clusters as a function of redshift to results from \cite{Chiu2018}, who study an SZ-selected sample of 91 galaxy clusters at $0.2 < z < 1.25$ detected in the 2500 deg$^2$ SPT-SZ survey.
Their halo mass estimates are obtained from the SZ signal using the best-fitting scaling relation for SPT-SZ clusters derived in \cite{Haan2016}.

\cite{Chiu2018} derive a best-fitting scaling relation linking the gas mass or stellar mass to the halo mass and redshift with (redshift-independent) log-normal intrinsic scatter at fixed halo mass.
We follow the same procedure, using the same functional form of the scaling relation:
\begin{equation}\label{eq:scaling_relation}
Y = A \left(\frac{M_{500}}{M_{\mathrm{piv}}}\right)^B \left(\frac{1 + z}{1 + z_{\mathrm{piv}}}\right)^C
\end{equation}
with log-normal intrinsic scatter, where $Y$ is the gas or stellar mass within $r_{500}$, $A$ is the normalisation at the pivot mass $M_{\mathrm{piv}}$ and redshift $z_{\mathrm{piv}}$, and $B$ and $C$ are the power law indices of the mass and redshift trends, respectively. We adopt $M_{\mathrm{piv}} = 4.8 \times 10^{14} M_{\odot}$ and $z_{\mathrm{piv}} = 0.6$ as in \cite{Chiu2018}. We have checked that the best-fitting parameters are not significantly affected by the choice of pivot mass or redshift.
We use the affine invariant Markov chain Monte Carlo (MCMC) ensemble sampler implemented in the \textsc{emcee} \textsc{python} package \citep{Foreman-Mackey2013} to sample the posterior  distribution of the parameters.
We perform the fitting in log-space with a Gaussian likelihood function and adopt flat priors of $\mathrm{log}_{10} A$ in $(11, 13)$, $B$ in $(0.1, 3.5)$, $C$ in $(-4, 4)$ and intrinsic scatter in $(10^{-3}, 1.5)$, as adopted in the \cite{Chiu2018} analysis.
We use an ensemble of $100$ walkers with $1000$ steps, excluding the first $200$ steps as burn-in.
The best-fitting parameters and their uncertainties are taken to be the median and $68$ per cent confidence interval about the median calculated from the marginalised posterior distributions for each parameter.

Following \cite{Chiu2018}, in Fig.~\ref{fig:baryon_mass_vs_z} we plot the gas mass (left-hand panel) and stellar mass (right-hand panel) as a function of redshift after removing the halo mass dependence of these quantities using the best-fitting scaling relations. As such, Fig.~\ref{fig:baryon_mass_vs_z} highlights the redshift trend of the gas mass and stellar mass at the pivot point of $M_{500} = 4.8 \times 10^{14} M_{\odot}$.
The \cite{Chiu2018} sample (grey diamonds), by virtue of its selection on the SZ signal, is approximately mass-limited, with a minimum mass of $M_{500} \approx 3 \times 10^{14} M_{\odot}$ across the full redshift range. For our comparison sample (blue diamonds) we therefore include all \fable\ clusters with a halo mass greater than $M_{500} = 3 \times 10^{14} M_{\odot}$ at each redshift. We have verified that this yields a similar range of masses as the observed sample at each redshift.
To complement the low-redshift comparison we also plot clusters with $M_{500} > 3 \times 10^{14} M_{\odot}$ from \cite{Gonzalez2013} (circles). Note however that we use the best-fitting mass slopes of their full cluster sample, as there are too few clusters in this mass range with which to derive the slope.

Upcoming SZ cluster surveys are expected to identify an unprecedented number of low-mass clusters and galaxy groups out to high redshift (e.g. \citealt{Benson2014, Henderson2016, Abazajian2016, Bender2018}).
For example, ongoing and future SZ-selected surveys such as SPT-3G \citep{Benson2014} and CMB-S4 \citep{Abazajian2016} are expected to be mass-limited at a level of $M_{500} \gtrsim 10^{14} M_{\odot}$.
With this in mind, we investigate the halo mass dependence of the gas and stellar mass redshift trends by deriving the best-fitting scaling relations for two lower mass samples (green solid lines in Fig.~\ref{fig:baryon_mass_vs_z}): low-mass clusters with $10^{14} M_{\odot} < M_{500} < 3 \times 10^{14} M_{\odot}$ and galaxy groups with $5 \times 10^{13} M_{\odot} < M_{500} < 10^{14} M_{\odot}$.
These mass limits are independent of redshift to mimic an SZ-selected sample.

\subsubsection{ICM mass redshift trend}\label{subsubsec:ICM_z}

The simulated clusters lie systematically above the mean relation of \cite{Chiu2018} but are in good agreement with the \cite{Gonzalez2013} clusters at low-redshift ($z \lesssim 0.2$). This is understandable given that the \cite{Gonzalez2013} data are based on X-ray hydrostatic masses, whereas the \cite{Chiu2018} mass estimates are effectively weak lensing-calibrated. As in Section~\ref{subsec:ICM_mass}, we have marked with a grey arrow the expected change to the observational data points when correcting for a fairly large X-ray hydrostatic mass bias of $30$ per cent.
As discussed in detail in Paper II, weak lensing-based constraints suggest that \fable\ clusters have somewhat high gas mass at fixed halo mass, consistent with the offset seen here.

\cite{Chiu2018} find that the gas mass at fixed halo mass varies mildly with redshift, with a best-fitting redshift trend that is close to zero ($C = -0.15 \pm 0.14$).
Our results appear to support this finding for the mass range in question, albeit with a relatively small sample size. Indeed, the best-fitting redshift trend parameter for the comparison sample, $C = 0.01 \pm 0.04$, is consistent with zero. In our simulations these massive clusters have baryon fractions close to the cosmic baryon fraction ($\gtrsim$ 95 \% of cosmic). The lack of evolution hence mostly reflects the fact that it is very difficult at any redshift to expel significant amounts of baryons from the deep potential wells and large collapsing Lagrangian regions of such massive objects.

For low-mass clusters with $M_{500} = [1-3] \times 10^{14} M_{\odot}$ we find that the gas mass at fixed halo mass increases with increasing redshift ($C = 0.29 \pm 0.04$).
Furthermore, galaxy groups with $M_{500} = [5-10] \times 10^{13} M_{\odot}$ present a slightly stronger redshift trend, with a best-fitting redshift trend of $C = 0.39 \pm 0.05$.
Qualitatively this agrees with \cite{Lin2012}, who find that the gas mass at fixed halo mass increases with increasing redshift as $(1 + z)^{0.41 \pm 0.14}$ at $z < 0.6$ for a sample spanning a wide mass range ($8 \times 10^{13} \lesssim M_{500} \lesssim 2 \times 10^{15} M_{\odot}$).

The positive redshift trends of the lower mass samples reflect the rapid decrease in the normalisation of the gas mass--halo mass relation at $z \lesssim 1$, as shown in Paper II. A similar trend was also found in the recent simulation studies of \cite{MACSIS} and \cite{Truong2018}. We attribute this evolution to the increasing effectiveness of gas expulsion by AGN feedback with decreasing redshift due to: a corresponding fall in the density (and therefore the binding energy) of haloes of fixed spherical overdensity mass in combination with a likely later occurring main growth phase of the AGN found in such low-redshift haloes, and (at least in \fable) an increasing fraction of radio-mode AGN, which are more effective than quasar-mode AGN at ejecting gas from group-scale haloes (see appendix~A of Paper~I).

\subsubsection{Stellar mass redshift trend}\label{subsubsec:stellar_z}
The stellar mass to halo mass relation derived by \cite{Chiu2018} suggests a weak redshift trend that is statistically consistent with zero ($C = 0.05 \pm 0.25$), albeit with large uncertainty.
This is consistent with \cite{Lin2012} and \cite{Lin2017}, who find no evidence for redshift evolution in the stellar mass--halo mass relation at $z \lesssim 0.6$ and $z \lesssim 1$, respectively.
In good agreement with the observational results, we find only a marginally significant change in stellar mass with redshift for \fable\ clusters with mass $M_{500} \gtrsim 3 \times 10^{14} M_{\odot}$ ($C = 0.12 \pm 0.08$; blue solid line in the right-hand panel of Fig.~\ref{fig:baryon_mass_vs_z}).

Naively one might expect that massive clusters form predominantly by the accumulation of lower mass haloes. However, this does not seem to be compatible with the shallow slope of the stellar mass--halo mass relation, which corresponds to lower mass haloes having higher stellar mass fractions (see Section~\ref{subsec:stellar_mass}). As such, when a cluster accretes a lower mass halo, its total stellar mass fraction is expected to increase.
\cite{Chiu2018} hypothesise that the accretion of lower mass haloes must therefore be balanced by a substantial infall of material from the surrounding environment, which has a substantially lower stellar mass fraction than the cluster itself, especially at $z \sim 1$ (see their fig.~6). Infall from these regions therefore counteracts the increase in the stellar mass fraction that would result purely from accretion of smaller objects.

There are a number of elements that are missing from this toy model however.
For example, while it seems difficult to assemble massive clusters from low-mass clusters and groups that have higher stellar mass fractions within $r_{500}$, clusters will also accrete mass belonging to these haloes that lies outside $r_{500}$, where the stellar mass fraction is smaller.
Furthermore, for our low-mass cluster sample the redshift trend is mildly negative ($C = -0.16^{+0.07}_{-0.08}$), implying that the stellar mass at fixed cluster mass increases slightly with decreasing redshift.
The trend is even more significant in haloes with $5 \times 10^{13} M_{\odot} < M_{500} < 10^{14} M_{\odot}$ for which we find a large negative redshift trend ($C = -0.51 \pm 0.08$).
This implies that the stellar mass fractions of galaxy groups were lower in the past, which goes some way to explaining the relatively low stellar mass fractions of massive clusters at the present day.

It is clear in the right-hand panel of Fig.~\ref{fig:baryon_mass_vs_z} that \fable\ clusters tend to lie on the upper end of the scatter in the SPT clusters.
This appears to be at odds with the $z \approx 0$ comparison in Fig.~\ref{fig:baryon_mass_vs_halo_mass} where we demonstrate a good match to observations of the total stellar mass at fixed halo mass. This offset may be the result of X-ray hydrostatic mass bias in the comparison samples, as we discussed in Section~\ref{subsec:stellar_mass}.
Indeed, we expect that the \cite{Chiu2018} mass estimates, which incorporate weak lensing information, are less biased than X-ray hydrostatic masses.
On the other hand, the stellar masses derived in \cite{Chiu2018} are likely underestimated given that the limited depth of their imaging data prohibits a measurement of the ICL.
We point out that \cite{Chiu2018} compare their stellar mass--halo mass relation to a number of other studies in the literature and find a fairly large variation in their normalisations ($\sim 40$ per cent), even after accounting for differences in the assumed initial mass function (IMF) and the method used to estimate the halo mass.
There are various sources of systematic error other than measurement of the ICL that may explain this variation, including sample selection biases (e.g. \citealt{Decker2019}), background subtraction (e.g. \citealt{Bernardi2007, Bernardi2013, VonderLinden2007, VanderBurg2014}), or the conversion from luminosity to stellar mass (e.g. \citealt{Conroy2009, Bernardi2017}).

\subsection{The stellar mass budget in groups and clusters}\label{subsec:stellar_mass_components}
In this section we quantify the fraction of the total stellar mass residing in the BCG, ICL and satellite galaxies (defined in Section~\ref{subsec:identification}) as a function of halo mass.
In Section~\ref{subsubsec:stellar_mass_components_z0} we compare the relations to observational constraints at $z \approx 0$ and in Section~\ref{subsubsec:stellar_mass_components_redshift} we look at the redshift evolution of the relations at $z < 3$.

\begin{figure}
  \centering
  \includegraphics[width=0.975\columnwidth]{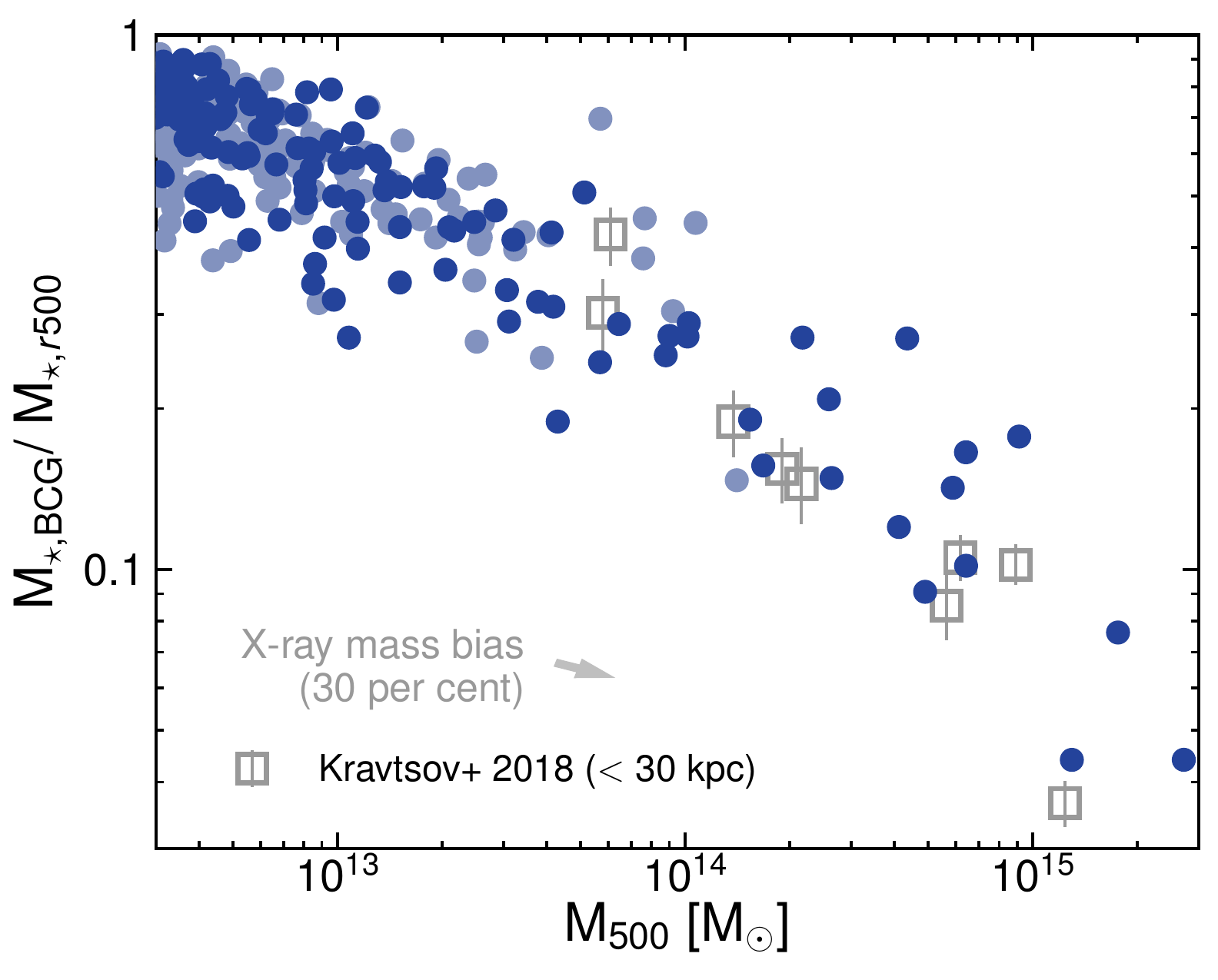}
  \includegraphics[width=0.975\columnwidth]{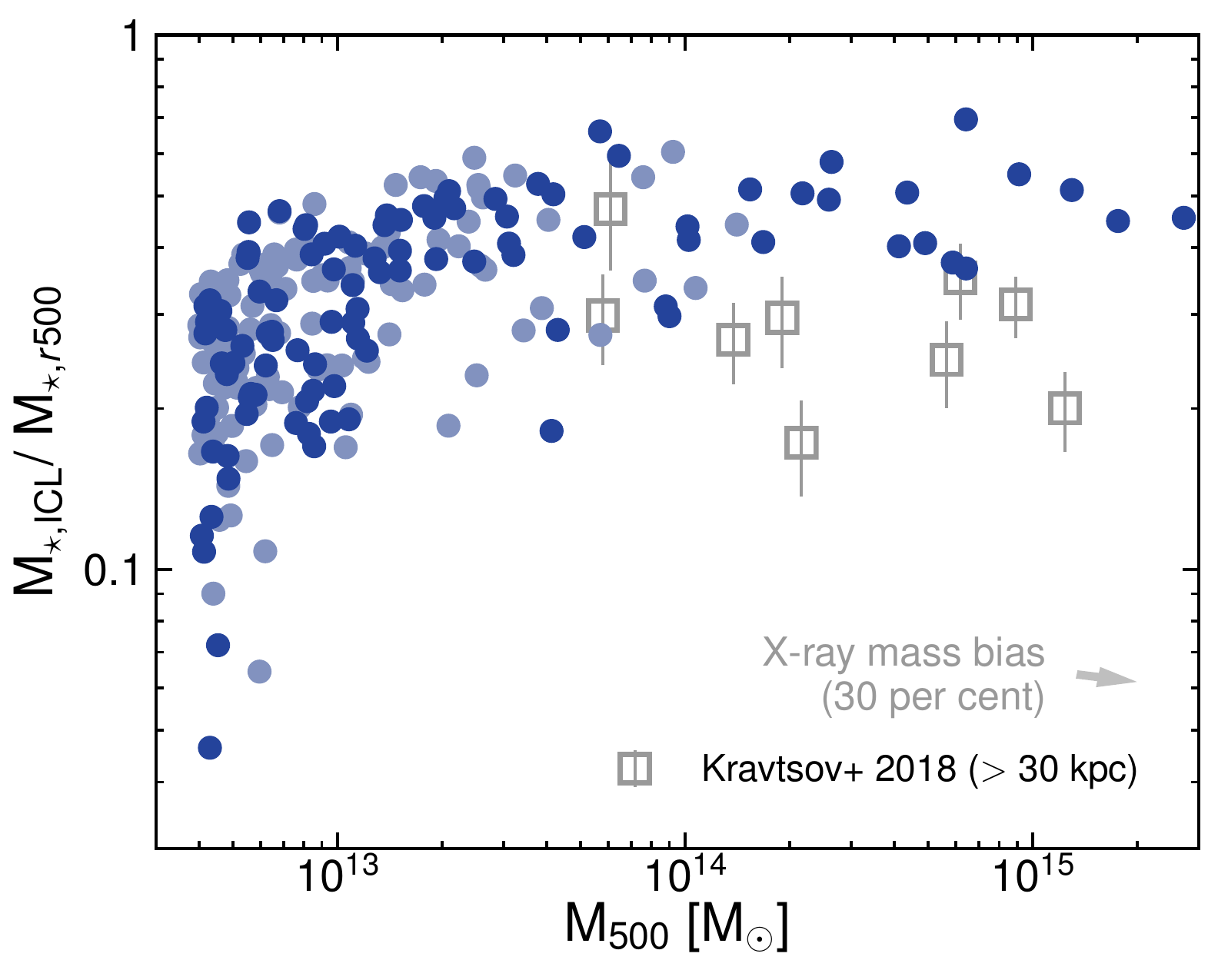}
  \includegraphics[width=0.975\columnwidth]{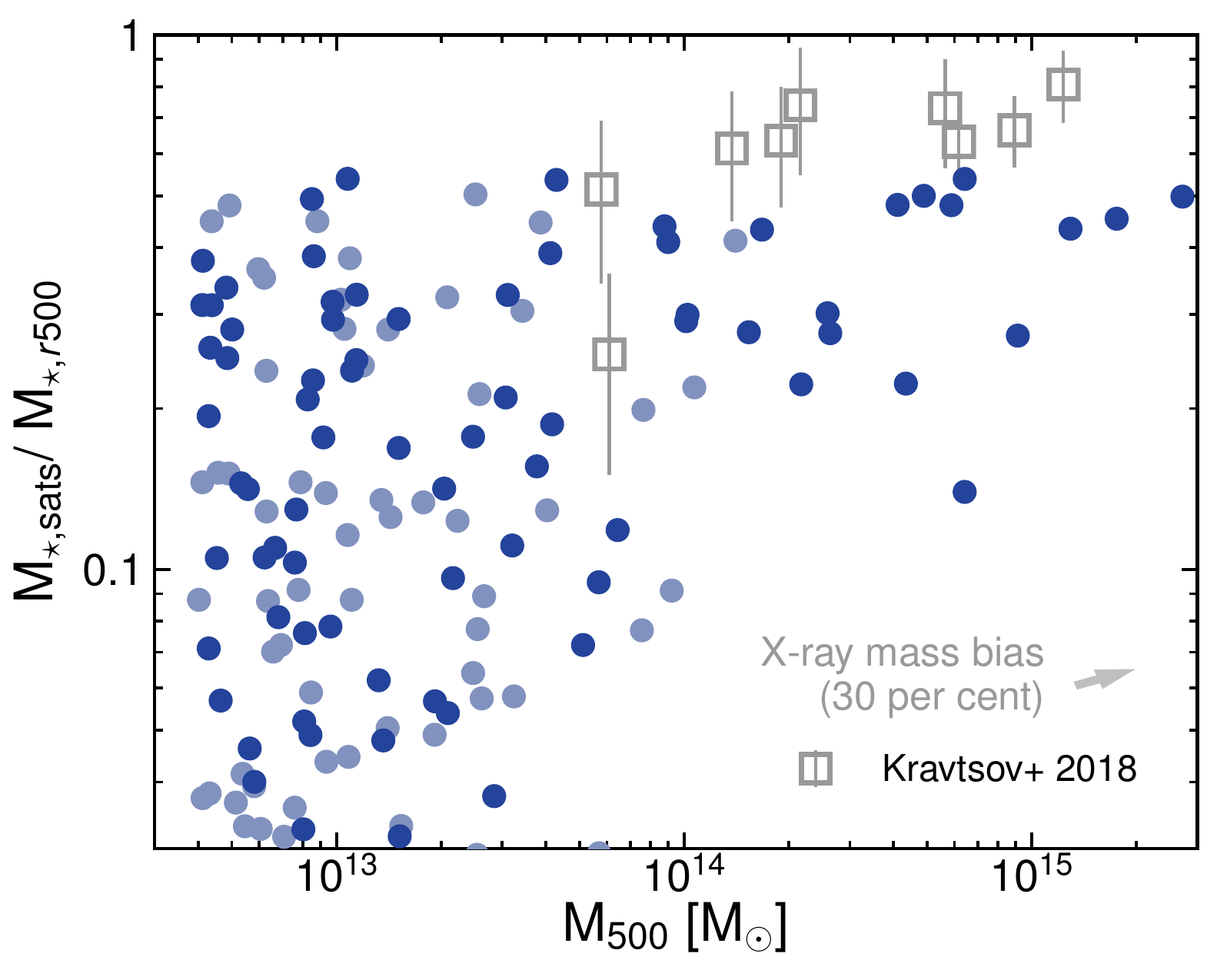}
  \caption{The stellar mass fraction in BCGs (top), the ICL (middle) and satellite galaxies (bottom) as a function of halo mass at $z=0$ compared with observational data from \protect\cite{Kravtsov2018}.
    Each quantity is shown as a fraction of the total stellar mass within $r_{500}$. BCG stellar mass is measured within $30$~pkpc and ICL stellar mass is counted outside this radius. Only stars within $r_{500}$ are counted in all cases.
    Grey arrows demonstrate how we would expect the \protect\cite{Kravtsov2018} data points to change if corrected for an X-ray hydrostatic mass bias in which X-ray masses underestimate the true mass by $30$ per cent.
  }
  \label{fig:stellar_mass_components_z0}
\end{figure}

\subsubsection{Stellar mass fractions halo mass dependence}\label{subsubsec:stellar_mass_components_z0}
Figure~\ref{fig:stellar_mass_components_z0} gives the fraction of the total stellar mass enclosed within $r_{500}$ in different stellar components as a function of halo mass at $z=0$.
We quantify: (1) the stellar mass in the BCG within a fixed spherical aperture of radius $30$ physical kiloparsecs (pkpc), (2) the stellar mass associated to the ICL, which is obtained by integrating from a fixed spherical aperture of radius $30$~pkpc out to $r_{500}$ (including all stars that are not bound to a satellite galaxy), and (3) the total stellar mass of satellite galaxies within $r_{500}$.
We compare our results to observational constraints from \cite{Kravtsov2018} (symbols with error bars).
\cite{Kravtsov2018} measure stellar mass profiles for their sample of nine BCGs and provide both `total' BCG stellar masses, which are obtained by integrating the best-fitting profiles to infinity, and BCG masses within $30$~pkpc (top panel). Subtracting the latter measurement from the former gives an estimate of the stellar mass in the ICL outside $30$~pkpc, as shown in the middle panel.
Since \cite{Kravtsov2018} employ X-ray hydrostatic halo mass estimates, we have estimated the change in the measured stellar mass that would occur if X-ray masses are biased low compared to the true halo mass by $30$ per cent. We find that, for \fable\ haloes with $M_{500} > 10^{14} M_{\odot}$ at $z=0$, correcting for such a bias leads to an increase in the measured ICL mass, satellite mass and total stellar mass of 3, 13 and 6 per cent in the median. We indicate the size of this effect via the grey arrows shown in each panel.

The fractional stellar mass in \fable\ central galaxies is in good agreement with the data (although in Section~\ref{subsec:BCG_mass_vs_mass} we show that the simulated BCGs are slightly too massive compared to observations).
For more massive haloes, the central galaxy represents an increasingly small fraction of the total stellar mass: for the most massive clusters with $M_{500} \gtrsim 10^{15} M_{\odot}$, only about $5$ ($15$) per cent of the total stellar mass lies within $30$ ($100$) pkpc of the cluster centres (top panel). Instead, for galaxy groups with $M_{500} \sim 5 \times 10^{13} M_{\odot}$ the inner $30$ ($100$) pkpc of their central galaxies account for approximately $30$ ($60$) per cent of the total stellar mass.
This halo mass dependence matches that of the \cite{Kravtsov2018} data and agrees qualitatively with the observations by \cite{Gonzalez2013} and \cite{Gozaliasl2018a}.

Conversely, the stellar mass situated outside the central regions has only a mild halo mass dependence, with more massive simulated haloes having a slightly higher ICL fraction on average.
The stellar mass bound in satellites becomes an increasingly important contributor to the total stellar mass for more massive objects.
For the richest \fable\ clusters, approximately half of their total stellar mass is bound in satellites galaxies, whereas for galaxy groups it is typically $\lesssim 20$ per cent, albeit with large scatter.
The increasing contribution of satellite galaxies to the total stellar mass in more massive haloes is understandable since, as host halo mass increases, the dynamical friction time for orbital decay of accreted galaxies increases as well. Hence, galaxies accreted onto more massive haloes survive longer before merging with the central galaxy or becoming tidally stripped, which would otherwise contribute stars to the BCG or ICL (see e.g. \citealt{Conroy2007}).

Whilst the observed halo mass trends of the ICL and satellite stellar mass fractions are broadly recovered, their exact values show some significant discrepancies compared to the data.
Whereas the large majority of the total stellar mass in observed clusters is locked up in satellite galaxies ($\approx 70$ per cent), satellites in our simulated clusters make up only $\sim 40$ per cent of the total stellar mass.
Given that the BCG stellar mass fractions are in reasonable agreement with the data, the `missing' satellite stellar mass seems to be found in the ICL, which accounts for a large fraction of the total stellar mass ($\sim 50$ per cent beyond $30$~pkpc), significantly higher than the \cite{Kravtsov2018} constraints ($\sim 25$ per cent beyond $30$~pkpc).
This is consistent with a picture in which the satellites are excessively stripped of their stars as they move through the cluster, either due to the cluster potential or interactions between galaxies.
We expect that this is a consequence of the sizes of our simulated galaxies, which are too large compared with observations. This is because the galaxy size--mass relation in \fable\ is relatively unchanged from the original Illustris galaxy formation model, which produced too-large galaxies for a given stellar mass (see Fig.~\ref{fig:size-mass} in Appendix~\ref{A:size-mass}).
Since satellites are preferentially stripped of stars on the outskirts of the galaxy where the gravitational binding strength is weakest (see e.g. \citealt{Puchwein2010}), more extended satellite galaxies of a given stellar mass will lose a larger fraction of their mass.
Hence, we expect that our too-large satellite galaxies lose too many of their stars by tidal stripping and that most of these stars are deposited into the ICL.

To shed some light on this issue we mimic the analysis of \textsc{c-eagle} clusters by \cite{Bahe2017} and compare the satellite galaxy stellar mass functions (SMFs) of our simulated clusters in different radial bins out to $5 \, r_{500}$, normalised to the halo mass within the respective volume (not shown).
We find a deficit of satellites at smaller cluster-centric radii across the full stellar mass range. This is consistent with our interpretation that enhanced tidal stripping and satellite disruption is responsible for the underestimate in total satellite stellar mass. The deficit is larger for low-mass or dwarf galaxies ($M_{\star} \sim 10^{8-10} M_{\odot}$) than for massive galaxies ($M_{\star} \sim 10^{10-11} M_{\odot}$) and is close to zero at the very high mass end ($M_{\star} \gtrsim 10^{11} M_{\odot}$). This is understandable since dwarf galaxies are more prone to disruption than massive galaxies.

\cite{Bahe2017} also find a deficit of low-mass galaxies ($M_{\star} \lesssim 10^{10} M_{\odot}$) within the virial radius of \textsc{c-eagle} clusters.
However, their deficit is significantly smaller than ours, likely due to the large difference in the sizes of \fable\ and \textsc{eagle} galaxies at fixed stellar mass ($\sim 0.4$ dex at $M_{\star} \sim 10^{10} M_{\odot}$; see Fig.~\ref{fig:size-mass}).
Currently the most viable solution to this problem is to calibrate the galaxy formation model -- in particular stellar feedback -- to reproduce the observed size--mass relation of galaxies, as was done for the \textsc{c-eagle} (\textsc{eagle}) model \citep{Crain2015} and IllustrisTNG \citep{Pillepich2018a}. We expect that this should reduce the degree of tidal stripping and dwarf disruption in our model and thus bring the stellar mass in satellites and the ICL into better agreement with the observations.

It is worth pointing out that observational estimates of the total stellar mass in the ICL show considerable variation, with some even approaching the fraction seen in our simulations.
Estimates of the ICL fraction range from $\approx 5-20$ per cent (e.g. \citealt{Krick2007, Burke2015, Montes2017, Jimenez-Teja2018}) to as much as $\approx 20-50$ per cent (e.g. \citealt{Gonzalez2007, Seigar2007, Zibetti2007, McGee2010, Gonzalez2013}).
Much of this variation can be attributed to the many varying definitions of ICL that have been employed by such studies. This is because the measured luminosity of the ICL depends sensitively on the method used to separate the light of the BCG from the diffuse ICL, which blend together smoothly in the outer regions of the BCG.
Moreover, the determination of the total luminosity of the ICL is notoriously difficult due to its diffuse nature, which requires very deep observations and careful consideration of contamination by foreground and background galaxies.
As such, it is possible that current observations underestimate the total light in ICL, which would somewhat alleviate the tension between the data and our simulations. In addition, the satellite and ICL fractions in our simulations may be sensitive to the details of the gravitational unbinding procedure that the \textsc{subfind} code uses to determine which particles are bound to a subhalo.

\subsubsection{Stellar mass fractions redshift dependence}
\label{subsubsec:stellar_mass_components_redshift}
\begin{figure}
  \centering
  \includegraphics[width=0.975\columnwidth]{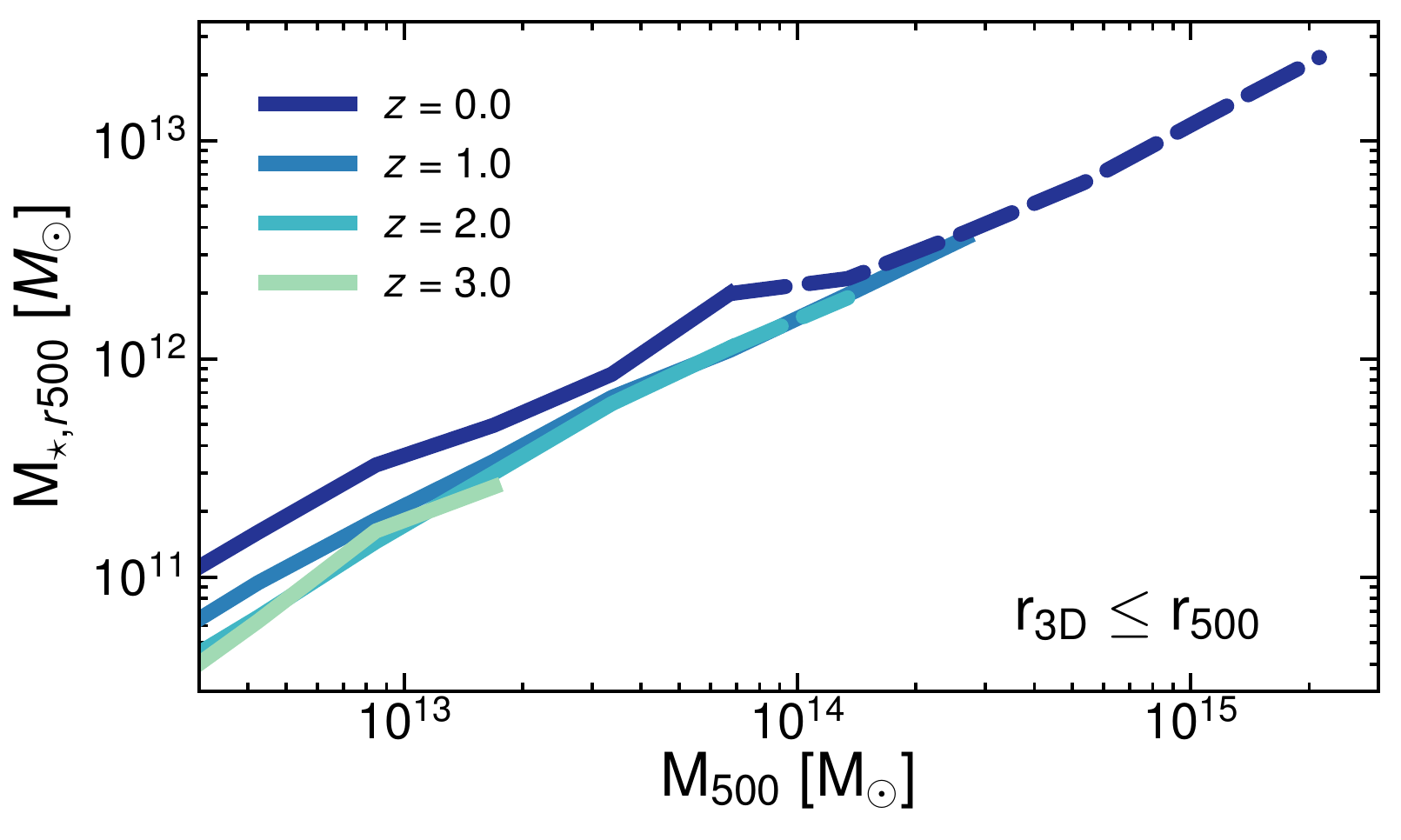}
  \includegraphics[width=0.975\columnwidth]{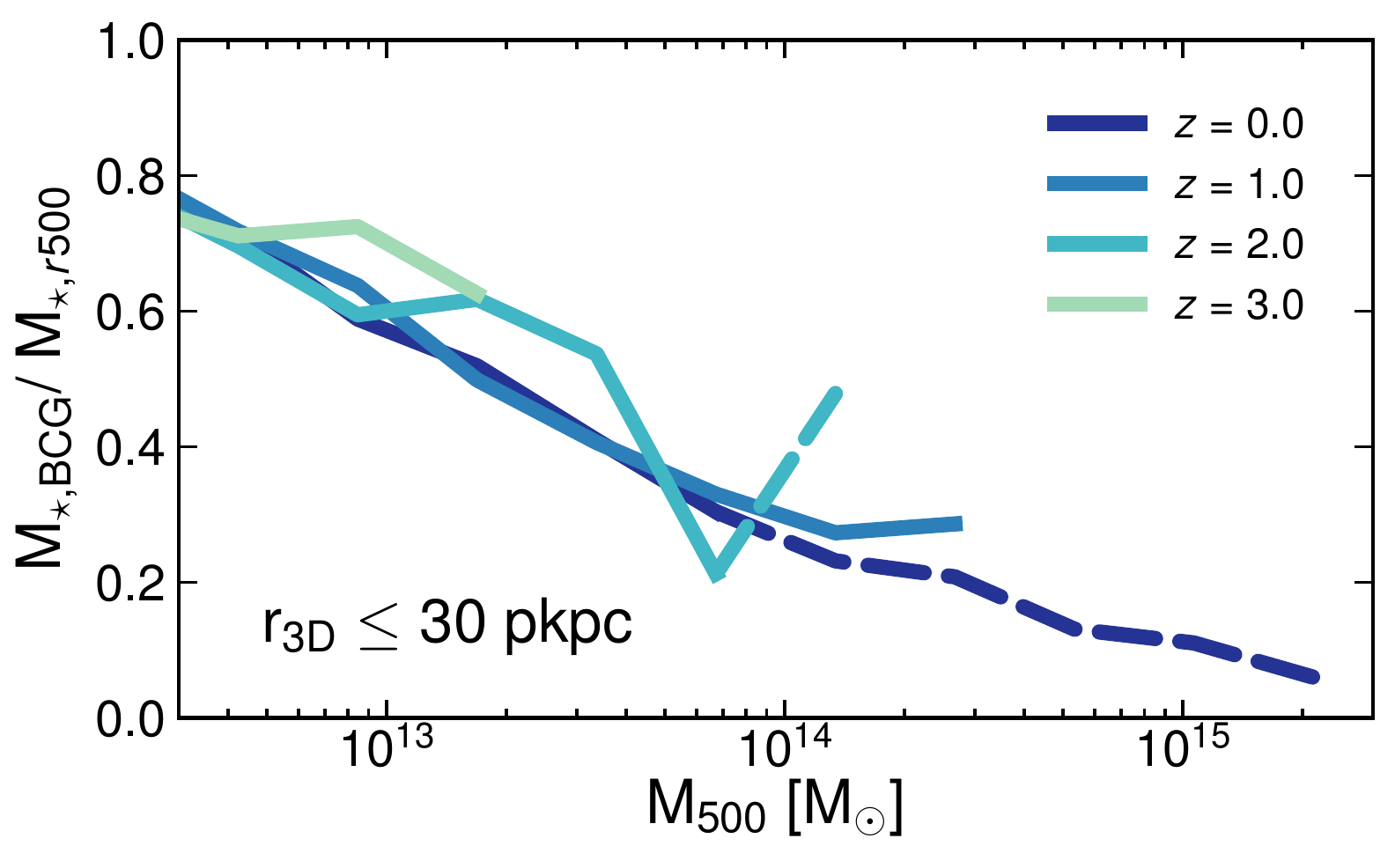}
  \includegraphics[width=0.975\columnwidth]{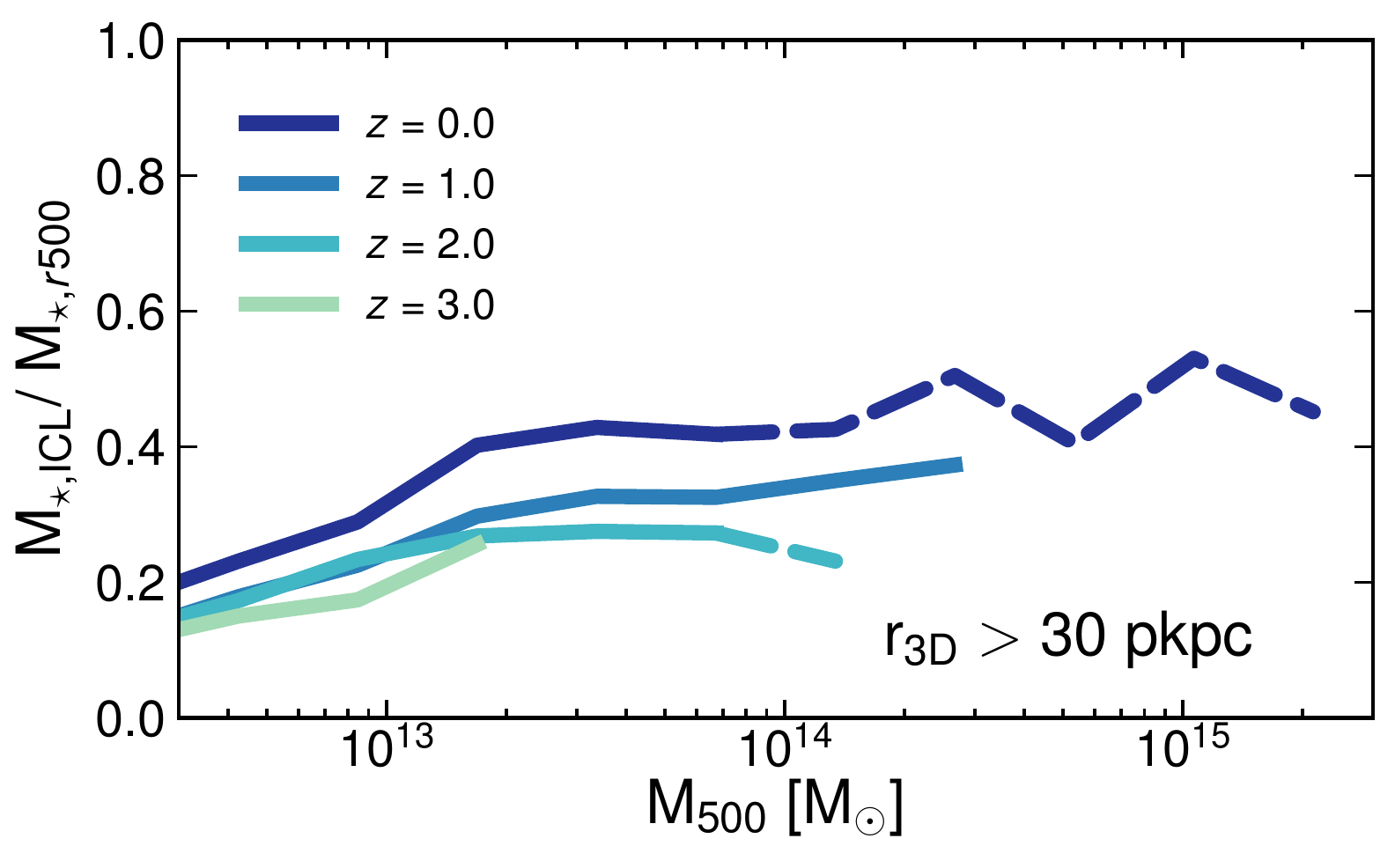}
  \includegraphics[width=0.975\columnwidth]{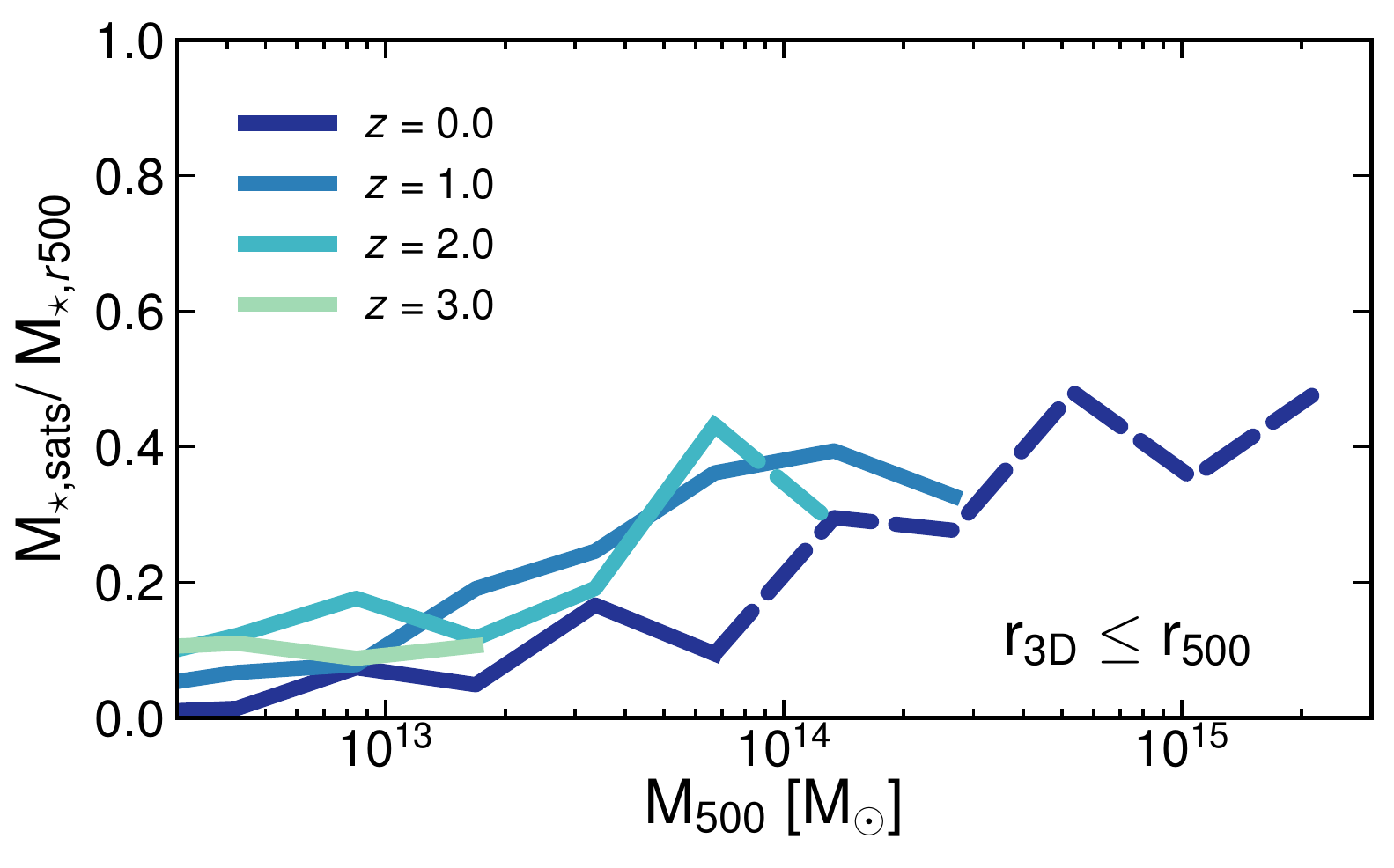}
  \caption{The redshift evolution of the total stellar mass within $r_{500}$ as a function of halo mass (top panel) and the fraction of this mass within the inner $30$~pkpc of the BCG (top-middle panel), the mass in the ICL beyond $30$~pkpc out to $r_{500}$ (bottom-middle panel), and the stellar mass bound in satellite galaxies within $r_{500}$ (bottom panel). Lines show the median in halo mass bins of width $0.3$~dex and become dashed when there are fewer than 10 objects per bin.
  }
  \label{fig:stellar_mass_components_redshift}
\end{figure}

Figure~\ref{fig:stellar_mass_components_redshift} shows the redshift evolution of the stellar mass to halo mass relation (top panel) and the stellar mass fractions in different stellar components up to $z=3$ (lower panels).

The total stellar mass at fixed halo mass remains invariant on cluster scales ($M_{500} \gtrsim 10^{14} M_{\odot}$) between $z=1$ and $z=0$, consistent with the results of Section~\ref{subsubsec:stellar_z}.
This implies that galaxy clusters build up their stellar mass at roughly the same pace as they assemble their total dark matter mass between $z \sim 1$ and $z=0$.
Conversely, on galaxy group scales the total stellar mass at fixed halo mass increases with decreasing redshift. On these scales, the BCG and ICL stellar mass fractions decrease and increase with decreasing redshift, respectively. This implies that BCG stellar growth at fixed halo mass is similar to, or slower than, the growth in total group stellar mass, whereas the opposite is true for the ICL mass, which grows more rapidly.
This result is in qualitative agreement with the predictions of IllustrisTNG \citep{Pillepich2018b} in addition to several other numerical works (e.g. \citealt{Rudick2011, Contini2014, Contini2018}).

Overall, the relative contribution of the various stellar components changes remarkably little from $z \sim 3$ to the present-day. In fact, the evolution of the relations is often comparable to, or smaller than, the differences between different mass definitions as seen in Fig.~\ref{fig:stellar_mass_components_z0}.
The strongest redshift evolution is that of the ICL stellar mass fraction, which increases significantly with decreasing redshift across the halo mass range probed.
The largest increase occurs between $z=1$ and $z=0$, which is consistent with a number of observational studies which find that the ICL is mainly formed at $z \lesssim 1$ (e.g. \citealt{Krick2007, Burke2012, Montes2014, Burke2015, Morishita2017}).
At the same time that the ICL fraction is increasing with cosmic time, the typical fraction of stars contained in satellite galaxies at fixed halo mass slightly decreases, albeit with large scatter about the median.
This supports the interpretation that a significant fraction of stars are tidally stripped from orbiting and incoming satellites and deposited in the ICL. As discussed above, tidal stripping and dwarf galaxy disruption may occur at an accelerated rate in our model and could explain the low satellite mass fraction and high ICL mass fraction compared to observations. The satellite mass fraction remains low even at high redshift, however we caution that we only count satellites within an evolving $r_{500}$ aperture and it is likely that satellites are influenced by the cluster environment (e.g. tidal stripping or star formation quenching) long before they reach this radius.

\section{Brightest Cluster Galaxies}\label{sec:BCGs}

\subsection{Observational data}
In recent years a number of observational studies have constructed large samples of BCGs spanning a wide range in mass and redshift with the aim of better understanding their growth history. We make use of several such studies in the following sections, specifically \cite{Lidman2012}, \cite{Bellstedt2016}, \cite{Zhang2016}, \cite{Kravtsov2018} and \cite{DeMaio2018}.

\cite{Lidman2012} and \cite{Bellstedt2016} each construct large samples of BCGs over a wide redshift range with new, published and archived data from various instruments.
The \cite{Lidman2012} sample contains 140 BCGs with host cluster mass estimates in the range $2.2 \times 10^{13} \, M_{\odot} \lesssim M_{500} \lesssim 3.7 \times 10^{15} \, M_{\odot}$ and redshifts $0.03 < z < 1.63$.
The \cite{Bellstedt2016} sample contains 132 BCGs (102 of which have host cluster mass estimates) spanning the redshift range $0.03 < z < 1.07$ and cluster mass range $1.3 \times 10^{14} \, M_{\odot} \lesssim M_{500} \lesssim 2.3 \times 10^{15} \, M_{\odot}$.
We have converted the cluster mass estimates reported in these studies from $M_{200}$ to $M_{500}$ by a factor of $0.72$ appropriate for a Navarro-Frenk-White profile \citep{Navarro1997} with a concentration parameter of $c=5$, which was performed in the reverse sense in \cite{Bellstedt2016}.
Masses are estimated from scaling relations between X-ray hydrostatic mass and X-ray luminosity, temperature or gas mass.
Both studies use the \texttt{MAG\_AUTO} magnitude estimate from {\sc SExtractor} to estimate total BCG luminosities and stellar masses.
We assume that the \texttt{MAG\_AUTO} aperture is comparable to a fixed $30$ pkpc radius aperture as suggested in \cite{Zhang2016} (see their appendix B4). We caution, however, that \texttt{MAG\_AUTO} uses an adaptively scaled aperture, which may introduce additional system-to-system scatter.
Individual uncertainties on their stellar mass measurements are not quoted but are expected to be on the order of $\approx 20$ per cent \citep{Bellstedt2016}.

\cite{Zhang2016} investigate BCG stellar mass growth using a sample of 106 X-ray selected groups and clusters at $0.07 < z < 1.26$ with deep Dark Energy Survey Science Verification (DES SV) data \citep{Sanchez2014}. Cluster masses are estimated from the weak lensing calibrated mass--temperature relation of \cite{Kettula2013a}. In the following we assume that their 32 pkpc radius aperture is equivalent to a 30 pkpc aperture. We convert $M_{200}$ to $M_{500}$ by the factor $0.72$.
\cite{Kravtsov2018} analyse SDSS data for nine nearby ($z < 0.1$) clusters. Three of the nine clusters have direct X-ray hydrostatic mass measurements. Cluster masses for the remaining six are estimated from the mass proxy $Y_{\mathrm{X}}$ \citep{Kravtsov2006} using the scaling relation derived in \cite{Vikhlinin2009a} based on X-ray hydrostatic masses. \cite{Kravtsov2018} do not quote individual errors on their stellar mass measurements, although the typical error due to uncertainties in the estimate of the background is small ($\lesssim 10$ per cent).
\cite{DeMaio2018} study 23 galaxy groups and clusters at $0.29 \leq z \leq 0.89$ with \textit{Hubble Space Telescope} imaging. Cluster masses are estimated from a scaling relation between X-ray temperature and X-ray hydrostatic mass \citep{Vikhlinin2009a}.

All of these studies assume the \cite{Chabrier2003} IMF in the conversion of luminosities to stellar masses. The same IMF was assumed in the \fable\ galaxy formation model.

\begin{figure*}
  \includegraphics[width=0.497\textwidth]{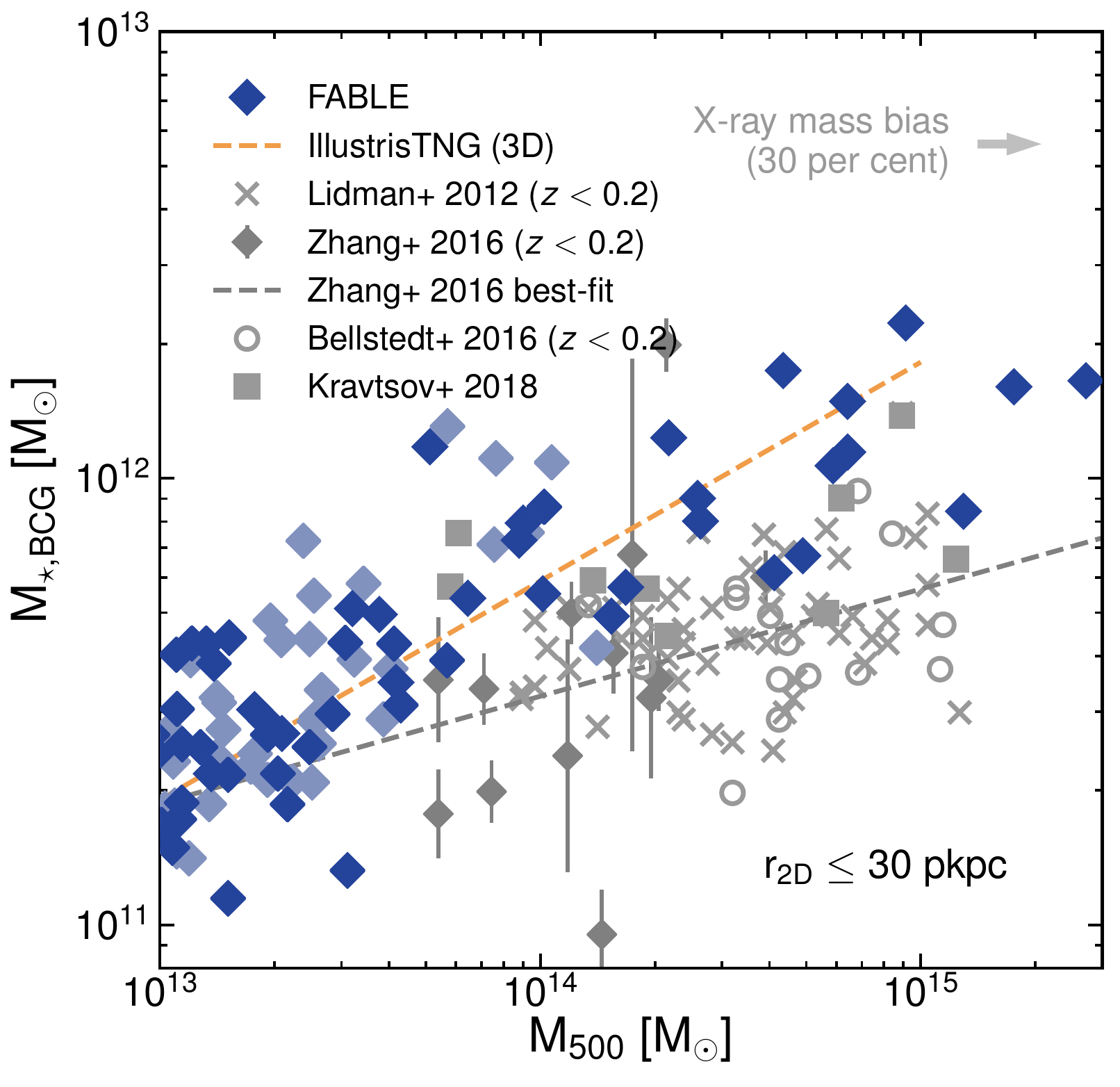}
  \includegraphics[width=0.497\textwidth]{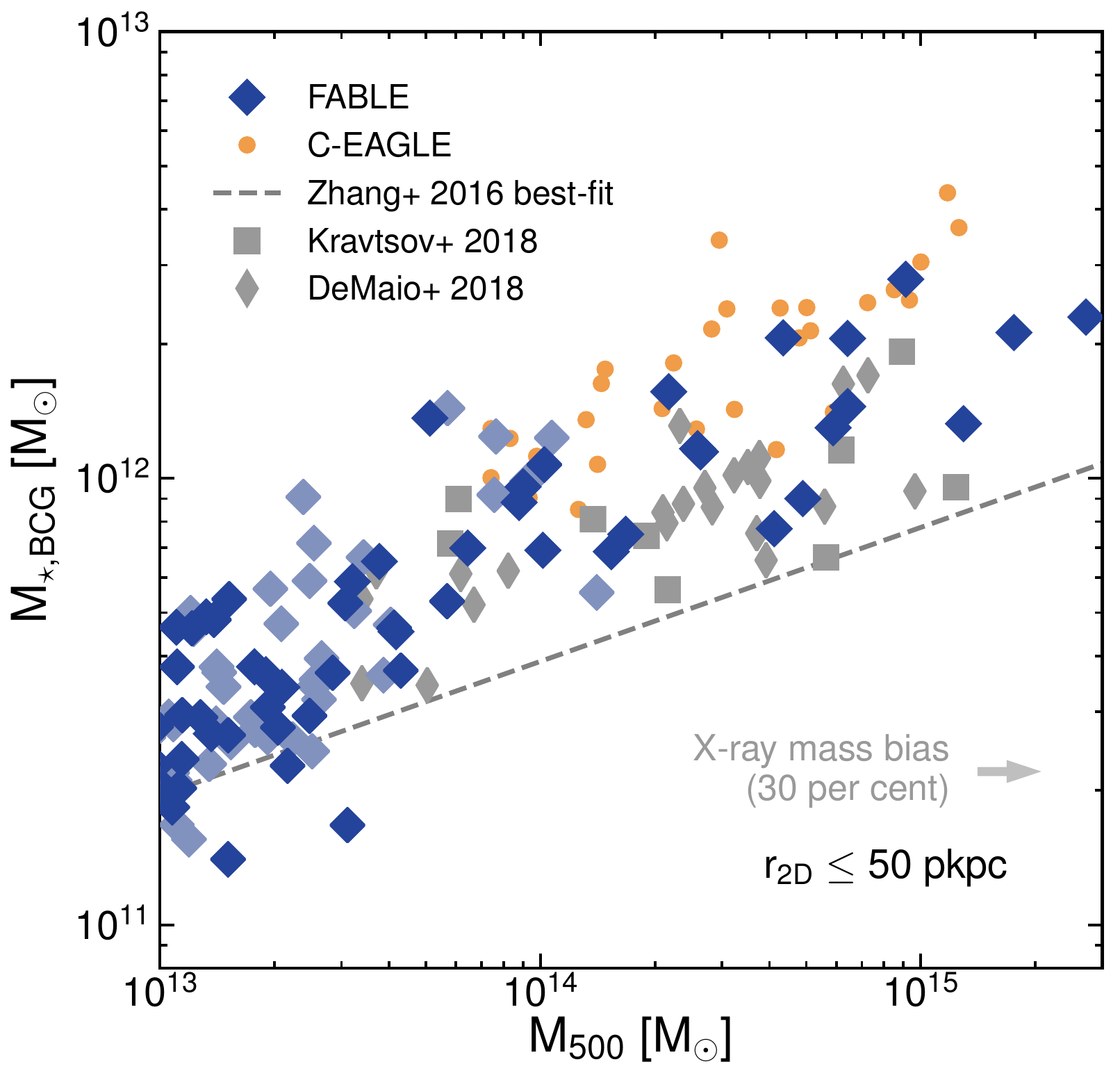}
  \caption{The stellar mass of \fable\ BCGs at $z=0$ (blue diamonds) measured within a projected aperture of radius 30~pkpc (left) and 50~pkpc (right) as a function of halo mass in comparison to observational data (grey points), C-EAGLE clusters (orange circles) and IllustrisTNG (orange dashed line). The best-fitting relation from IllustrisTNG is based on stellar masses measured within a spherical rather than a circular aperture. For clarity we neglect to show error bars on the observed halo mass estimates. Grey arrows demonstrate the expected change to the observational data points if corrected for a possible X-ray hydrostatic mass bias in which X-ray masses underestimate the true mass by $30$ per cent.
  }
  \label{fig:BCG_mass_vs_mass}
\end{figure*}

\subsection{BCG stellar mass to halo mass relation}\label{subsec:BCG_mass_vs_mass}
Figure~\ref{fig:BCG_mass_vs_mass} shows the stellar mass of \fable\ BCGs as a function of their host cluster mass at $z=0$ in comparison to data from the aforementioned studies, as well as the recent \textsc{c-eagle} and IllustrisTNG simulations. We calculate stellar masses within a 2D radius of 30 pkpc (left-hand panel) and 50 pkpc (right-hand panel) integrated through the entire simulation volume, which mimics observed BCG luminosities measured within a circular aperture.
For the \cite{Lidman2012}, \cite{Bellstedt2016} and \cite{Zhang2016} samples we restrict our comparison to low-redshift BCGs at $z < 0.2$.\footnote{For \cite{Zhang2016} individual stellar mass measurements are available only for the 32 pkpc aperture.}
We also plot the best-fitting BCG mass--halo mass relation derived by \cite{Zhang2016} for both apertures (dashed line).
We compare to the full sample of BCGs from \cite{DeMaio2018} but caution that these are situated at somewhat higher redshifts ($0.29 \leq z \leq 0.89$).

The mean \fable\ relation is systematically higher than the observed relations by $\sim 0.2-0.3$ dex, although there is some overlap in the scatter.
Still, it is worth pointing out that the \fable\ BCGs are much closer to the data than many simulations have been in the past (see e.g. \citealt{Puchwein2010}, \citealt{Ragone-Figueroa2013} and \citealt{Martizzi2014}).
The discrepancy with the observations is smaller in comparison to the \cite{Kravtsov2018} and \cite{DeMaio2018} constraints than those of \cite{Lidman2012}, \cite{Bellstedt2016} and \cite{Zhang2016}. The difference between these studies is most likely due to different assumptions about the mass-to-light ratio, which is the dominant source of uncertainty in the BCG stellar masses.

The high stellar masses of \fable\ BCGs may reflect a slight overestimate in the abundance of massive field galaxies ($M_{\star} \gtrsim 10^{11} M_{\odot}$) relative to observations at $z \approx 0$ (see fig.~2 in Paper I).
Part of the discrepancy may also result from a possible excess of in-situ star formation in the simulated BCGs, as discussed in Section~\ref{subsec:BCG_SFR}.
In both cases, further improvements to the modelling of AGN feedback may aid in reducing the stellar masses of the simulated BCGs.
In addition, one of the dominant mechanisms by which BCGs are expected to grow in mass is via ``galactic cannibalism''. This describes the process in which smaller satellite galaxies sink towards the cluster centre via dynamical friction, eventually merging with the BCG. Along the way, a significant mass of stars can accumulate in the ICL or BCG via tidal stripping. As discussed in Section~\ref{subsec:stellar_mass_components} (see also Appendix~\ref{A:size-mass}), our results suggest that this process occurs at too high a rate in \fable\ clusters. It is, however, unclear how many of these stars end up in the BCG (e.g. within 30 pkpc) rather than in the ICL. Furthermore, for artificially boosting the BCG mass, they would need to do so on a timescale that is significantly shorter than the merger timescale of their former host galaxy.

The IllustrisTNG and \textsc{c-eagle} clusters (orange dashed line and orange circles in Fig.~\ref{fig:BCG_mass_vs_mass}) also possess significantly more massive BCGs than observed despite their more realistic satellite galaxy sizes. For \textsc{c-eagle}, \cite{Bahe2017} find that most of the stellar mass in BCGs is already in place at $z \sim 1$, which implies that further work is needed in modelling their high-redshift progenitors.
We explore this possibility in our own simulations in Section~\ref{subsubsec:BCG_mass_vs_z} where we study the redshift evolution of BCG stellar mass.

\begin{figure*}
  \includegraphics[width=0.994\textwidth]{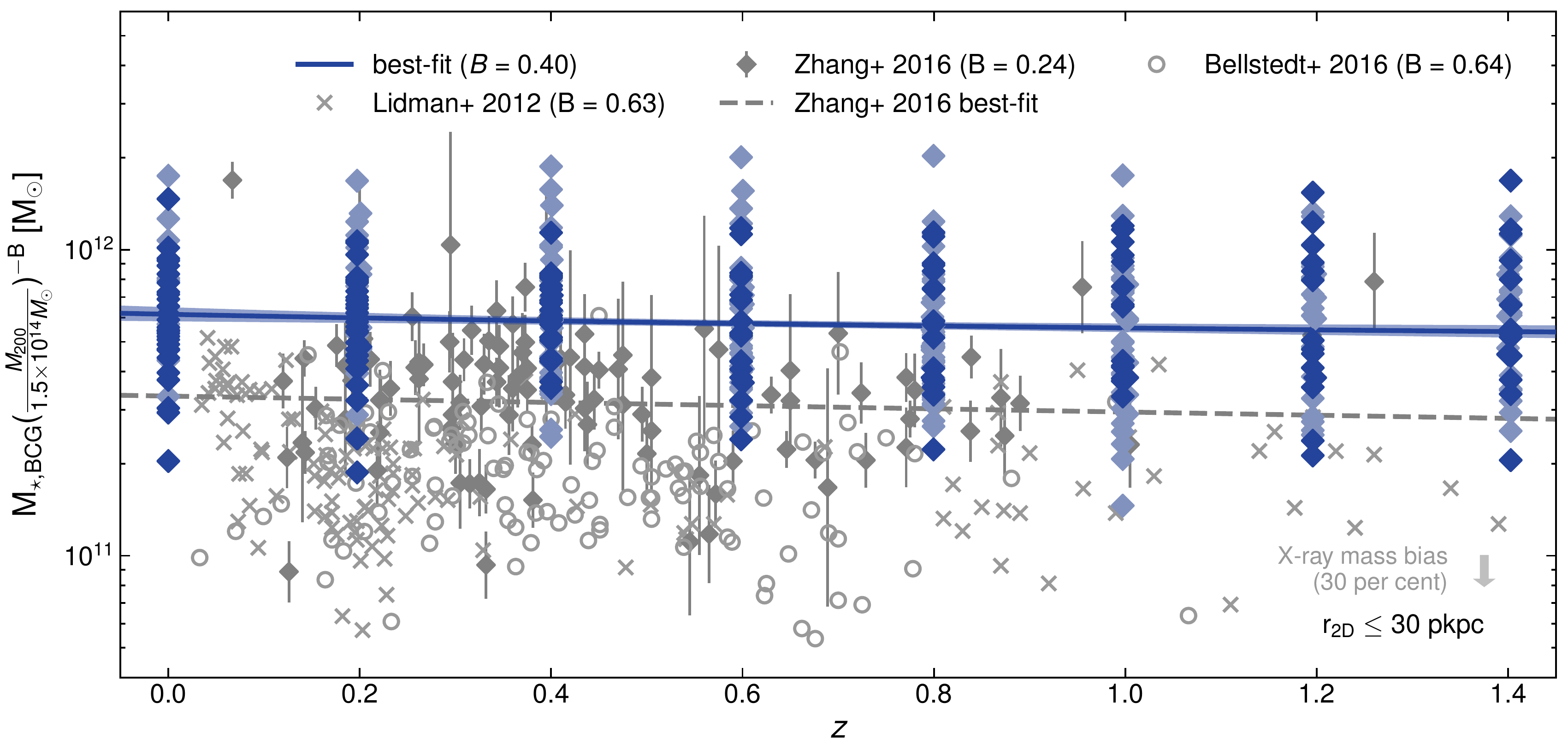}
  \caption{The stellar mass of \fable\ BCGs (blue diamonds) normalised to a halo mass of $M_{200} = 1.5 \times 10^{14} M_{\odot}$ as a function of redshift compared to observational data from \protect\cite{Lidman2012} (grey crosses), \protect\cite{Zhang2016} (grey diamonds) and \protect\cite{Bellstedt2016} (grey circles). We remove the halo mass dependence of the BCG stellar mass using the best-fitting power law index of each sample as indicated in the legend.
  We select BCGs from \fable\ haloes with $M_{200} > 5 \times 10^{13} M_{\odot}$ at each redshift to approximately match the \protect\cite{Zhang2016} sample selection.
  The blue solid line shows the redshift trend of the best-fitting BCG--cluster mass relation as calculated from the posterior median of the parameters. Shaded regions enclose the $16$th to $84$th percentile range of the posterior samples at each redshift. The grey dashed line shows the redshift trend of the best-fitting BCG--cluster mass relation from \protect\cite{Zhang2016}.
  The BCG stellar mass is measured within a projected aperture of radius 30~pkpc to approximately match the 32 pkpc radius aperture of \protect\cite{Zhang2016} and the \texttt{MAG\_AUTO} apertures of \protect\cite{Lidman2012} and \protect\cite{Bellstedt2016} (see text).
    All stellar masses are based on a \protect\cite{Chabrier2003} IMF.
    The grey arrow in the lower right shows the expected change to the observational data points if corrected for a possible X-ray hydrostatic mass bias in which X-ray masses underestimate the true mass by $30$ per cent, assuming a halo mass slope of $B=0.64$.
  }
  \label{fig:BCG_mass_vs_z}
\end{figure*}

Despite the offset in the normalisation, the slope of the simulated relation appears to be in good agreement with observations in the cluster regime.
To make a quantitative comparison we fit a power law relation to \fable\ haloes with $M_{500} > 6 \times 10^{13} M_{\odot}$, which is the approximate lower mass limit of the data at $z < 0.2$. This yields best-fitting slopes of $0.27 \pm 0.07$ and $0.30 \pm 0.05$ for the $30$ and $50$ pkpc apertures, respectively.
These are in very good agreement with the values of $0.24 \pm 0.08$ and $0.30 \pm 0.08$ derived in \cite{Zhang2016} for their full sample.
\cite{Kravtsov2018} measure a slope of $0.39 \pm 0.17$ using `total' BCG stellar masses derived from triple S\'ersic fits to the light profiles, which is also consistent with our results.
On the other hand, \cite{Bellstedt2016} and \cite{Lidman2012} derive significantly steeper slopes of $0.64 \pm 0.03$ and $\approx 0.63$, respectively. This difference may be due to (possibly redshift-dependent) selection effects, as noted in \cite{Lidman2012}. It may also indicate a redshift-dependent slope, however we lack a sufficient number of high-redshift clusters in order to constrain this possibility from the simulations.

The stellar mass of \fable\ BCGs has substantial scatter at fixed halo mass, at a similar level to the observations. The intrinsic scatter about the best-fitting relation is $0.15^{+0.02}_{-0.01}$ dex and $0.14^{+0.02}_{-0.01}$ dex for the 30 pkpc and 50 pkpc apertures, respectively, which is comparable to the values of $0.18 \pm 0.02$ and $0.19 \pm 0.02$ derived in \cite{Zhang2016} and the scatter of $0.21 \pm 0.09$ measured in \cite{Kravtsov2018} for their `total' BCG masses.

\subsection{BCG stellar mass evolution}\label{subsec:BCG_mass_evol}

\subsubsection{BCG stellar mass evolution at fixed cluster mass}\label{subsubsec:BCG_mass_vs_z}

In Fig.~\ref{fig:BCG_mass_vs_z} we plot the stellar mass of \fable\ BCGs at fixed halo mass as a function of redshift compared to data from \cite{Zhang2016}, \cite{Lidman2012} and \cite{Bellstedt2016}.
As motivated in Section~\ref{subsec:BCG_mass_vs_mass}, we compare to these data using BCG stellar masses measured within a 30~pkpc radius aperture.
We remove the halo mass dependence of the BCG stellar mass by modelling a redshift-dependent BCG--cluster mass relation. Following \cite{Zhang2016}, we adopt the form of Equation~\ref{eq:scaling_relation} using $M_{200}$ in place of $M_{500}$ and with pivot mass $M_{\mathrm{piv}} = 1.5 \times 10^{14} M_{\odot}$ and redshift $z_{\mathrm{piv}} = 0$.
We constrain the best-fitting parameters using \textsc{emcee} via the approach described in Section~\ref{subsec:ICM_stellar_mass_z}.
We adopt flat priors of $\mathrm{log}_{10} A$ in $(10, 13)$, $B$ in $(-0.5, 0.8)$, $C$ in $(-2, 2)$ and intrinsic scatter in $(10^{-3}, 1)$, as used in \cite{Zhang2016}.
For our comparison sample (blue diamonds in Fig.~\ref{fig:BCG_mass_vs_z}) we include all \fable\ BCGs in clusters with $M_{200} > 5 \times 10^{13} \, M_{\odot}$ at each redshift, which is the approximate lower halo mass limit of the observational sample.
The best-fitting parameters for this sample are $\mathrm{log}_{10} A = 11.79 \pm 0.02$, $B = 0.40 \pm 0.03$ and $C = -0.15 \pm 0.10$ with an intrinsic scatter of $0.208 \pm 0.008$ dex.
In Fig.~\ref{fig:BCG_mass_vs_z} we factor out the halo mass dependence to highlight the redshift trend of the BCG stellar mass at the pivot mass.
For the \cite{Lidman2012} and \cite{Bellstedt2016} data we adopt the halo mass dependence from their best-fitting BCG--cluster mass relations ($B = 0.63$ and $B = 0.64$, respectively).

The best-fitting redshift trend of the \fable\ BCG--cluster mass relation ($C = -0.15 \pm 0.10$; solid line) is similar to that of \cite{Zhang2016} ($C = -0.19 \pm 0.34$; dashed line). This implies that the average stellar mass of simulated and observed BCGs in clusters of fixed mass increases only mildly with decreasing redshift (e.g. an increase of less than $15$ per cent from $z=1$ to $z=0$).
\cite{Zhang2016} notice a stronger redshift evolution ($C = -0.62 \pm 0.34$) when restricting their sample to systems with $M_{200} \gtrsim 7 \times 10^{13} M_{\odot}$. However, they note that this shift may be caused by the inaccuracy of their X-ray observable--mass scaling relations at the low-mass end.
We find no evidence for such a dependence in our simulated sample. For example, including only \fable\ clusters with $M_{200} > 10^{14} M_{\odot}$ leads to a similarly shallow redshift trend ($C = -0.01^{+0.12}_{-0.13}$) with a similar halo mass dependence ($B = 0.38 \pm 0.04$).

The normalisation of the \fable\ BCG--cluster mass relation is systematically high compared to the data by $\sim 0.2-0.3$~dex, similar to the offset found at $z = 0$ (see Fig.~\ref{fig:BCG_mass_vs_mass}). This offset is exacerbated if the X-ray hydrostatic mass estimates are indeed biased low, as demonstrated by the grey arrow in the lower right corner of the figure, which shows the expected decrease in the normalisation of the observed relations if X-ray masses underestimate the true mass by $30$ per cent. The fact that an offset persists to high redshift suggests that the high masses of \fable\ BCGs at $z = 0$ are largely due to stellar mass build-up at $z \gtrsim 1$. A similar conclusion was reached by \cite{Bahe2017} for BCGs in the \textsc{c-eagle} simulations.
It remains unclear whether this results from an excess of in-situ star formation in the main progenitors or an excess of stellar mass in accreted galaxies at high-redshift.
We investigate the level of in-situ star formation in the simulated BCGs in Section~\ref{subsec:BCG_SFR}. We will show that, whilst the BCGs may be forming too many stars at low redshift ($z \lesssim 0.3$), the median SFR increases with increasing redshift in good agreement with the redshift evolution observed by \cite{McDonald2016} and \cite{Bonaventura2017} at $0.25 < z < 1.25$ and $0 < z < 1.8$, respectively, with little evidence of an increase at higher redshifts.
Furthermore, it is unlikely that BCGs are accreting overly-massive satellite galaxies given that the galaxy stellar mass function is in good agreement with observations of field galaxies at $z > 1$ (see fig. 3 in Paper I). The exception is an apparent surplus of galaxies at $M_{\star} \lesssim 10^{10} M_{\odot}$, however these are expected to contribute only a small fraction ($\lesssim 20$ per cent) of the final stellar mass of the BCG (see e.g. fig. 4 of \citealt{DeLucia2007}).\\

\begin{figure*}
  \includegraphics[width=0.497\textwidth]{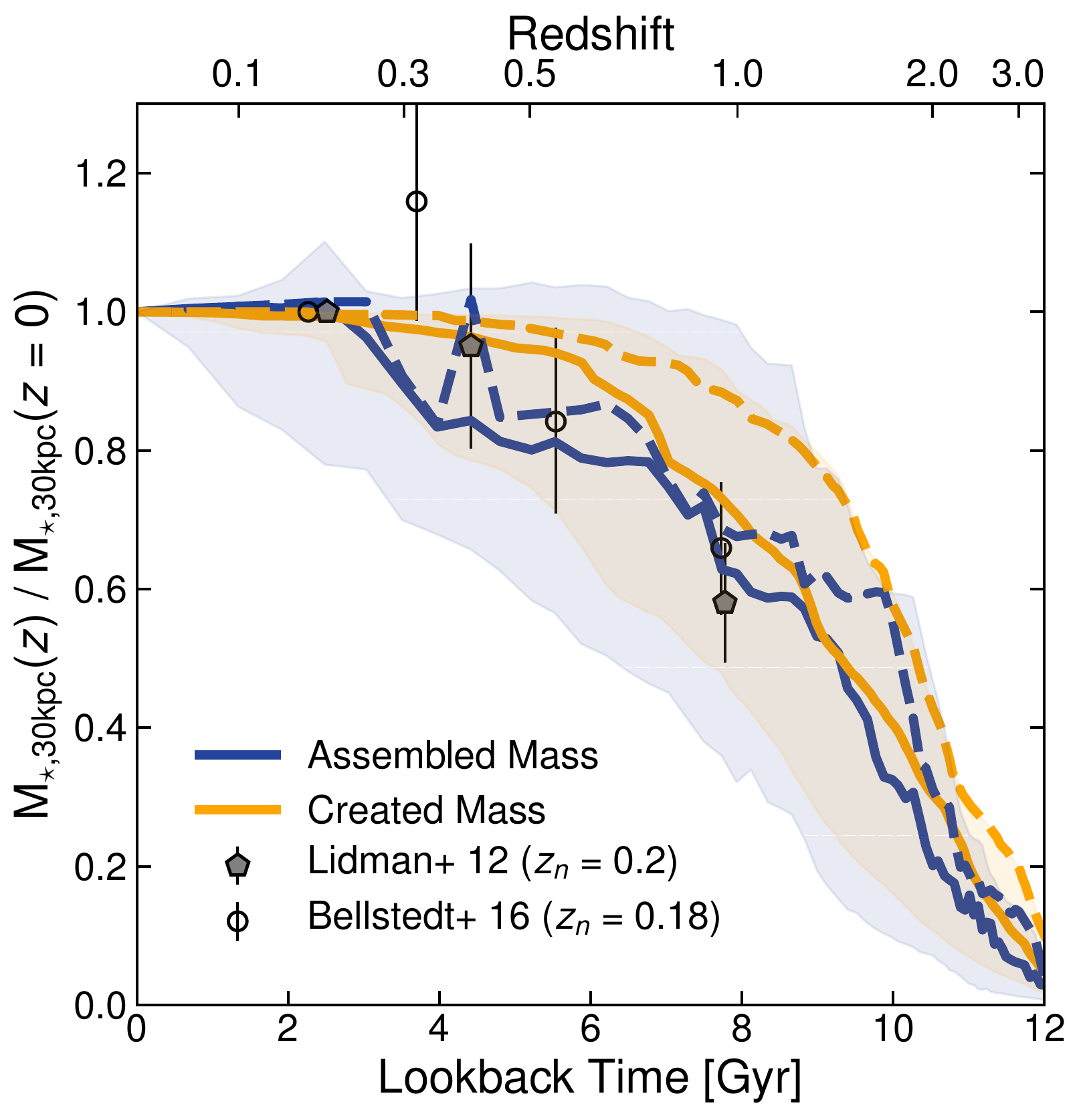}
  \includegraphics[width=0.497\textwidth]{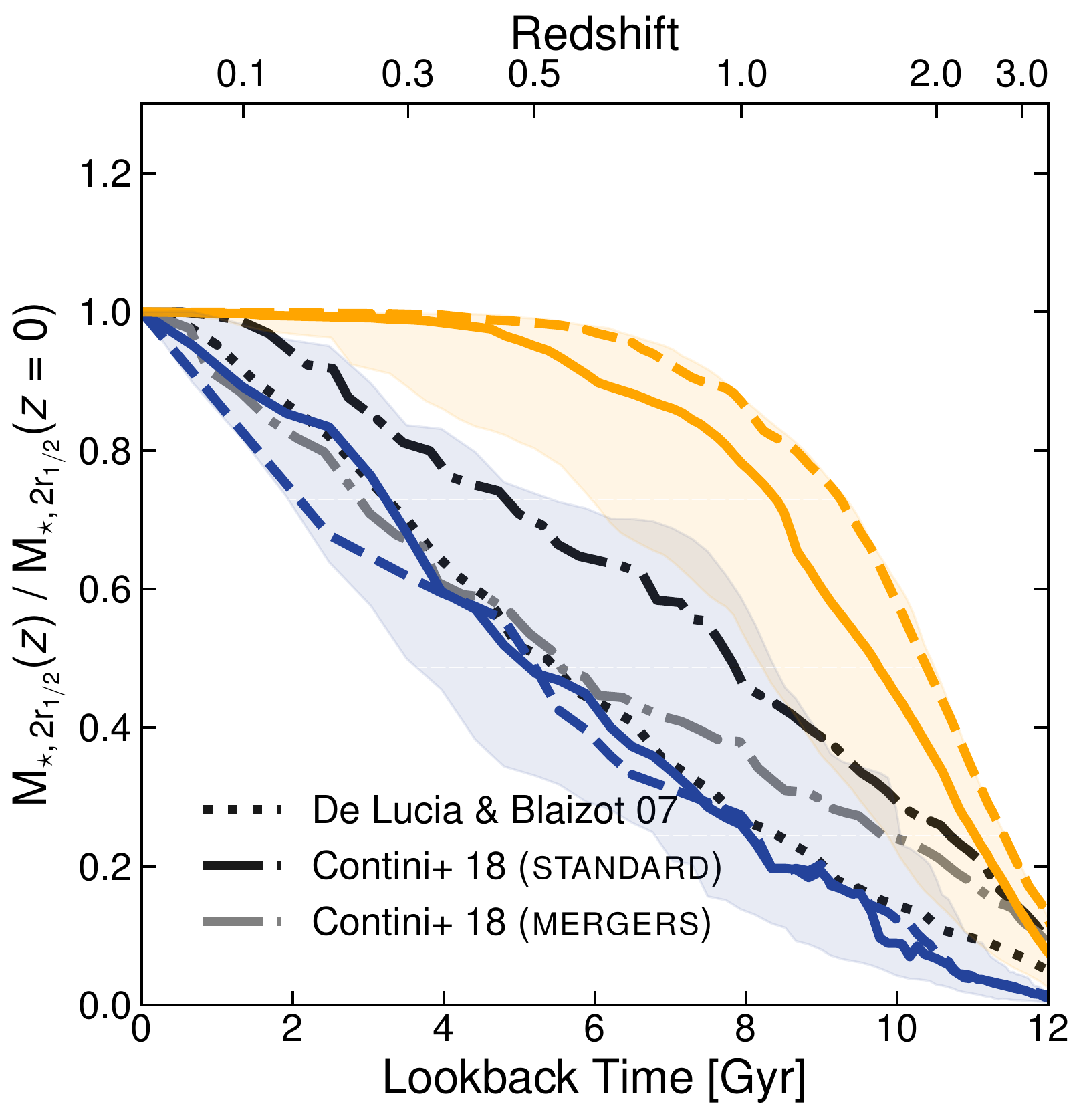}
  \caption{The median assembly and creation histories of the main progenitors of a sample of BCGs located in \fable\ haloes with $M_{200} > 10^{14} M_{\odot}$ ($23$ BCGs; solid lines) and $M_{200} > 5 \times 10^{14} M_{\odot}$ ($10$ BCGs; dashed lines) at $z=0$ normalised to the BCG mass at $z=0$. In the left-hand panel the BCG stellar mass is measured within a fixed $30$ pkpc radius aperture, while in the right-hand panel it is the mass within twice the stellar half-mass radius. The shaded regions show the $16$th to $84$th percentile range of the distribution at each redshift for the lower mass sample.
    The `assembled' mass curve shows the stellar mass of the main progenitor within the aperture at each point in time relative to its mass at $z=0$. The `created' mass curve shows the mass fraction of stars located within the aperture at $z=0$ that have already formed somewhere in the simulation at each point in the galaxy's history.
    Error bars represent observational constraints on BCG mass growth from \protect\cite{Bellstedt2016} and \protect\cite{Lidman2012} normalised to their lowest redshift bins ($z=0.18$ and $z=0.2$, respectively). These data should be compared to the assembled mass.
    The dotted and dashed-dotted lines in the right-hand panel shows the median assembly history predicted by the semi-analytic models of \protect\cite{DeLucia2007} and \protect\cite{Contini2018}.
  }
  \label{fig:BCG_mass_growth}
\end{figure*}

\subsubsection{Progenitor assembly and creation history}
In the previous section we found that the BCG stellar mass at fixed halo mass increases only mildly with decreasing redshift at $z \lesssim 1.5$.
In this section we study the history of stellar mass growth in individual BCGs, including when the stars were first created and when they were finally assembled into the BCG.
To do this we construct merger trees using the \textsc{sublink} code \citep{Rodriguez-Gomez2015}, which finds the descendants and progenitors of subhaloes based on their baryonic content. In the following we track each BCG backwards through cosmic time along their main progenitor branch, which is the branch that accounts for most of the mass of the final system for the longest period following the method described in \citealt{DeLucia2007}.

Figure~\ref{fig:BCG_mass_growth} shows the median `assembled' and `created' mass histories of our simulated BCGs (blue and orange lines, respectively) as measured within spherical radii of $30$ pkpc (left-hand panel) and twice the stellar half-mass radius (right-hand panel).
The `assembled' mass history refers to the stellar mass of the main progenitor at each redshift within the given aperture, while the `created' mass history describes the redshift at which the star particles that belong to the BCG at $z=0$ were originally created. Specifically, the `created' mass at a given redshift is the mass of stars that have already formed somewhere in the simulation and that are due to end up in the specified aperture at $z=0$.
We consider the main progenitor branches of two samples of BCGs: those in haloes with $M_{200} > 10^{14} M_{\odot}$ at $z=0$ ($23$ BCGs; solid lines), and those in more massive haloes with $M_{200} > 5 \times 10^{14} M_{\odot}$ at $z=0$ ($10$ BCGs; dashed lines).
Note that we have normalised the assembled and created mass to the final BCG mass at $z=0$ to reduce the scatter due to the wide mass range of our sample.

In the left-hand panel of Figure~\ref{fig:BCG_mass_growth} we compare the assembled mass within $30$~pkpc to observational inferences of BCG mass growth from \cite{Lidman2012} and \cite{Bellstedt2016}.
These studies estimate the stellar mass growth of BCGs by comparing the median mass of BCGs in pairs of redshift bins. The sub-samples of clusters in each bin were chosen to have a matching distribution of cluster masses, which were extrapolated to $z=0$ using mean accretion rates derived from simulations. As such, the cluster samples are chosen to approximate an evolutionary sequence as closely as possible.
The data points in Fig.~\ref{fig:BCG_mass_growth} show the median BCG stellar mass in each redshift bin relative to their lowest redshift bin. Error bars indicate the uncertainties on these mass ratios. The data should be compared to the median assembled BCG mass in the simulations (blue lines).

Within $30$~pkpc, the simulated BCGs show little mass growth at low redshift below $z \sim 0.5$, consistent with the \cite{Lidman2012} and \cite{Bellstedt2016} results, among others (e.g. \citealt{Lin2013, Inagaki2015}). In fact, mass growth in the simulated BCGs appears to halt below $z \sim 0.3$, in agreement with \cite{Oliva-Altamirano2014} and \cite{Cerulo2019} who find little to no stellar mass growth at $z \lesssim 0.35$.
We find a similar lack of growth at $z < 0.3$ when considering the stellar mass within larger apertures of $50$ or $100$~pkpc, in qualitative agreement with recent observational results from \cite{DeMaio2019}. In contrast, the right-hand panel of Fig.~\ref{fig:BCG_mass_growth} shows that the stellar mass within twice the stellar half-mass radius continues to grow up to the present day. This implies that most of the stars that reach the cluster core below $z=0.3$ end up in the outskirts of the BCG ($\gtrsim 100$~pkpc), potentially contributing to the ICL. This interpretation is consistent with the findings of \cite{Burke2015} and \cite{Morishita2017}, who show that the ICL grows substantially below a redshift of $z \sim 0.4$.

The observations suggest a moderate BCG mass growth below $z \sim 1$ that is reproduced well by the simulations.
Between $z \sim 0.9$ and $z \sim 0.2$, \cite{Lidman2012} and \cite{Bellstedt2016} find that the stellar mass of BCGs increases by a factor of $1.8 \pm 0.3$ and $1.52 \pm 0.22$, respectively. This is in good agreement with the simulated BCGs, which grow in mass by a factor of $\sim 1.6$ and $\sim 1.5$ between $z=1$ and $z=0$ for the low- and high-mass samples, respectively.
This level of mass growth is consistent with some other recent results (e.g. \citealt{Lin2013, Gozaliasl2018a}) but contrasts with some earlier studies, such as \cite{Brown2008}, \cite{Whiley2008}, \cite{Collins2009} and \cite{Stott2010}, which find no evidence of BCG stellar mass growth since $z \sim 1$.
The reason for the lack of consensus in the literature remains unclear.
Evolutionary studies are complicated by the correlation between BCG mass and host cluster mass and how the cluster mass varies with time. Different methods for accounting for this dependence may be responsible for some of the inconsistencies. Furthermore, deriving BCG stellar masses from imaging data is not a straightforward problem and inconsistent measurements might explain some of the difference. Indeed, the inferred mass growth can depend quite strongly on the adopted mass definition, as we discuss further below.

Of the star particles that end up in the BCG at $z=0$, the large majority are created at $z \gtrsim 1$.
Hence, the moderate mass growth of the simulated and observed BCGs at $z \lesssim 1$ largely reflects the accretion of pre-existing stars in satellite galaxies.
At $z \gtrsim 1$ the simulated BCGs rapidly gain stellar mass inside $30$~pkpc, assembling half of their final stellar mass in the $\sim 4$~Gyr prior to $z \sim 1.6$.
The closeness of the assembled and created mass curves at these redshifts implies that the majority of this growth is due to star formation that occurs either in situ or in galaxies that are soon accreted. Indeed, in Section~\ref{subsubsec:in-situ_mass_growth} we find that in-situ star formation makes a significant contribution to the BCG stellar mass at high redshift.

In the right-hand panel of Figure~\ref{fig:BCG_mass_growth} we show the assembled stellar mass within twice the stellar half-mass radius, which we take as an estimate of the `total' BCG mass. The median radius of this aperture increases from $\sim 60$~pkpc at $z=3$ and $z=2$ to $\sim 120$~pkpc at $z=1$ and $\sim 200$~pkpc at $z=0$.
For reference we compare to the predictions of the semi-analytic models presented in \cite{DeLucia2007} and \cite{Contini2018}.
The model of \cite{DeLucia2007} does not consider an ICL component, while \cite{Contini2018} implement two independent models for the formation of the ICL: the \textsc{standard} model that considers both mergers and stellar stripping, and the \textsc{mergers} model that considers only mergers. We present the results from their most massive sample of BCGs ($M_{\mathrm{BCG}} \gtrsim 3 \times 10^{11} M_{\odot}$ at $z=0$) for fairer comparison with our sample ($M_{\mathrm{BCG}} \gtrsim 5 \times 10^{11} M_{\odot}$) and that of \cite{DeLucia2007} ($M_{\mathrm{BCG}} \sim 10^{12} M_{\odot}$). Our predicted BCG mass growth is overall in good agreement with the semi-analytical models. It is somewhat closer to that of \cite{DeLucia2007} and the \textsc{mergers} model than for the \textsc{standard} model. Given the many differences in the modelling, one should however not draw conclusions about the importance of stripping in our simulations from this.

A comparison of the left- and right-hand panels of Figure~\ref{fig:BCG_mass_growth} demonstrates the dependence of the inferred BCG history on the adopted mass definition.
On the one hand, the creation histories are almost identical between the different apertures, which implies a fairly mild stellar age gradient in the simulated BCGs.
On the other hand, the assembled mass history is quite sensitive to the choice of aperture (see also \citealt{Ragone-Figueroa2018}). For example, the mass growth factor from $z=1$ to $z=0$ is larger for bigger apertures, increasing from $1.6$ to $2.1$ to $4.0$ for apertures of $30$~pkpc, $100$~pkpc and twice the stellar half-mass radius, respectively. This suggests an inside-out growth scenario for the simulated BCGs, consistent with a number of observational studies (e.g. \citealt{VanDokkum2010, Bai2014, DeMaio2019}).
However, it should also be borne in mind that part of this aperture dependence is due to the fact that an aperture of fixed size will enclose a smaller and smaller fraction of the galaxy's halo mass as redshift decreases and the galaxy's size increases.

This aperture dependence may explain some of the lack of consensus in the literature. Some authors advocate for the use of fixed apertures for simplicity, and for enabling cleaner comparisons between models and observations (e.g. \citealt{Lin2013, Zhang2016, Kravtsov2018}). Others employ an adaptively-scaled aperture, such as the \texttt{MAG\_AUTO} aperture used in \cite{Lidman2012} and \cite{Bellstedt2016}, which adjusts in size depending on the observed surface brightness with the aim of deriving the best possible estimate of the total BCG luminosity. \cite{Lin2017} find that the typical characteristic radius of the \texttt{MAG\_AUTO} aperture decreases from 36 pkpc to 22 pkpc from $z \sim 0.4$ to $z \sim 1.0$. As such, \cite{Lidman2012} and \cite{Bellstedt2016} may measure on average lower BCG masses at high redshift compared to studies that employ a fixed physical aperture and this could explain some of the difference between their inferred growth rates. Indeed, \cite{Zhang2016} find that using the \texttt{MAG\_AUTO} aperture increases the inferred BCG mass growth from $z=1$ to $z=0$ by 1-sigma from $\sim 35$ to $\sim 70$ per cent, which better matches the level of mass growth inferred by \cite{Lidman2012}, \cite{Lin2013} and \cite{Bellstedt2016}.
To fully investigate the impact of these biases will require synthetic images produced from the simulations and to analyse them similarly to the data. This could also circumvent some of the biases in the comparison between the simulations and real data.

\subsection{BCG in-situ star formation}\label{subsec:BCG_SFR}

\begin{figure}
  \includegraphics[width=\columnwidth]{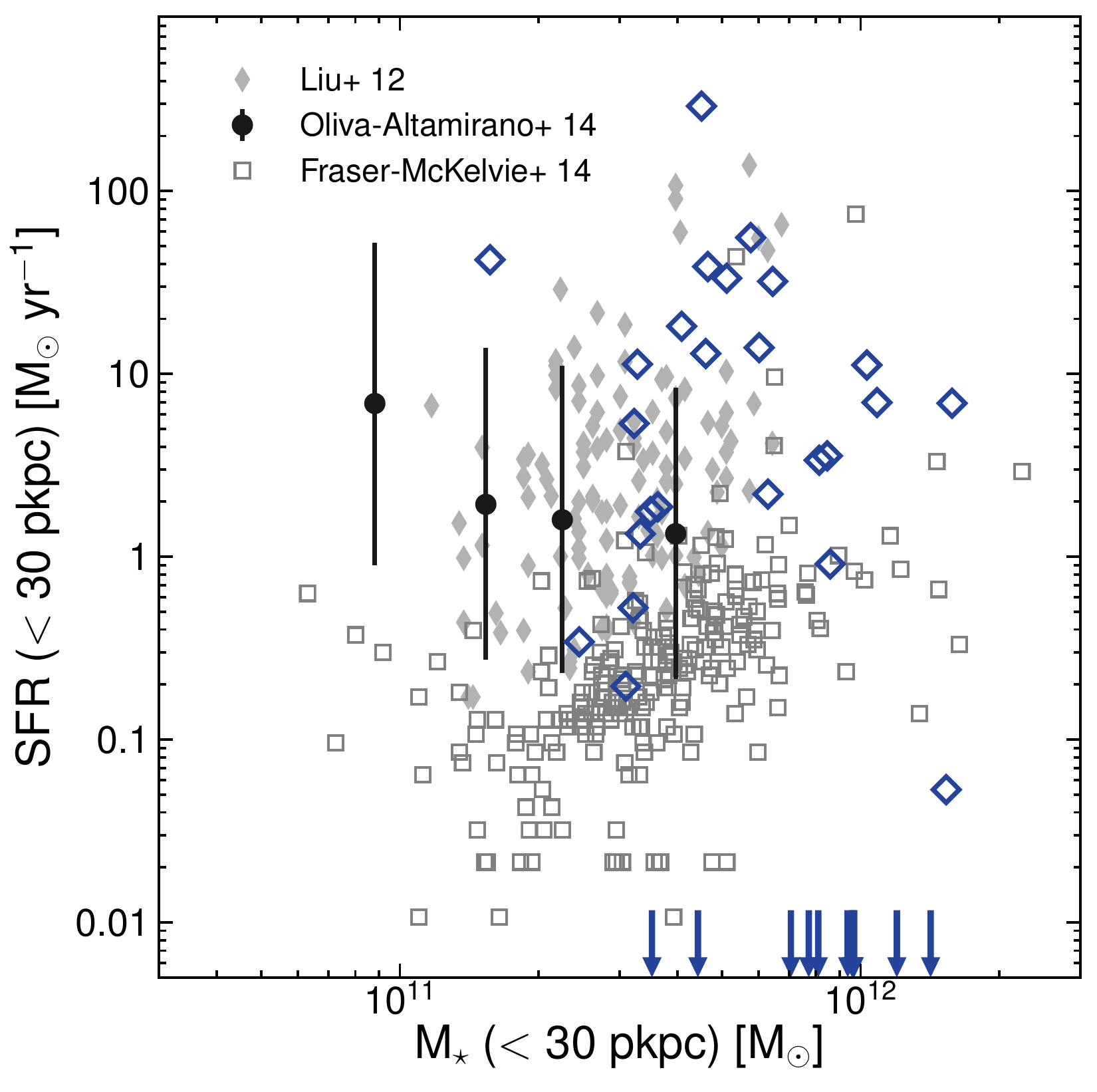}
  \caption{The instantaneous star formation rate of \fable\ BCGs as a function of stellar mass at $z=0.2$ (blue diamonds) compared to the median relation from \protect\cite{Oliva-Altamirano2014} (circles) and individual BCGs from \protect\cite{Fraser-McKelvie2014} (squares) and \protect\cite{Liu2012} (diamonds). Only simulated BCGs in haloes of mass $M_{200} > 10^{14} M_{\odot}$ are considered. The SFR and stellar mass of the simulated BCGs are measured within a spherical aperture of radius $30$~pkpc. The arrows along the horizontal axis indicate the stellar masses of those objects with an SFR of zero within this aperture.
  }
  \label{fig:BCG_SFR_vs_stellar_mass}
\end{figure}

\subsubsection{Star formation rate to stellar mass relation}\label{subsubsec:SFR_vs_mass}
Figure~\ref{fig:BCG_SFR_vs_stellar_mass} shows the instantaneous star formation rate as a function of stellar mass for \fable\ BCGs at $z=0.2$ compared to data from \cite{Liu2012}, \cite{Oliva-Altamirano2014} and \cite{Fraser-McKelvie2014}. These data derive SFRs from H$\alpha$ emission, which is an indicator of the nearly instantaneous SFR.
We measure the SFRs and stellar masses of our simulated galaxies within a spherical aperture of radius $30$ pkpc.
The exact aperture definition does not change our conclusions, which are merely qualitative due to the large scatter in the observed SFRs at fixed stellar mass.

\cite{Liu2012} identify a sample of 120 highly star-forming, early-type BCGs at $0.1 < z < 0.4$ selected from the SDSS with strong H$\alpha$ line emission (equivalent width greater than $3 \mathrm{\AA}$).
\cite{Oliva-Altamirano2014} select 883 group and cluster haloes from the GAMA Galaxy Group Catalogue \citep{Robotham2011} with five or more member galaxies at $0.09 < z < 0.27$, of which $235$ BCGs show evidence of star formation above their detection limit of $0.1 \, M_{\odot} \mathrm{yr}^{-1}$.
The error bars in Fig.~\ref{fig:BCG_SFR_vs_stellar_mass} show the median SFR and scatter in stellar mass bins for their star-forming galaxy sample.
\cite{Fraser-McKelvie2014} employ an X-ray selected sample of 267 clusters at $z < 0.1$, of which 245 contain unambiguously identified BCGs with uncontaminated infra-red photometry from the \textit{Wide-Field Infrared Survey Explorer (WISE)}. The selection on cluster X-ray luminosity corresponds to an approximate cluster mass limit of $M_{200} \gtrsim 2 \times 10^{14} M_{\odot}$.

The star-forming \fable\ BCGs and the \cite{Liu2012} BCGs span a very similar range of SFRs at fixed stellar mass. We point out however that the \cite{Liu2012} sample represents a unique population of highly star-forming BCGs, constituting just $\sim 0.5$ per cent of their full sample of early-type BCGs selected from the SDSS cluster catalogue.
The \cite{Oliva-Altamirano2014} star-forming sample ($\mathrm{SFR} > 0.1 \, M_{\odot} \, \mathrm{yr}^{-1}$) represents a less extreme population than the \cite{Liu2012} sample, constituting $27$ per cent of their whole sample. The fraction of \fable\ BCGs with $\mathrm{SFR} > 0.1 \, M_{\odot} \, \mathrm{yr}^{-1}$ is larger than observed, constituting 67 per cent of BCGs in haloes of mass $M_{200} > 10^{14} M_{\odot}$. The SFRs are in reasonable agreement with the data however, even if the average SFR is somewhat high.
The \cite{Fraser-McKelvie2014} BCGs were selected regardless of their SFR and, although several of their BCGs show similarly high SFRs as the \cite{Liu2012} population, the majority of their sample lie at modest SFRs below $1 \, M_{\odot} \, \mathrm{yr}^{-1}$. Based on the assumption that the \cite{Fraser-McKelvie2014} sample is more representative of the general population, this would imply that the simulated BCGs that continue to form stars at low-redshift are doing so too rapidly.
On the other hand, whilst \cite{Liu2012} and \cite{Oliva-Altamirano2014} measure the H$\alpha$ emission of their BCGs directly, \cite{Fraser-McKelvie2014} estimate SFR from a relation between H$\alpha$-derived SFR and infra-red luminosity that shows considerable scatter, and may underestimate SFRs for low-redshift, low infra-red luminosity galaxies ($\mathrm{SFR} \lesssim 5 \, M_{\odot} \, \mathrm{yr}^{-1}$ at $z \lesssim 0.05$; see \citealt{Cluver2014}). Indeed, \cite{Fraser-McKelvie2014} find considerable scatter between their SFR estimates and those of \cite{Hoffer2012} and \cite{Crawford1999} for the same BCGs, with differences of up to 1 dex for SFRs $< 3 \, M_{\odot} \, \mathrm{yr}^{-1}$.

Overall, the data hint that the star-forming simulated BCGs are forming too many stars at low redshift, although differences in sample selection and methodology mean that this is not conclusive.
Conversely, a significant proportion of the simulated BCGs are no longer forming stars at $z=0.2$, as indicated by the arrows in Fig.~\ref{fig:BCG_SFR_vs_stellar_mass}.
Indeed, the majority of BCGs in haloes of mass $M_{200} > 10^{14} M_{\odot}$ have either zero SFR (31 per cent) or are forming stars at least at a moderate rate of $\gtrsim 1 \, M_{\odot} \mathrm{yr}^{-1}$ (56 per cent), with few BCGs occupying the range in between.
This apparent dichotomy may result from intermittent AGN feedback from the central black hole of the BCGs. That is, if the black hole has recently injected a large amount of energy into the surrounding gas, star formation in the BCG may be quenched. However, if the time taken for the gas to cool and begin forming stars is less than the time between feedback events, one might expect to see a population of non-star-forming BCGs that have experienced a recent feedback event and a second population of highly star-forming BCGs that have not.
We return to this point in the following section, where we study the SFR as a function of redshift in the main progenitors of these BCGs.

\begin{figure}
  \centering
  \includegraphics[width=0.93\columnwidth]{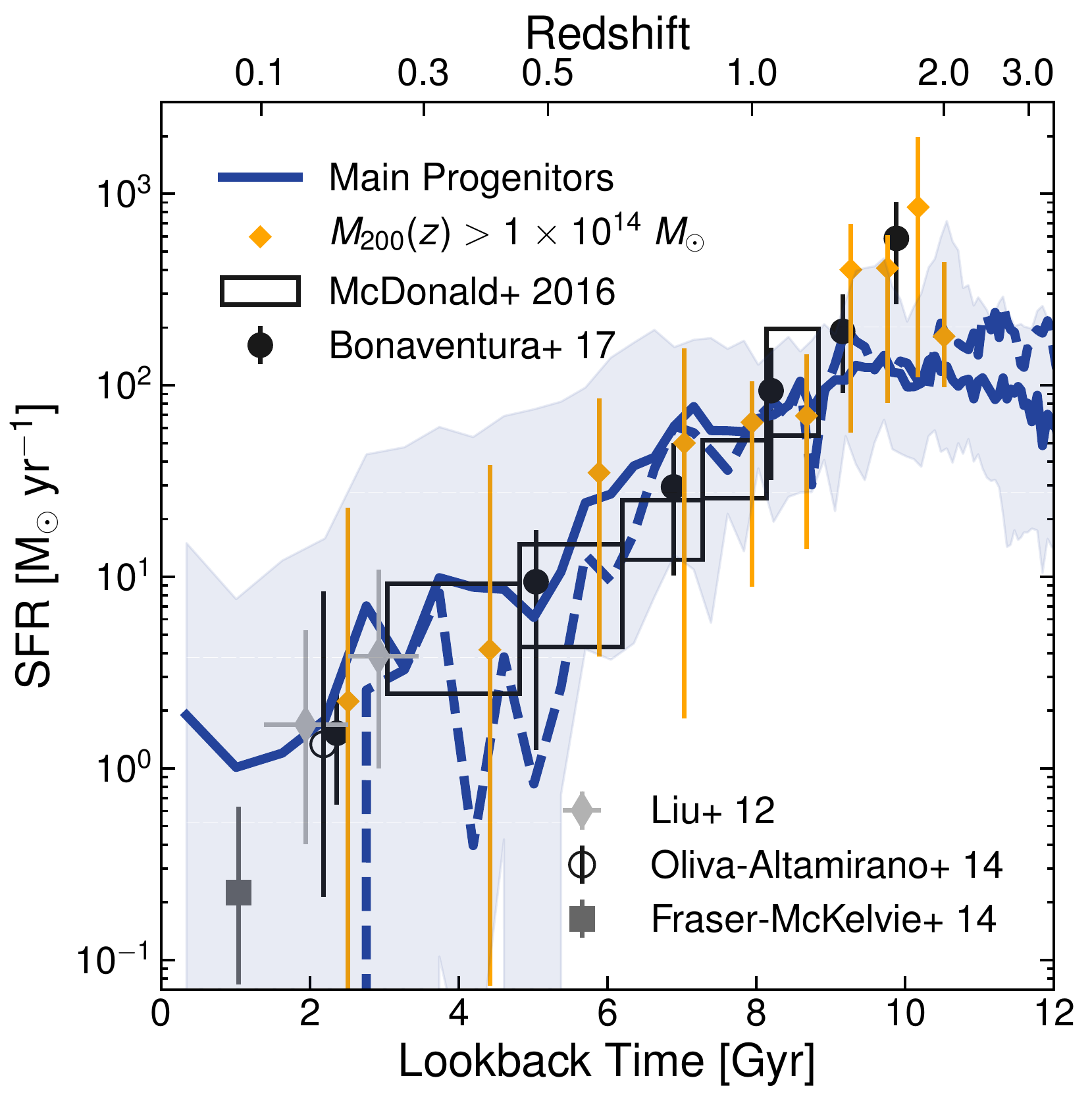}
  \includegraphics[width=0.98\columnwidth]{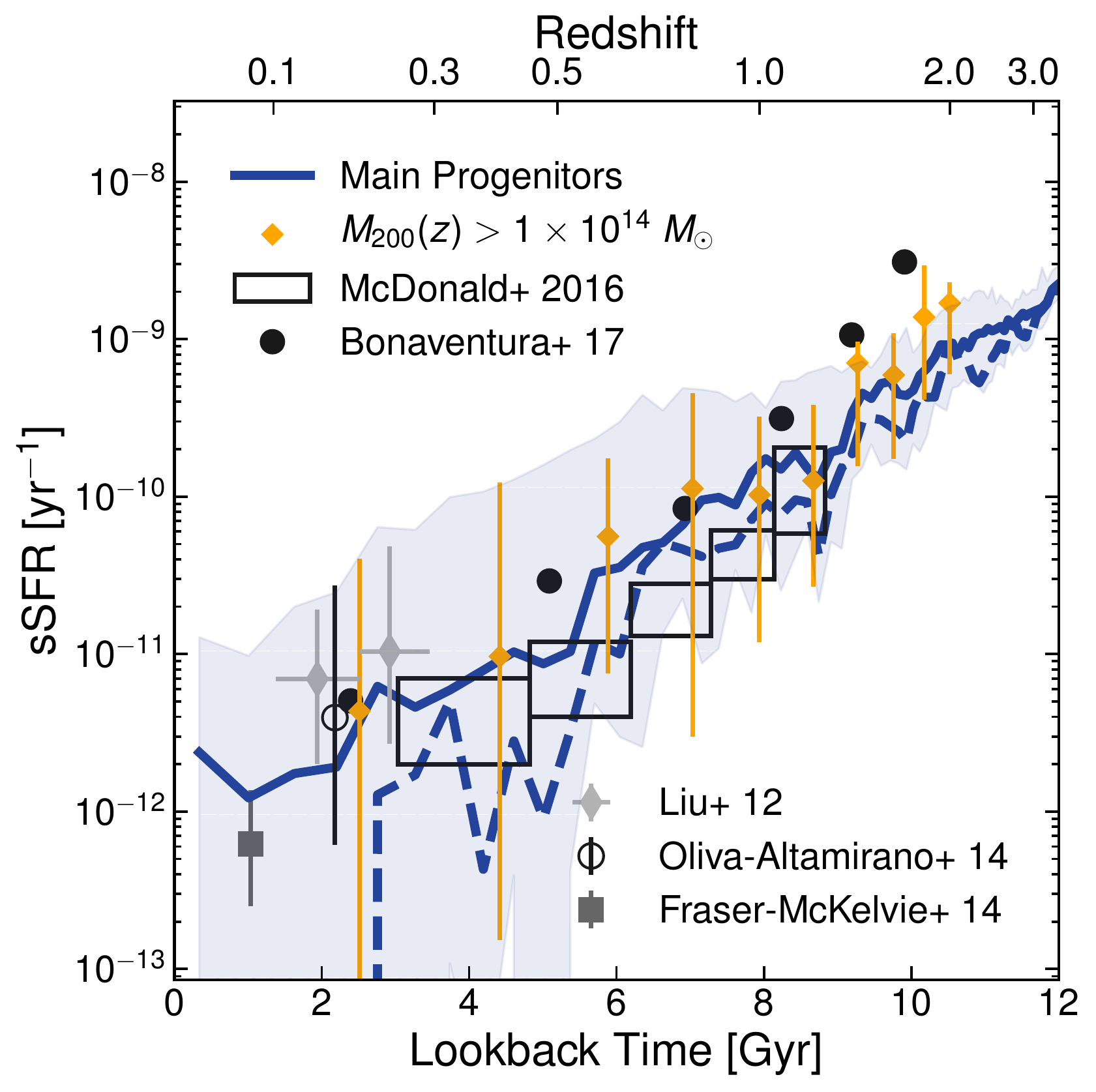}
  \caption{The time evolution of the absolute and specific star formation rate (upper and lower panels, respectively) inside $50$~pkpc for the main progenitors of \fable\ BCGs in haloes with $M_{200} > 10^{14} M_{\odot}$ (solid lines) and $M_{200} > 5 \times 10^{14} M_{\odot}$ (dashed lines) at $z=0$ compared to observational results from \protect\cite{McDonald2016} and \protect\cite{Bonaventura2017}. Lines show the median SFR and sSFR while the shaded region encloses the $16$th to $84$th percentiles of the lower mass sample.
    Orange error bars show the median SFR or sSFR and $16$th to $84$th percentile range for BCGs in haloes with $M_{200} > 10^{14} M_{\odot}$ at each redshift, which mimics the approximately halo mass-selected observed samples.
    Black rectangles show the average SFR or sSFR in different redshift bins from \protect\cite{McDonald2016}, where the height represents the combined statistical uncertainty and uncertainty due to non-detections.
    We also show the median SFR of the \protect\cite{Liu2012}, \protect\cite{Oliva-Altamirano2014} and \protect\cite{Fraser-McKelvie2014} samples at their median redshifts. The \protect\cite{Liu2012} BCGs have been split into two redshift bins. Error bars correspond to the $16$th and $84$th percentiles.
  }
  \label{fig:BCG_SFR_vs_z}
\end{figure}

\subsubsection{Star formation rate history}\label{subsubsec:SFR_vs_z}
Figure~\ref{fig:BCG_SFR_vs_z} shows the evolution of the median instantaneous star formation rate in the main progenitors of \fable\ BCGs found in haloes of mass $M_{200} > 10^{14} M_{\odot}$ at $z=0$.
We compare to observational constraints from \cite{McDonald2016} and \cite{Bonaventura2017}.
\cite{McDonald2016} study the SFRs of 90 BCGs from a sample of SZ-selected clusters at $0.25 < z < 1.2$ derived from a combination of three SFR indicators: UV continuum, [O\,\textsc{ii}] line emission and IR continuum.
\cite{Bonaventura2017} derive the median SFR in redshift bins between $0 < z < 1.8$ from stacked IR data for an optically-selected sample of 716 BCGs drawn from the \textit{Spitzer} \citep{Ashby2013} adaption of the Red-Sequence Cluster Survey.
We point out that the observed BCGs were selected from an approximately halo mass-selected sample of clusters and thus do not necessarily represent an evolutionary sequence. Thus, to better compare with the data, in Fig.~\ref{fig:BCG_SFR_vs_z} we also show the median SFR of BCGs in haloes with $M_{200} > 10^{14} M_{\odot}$ at each redshift. This mass limit is equal to the estimated mass limit of the \cite{Bonaventura2017} sample, although it is somewhat lower than that of \cite{McDonald2016} ($M_{200} \sim 5 \times 10^{14} M_{\odot}$; \citealt{Bleem2015}).

The SFR of the simulated and observed BCGs decrease rapidly with cosmic time, consistent with observations of massive galaxies in general (e.g. \citealt{Daddi2007, VanDokkum2010, Ownsworth2012, Ownsworth2014}).
The median SFR history of the simulated BCGs is in remarkable agreement with the observational constraints out to $z \sim 1$, and to $z \sim 1.8$ when employing a similar sample selection (orange error bars). We find that the median SFR is fairly insensitive to the mass of the sample, which follows from the weak correlation between SFR and stellar mass seen in Fig.~\ref{fig:BCG_SFR_vs_stellar_mass}.
At $z \sim 0.2$ the median SFR of the \fable\ sample is in good agreement with that of the \cite{Bonaventura2017}, \cite{Liu2012} and \cite{Oliva-Altamirano2014} samples, although at $z \sim 0.1$ it is slightly higher than the median SFR of the \cite{Fraser-McKelvie2014} sample, albeit with large scatter.
The star formation rate per unit stellar mass, or specific star formation rate (sSFR), is in better agreement with \cite{Fraser-McKelvie2014} because our simulated BCGs tend to be slightly too massive for a given host halo mass.
At higher redshift, the median sSFR history lies in between those of \cite{McDonald2016} and \cite{Bonaventura2017}, which are offset from one another due to an offset in the average stellar mass of the two samples.

The scatter in the SFRs of the simulated BCGs is somewhat larger than observed. This is particularly true at low redshift ($z \lesssim 0.5$), where the simulated BCGs tend to be either considerably star-forming or not forming stars at all, as seen in Fig.~\ref{fig:BCG_SFR_vs_stellar_mass} at $z=0.2$.
This may be partially explained by the fact that the SFRs measured from the simulation are truly instantaneous, whereas the observed SFRs are averaged over a time-scale that is dependent on the SFR indicator used.
On the other hand, the fact that the scatter in SFR appears to increase with decreasing redshift could be explained by the switching of black holes from the quasar-mode, which efficiently suppresses star formation, to the radio-mode, which is more concerned with ejecting gas from massive haloes as mentioned in Section~\ref{subsubsec:ICM_z}.
The ability of the radio-mode to eject gas from massive haloes is partly the result of its duty cycle, which causes the AGN to store feedback energy before releasing it in a single, powerful event. The thermal energy deposited in such an event is likely to stop star formation in the BCG, leading to a population of quenched BCGs. However, if the time between events is too long, then the gas that is heated by an event may have enough time to cool and begin to form stars before another feedback event occurs, leading to a secondary population of considerably star-forming BCGs.
Increasing the frequency of radio-mode feedback events may help to resolve this issue. However, this would also reduce the energy of each event and would likely lead to too-high gas mass fractions in groups and clusters.
This motivates the need for a new AGN feedback scheme that is capable of suppressing star formation in BCGs more consistently, as well as preventing or slowing the accumulation of gas onto its host halo.

\begin{figure}
  \centering
  \includegraphics[width=\columnwidth]{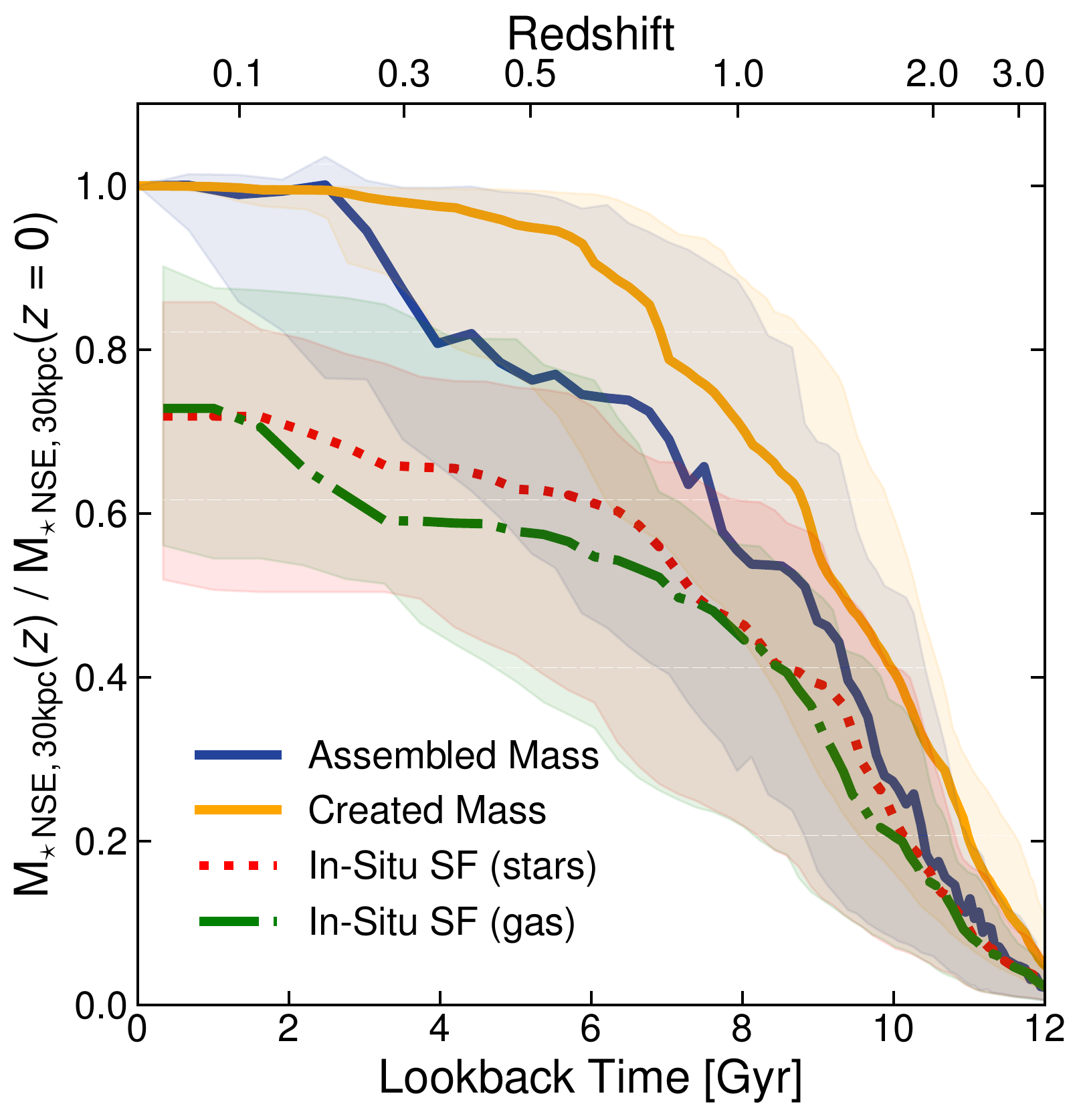}
  \caption{The median cumulative mass gain due to in-situ star formation within $30$~pkpc compared to the median assembly and creation histories for the main progenitors of \fable\ BCGs in haloes with $M_{200} > 10^{14} M_{\odot}$ at $z=0$.
  The in-situ mass gain is calculated either from the mass of stars formed between one snapshot and the next (red dotted line), or from the instantaneous SFR of the gas assuming it remains constant from one snapshot to the next (green dashed-dotted line). Both methods give largely consistent results.
    The assembled and created mass curves are similar to that shown in Fig.~\ref{fig:BCG_mass_growth} but, for fairer comparison with the in-situ mass gain, do not take into account mass loss due to stellar evolution.
    All curves are normalised to the total \textit{assembled} mass at $z=0$. Shaded regions enclose the $16$th to $84$th percentile range.
  }
  \label{fig:BCG_mass_growth_insitu}
\end{figure}

\subsubsection{In-situ mass growth}\label{subsubsec:in-situ_mass_growth}
Figure~\ref{fig:BCG_mass_growth_insitu} illustrates the contribution of in-situ star formation to the final stellar mass of the BCG within $30$~pkpc.
Overall, almost $70$ per cent of the final stellar mass is due to in-situ star formation in the median.
Although, for larger apertures the fractional contribution of in-situ star formation to the final mass is somewhat smaller. Within $50$~pkpc and $100$~pkpc it is approximately $60$ per cent and $50$ per cent, respectively.
The bulk of the in-situ mass gain occurs at $z \gtrsim 1$ where the average SFR is highest.
Between $z=2$ and $z=1$, the median stellar mass gain due to in-situ star formation is $30$ and $23$ per cent of the final stellar mass for a $30$ and $50$ pkpc aperture, respectively. These values fall in between the observational estimates by \cite{Bonaventura2017} ($\sim 35$ per cent) and \cite{Zhao2017} ($7 - 18$ per cent).
These studies, among others (e.g. \citealt{Collins2009, Stott2011, Lidman2013, Webb2015}), suggest that in-situ star formation plays an important, even dominant, role in BCG mass assembly above $z \sim 1$, contributing at least as much to the mass growth of BCGs at $z \gtrsim 1$ as the mass assembled through mergers. This is borne out by the simulations, which imply an increasing contribution from in-situ star formation with increasing redshift.

Although the median SFR falls rapidly below $z \sim 1$, the accumulated mass gain due to in-situ star formation remains significant, contributing $\sim 25$ per cent of the final assembled mass.
This represents approximately half of the assembled mass gain between $z=1$ and $z=0$.
This is consistent with the mean SFR-redshift relation derived in \cite{Gozaliasl2018a}, which suggests that up to $\sim 45$ per cent of the mean stellar mass growth since $z \sim 1.2$ can be due to in-situ star formation. On the other hand, for a typical BCG that follows their median SFR-redshift relation the in-situ mass growth is negligible.
Furthermore, a number of studies find that BCG mass growth at $z \lesssim 1$ is likely dominated by dry mergers, with a sub-dominant contribution from in-situ star formation (e.g. \citealt{Liu2009, Edwards2012, Burke2013, Lidman2013, Webb2015, Zhao2017}).
As such, excess star formation in our simulated BCGs at low redshift may be responsible for the excess stellar mass of our simulated BCGs at $z=0$ compared to observations (Fig.~\ref{fig:BCG_mass_vs_mass}).
In this case, better agreement with observed BCG masses may be attained through improvements to the efficiency of star formation suppression in massive galaxies via AGN feedback.

The fractional mass gain due to in-situ star formation is larger than that found in the cosmological hydrodynamical simulations by \cite{Ragone-Figueroa2018}, who find a fractional contribution of approximately $30$ per cent within a $50$~pkpc aperture (compared to our $\sim 60$ per cent).
We caution, however, that \cite{Ragone-Figueroa2018} employ an order of magnitude more massive cluster sample than the \fable\ sample shown in Fig.~\ref{fig:BCG_mass_growth_insitu} ($M_{200} > 1.1 \times 10^{15} M_{\odot}$ versus $M_{200} > 10^{14} M_{\odot}$ at $z=0$). We lack a sufficiently large sample of massive clusters to make a direct comparison with their results, however, we can make a rough approximation by restricting the sample to massive \fable\ clusters with $M_{200} > 8 \times 10^{14} M_{\odot}$ at $z=0$. In this case, the fractional contribution of in-situ star formation to the final stellar mass is reduced by $\sim 10 - 15$ per cent (e.g. from $\sim 60$ to $\sim 45$ per cent within $50$~pkpc). The remaining difference is likely because the \cite{Ragone-Figueroa2018} BCGs show slightly less intense star formation at $z \gtrsim 1$ compared with \fable, where the bulk of the in-situ mass gain occurs. This suggests that the star formation histories of massive galaxies such as BCGs may provide an important diagnostic for distinguishing between AGN feedback models in the future.

\subsection{BCG stellar mass profiles}\label{subsec:BCG_mass_profiles}
The distribution of stellar mass within a galaxy and how this evolves with time can greatly inform our understanding of the dominant growth pathways of BCGs (e.g. \citealt{Bernardi2009, Ascaso2011, Bai2014, Furnell2018}) and massive galaxies in general (e.g. \citealt{Hopkins2009, Hopkins2010, Wuyts2010, Sonnenfeld2013, Zahid2019}), including the relative contribution of stars from dry mergers versus in-situ star formation and the importance of feedback processes.

Observational studies of this type have found conflicting results.
For example, \cite{Stott2011} find at most a small increase in the size of BCGs between $z \sim 1$ and $z \sim 0.2$ and no evidence for a change in the shape of their light profiles. These results are generally confirmed by \cite{Bai2014} for a sample of BCGs at $0.3 < z < 0.9$ with HST imaging data and a local sample from \cite{Gonzalez2005}. Recent results from \cite{Furnell2018} for an X-ray selected sample of 329 clusters at $0.05 < z < 0.3$ also show little evolution in the size of BCGs, consistent with \cite{Stott2011}.
Conversely, \cite{Bernardi2009} find that BCGs at $z \sim 0.25$ are up to 70 per cent smaller than their local counterparts and \cite{Ascaso2011} find an increase in the size of BCGs by a factor of $\sim 2$ between $z \sim 0.5$ and $z \sim 0$ but no change in the shape of their light profiles.
These conflicting results largely reflect the difficulty in measuring accurate sizes for BCGs, which depend sensitively on the profile modelling and measured sky background level (see e.g. discussion in \citealt{Stott2011}).
In this section we study the stellar mass profiles of our simulated BCGs to gain some insight on these issues from our model predictions.

\begin{figure}
  \includegraphics[width=\columnwidth]{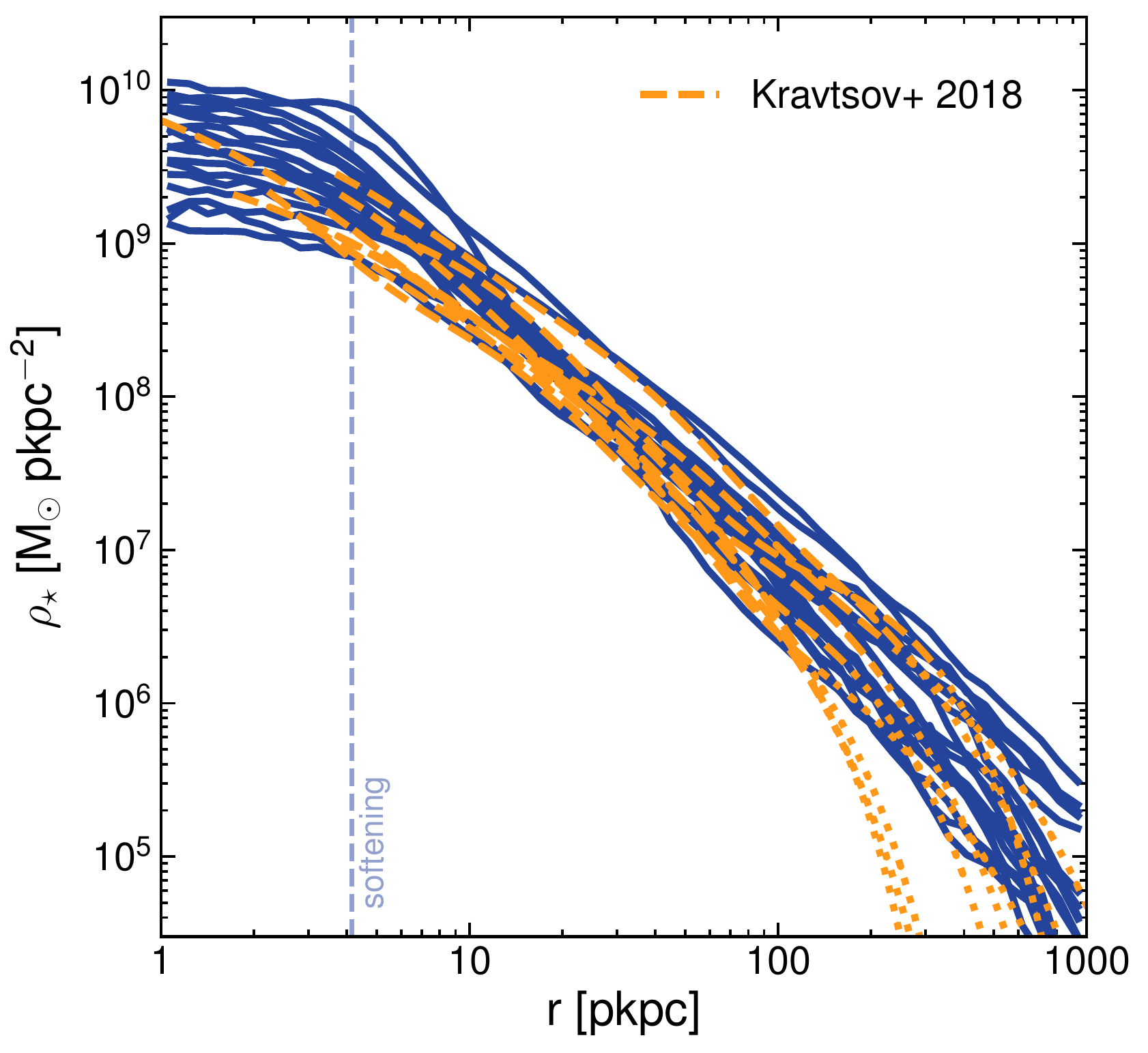}
  \caption{Radial profiles of stellar mass surface density for \fable\ BCGs (solid lines) at $z=0$ compared to the best-fitting profiles of observed BCGs from \protect\cite{Kravtsov2018} (dashed lines).
    Dotted lines indicate extrapolation of the best-fitting triple S\'ersic model.
    The vertical dashed line indicates the gravitational softening length of the simulation below which the simulation predictions are no longer reliable.
  }
  \label{fig:BCG_prof_Kravtsov}
\end{figure}

\subsubsection{Comparison with observations at $z \approx 0$}
As a first step we evaluate the consistency of the simulations with observations at $z \approx 0$.
In Fig.~\ref{fig:BCG_prof_Kravtsov} we compare the (projected) stellar mass surface density profiles of \fable\ BCGs at $z=0$ with observations of BCGs in local clusters and groups from \cite{Kravtsov2018}. The observed profiles are fit with a triple S\'ersic model, which are shown in Fig.~\ref{fig:BCG_prof_Kravtsov} as orange dashed lines.
The observed sample spans the mass range $5.6 \times 10^{13} M_{\odot} < M_{500} < 1.2 \times 10^{15} M_{\odot}$ with a median mass of $M_{500} = 2.1 \times 10^{14} M_{\odot}$. Our comparison sample consists of all $z=0$ \fable\ haloes with $8 \times 10^{13} M_{\odot} < M_{500} < 1 \times 10^{15} M_{\odot}$ and has a median mass of $M_{500} = 2.2 \times 10^{14} M_{\odot}$ similar to the observed sample.
In constructing the profiles we include only stars that are gravitationally bound to the main halo, which includes both the BCG and ICL stellar mass but excludes satellites and unbound stars.
As mentioned previously, the distinction between BCG and ICL is highly ambiguous, both in simulations and observations. As such, we present the total BCG+ICL stellar mass profiles centred on each BCG.

The \fable\ BCGs are slightly too massive at fixed cluster mass compared with the \cite{Kravtsov2018} sample (see Fig.~\ref{fig:BCG_mass_vs_mass}). In Fig.~\ref{fig:BCG_prof_Kravtsov}, this is reflected by the majority of the observed profiles falling closer to the lower end of the scatter of the simulated profiles.
Nevertheless, there is significant overlap between the profiles and excellent agreement in terms of shape across a wide range of scales.
We find a similar level of agreement when comparing the three-dimensional simulated profiles with the deprojected stellar mass density profiles derived in \cite{Kravtsov2018}.
Remarkably, the simulated and observed profiles show a very similar level of scatter in stellar mass surface density at fixed radius. Indeed, we showed in Section~\ref{subsec:BCG_mass_vs_mass} that the intrinsic scatter in the BCG stellar mass--cluster mass relation is in good agreement with observational constraints.
This suggests that the sources of intrinsic scatter in observed BCGs, for example feedback processes and merger histories, are adequately modelled in our simulations.

The majority of the extrapolated profiles match the simulated profiles out to several hundred pkpc. This suggests that the best-fitting triple S\'ersic profile gives a robust measurement of the total stellar mass even beyond the radius of the fit. Indeed, in a recent study of $\sim 300$ clusters at $z \sim 0.2$, \cite{Zhang2018} find that the BCG+ICL light profile is well described by a triple S\'ersic model using high-resolution DES data.
On the other hand, for three of the observed BCGs the extrapolated profiles demonstrate a sharper drop in density at large radii ($\gtrsim 100$~pkpc) than predicted by the simulations. These three objects have the smallest extraction radius (the outer radius used in the fitting procedure), which suggests that the triple S\'ersic model may underestimate the stellar surface density outside the fitting region if the fitting radius is not sufficiently large ($\gtrsim 150$ pkpc in this case).
This result appears to conflict with \cite{Martizzi2014} who find that the extrapolated profiles in the observed sample closely match their simulated BCG profiles at large radii. This may be a sample selection effect however, as their comparison sample extends to lower masses than the \cite{Kravtsov2018} sample. Since lower mass BCGs typically show less extended profiles (i.e. a smaller S\'ersic index), this may bias the \cite{Martizzi2014} sample towards profiles with a steeper drop off in density at large radii.

\begin{figure*}
  \includegraphics[width=0.497\textwidth]{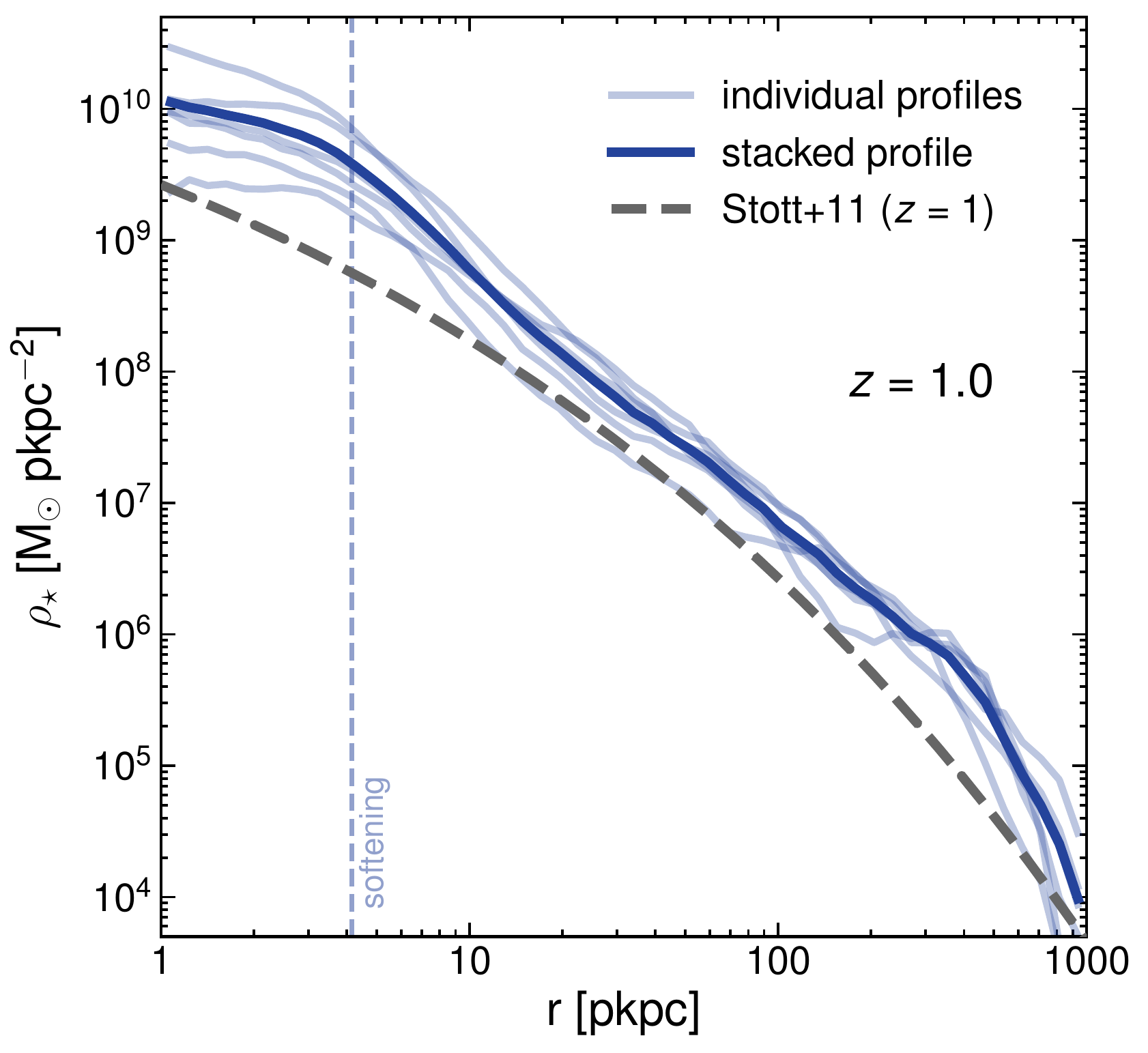}
  \includegraphics[width=0.497\textwidth]{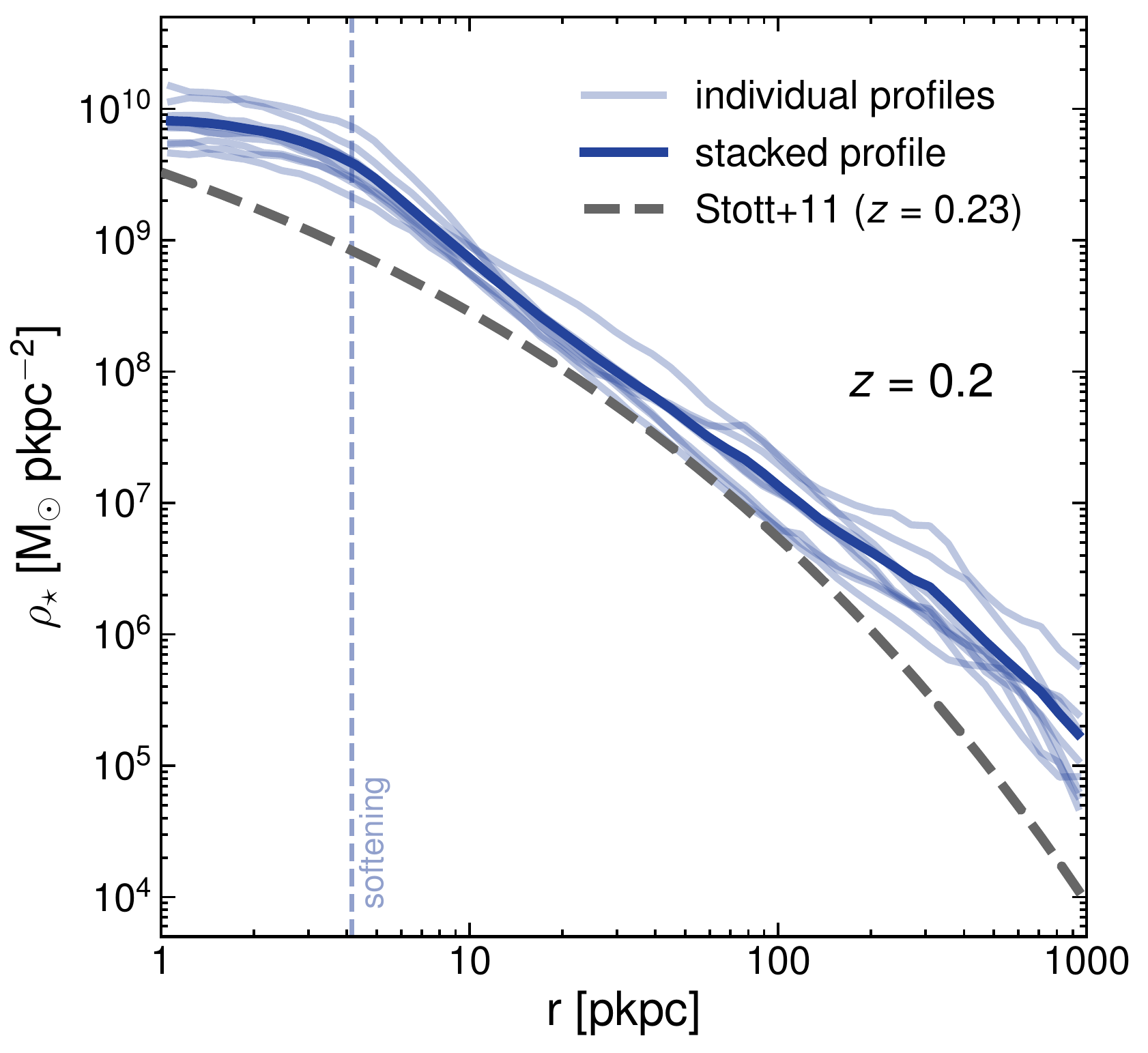}
  \caption{Stacked radial profiles of stellar mass surface density for \fable\ BCGs (solid lines) compared to the stacked profiles from \protect\cite{Stott2011} (dashed lines) at $z=1$ (left) and $z \simeq 0.2$ (right).
    \protect\cite{Stott2011} stack radial surface brightness profiles for a high- and low-redshift sample of BCGs, correcting the profiles to $z=1$ and $z=0.23$, respectively. We have converted the best-fitting S\'ersic profile for each stack into a stellar mass surface density profile (grey dashed lines) assuming a constant mass-to-light ratio as outlined in the text. The fits shown are those with the S\'ersic index allowed to vary during the fit (see text for the best-fitting parameters).
    Faint blue lines show the individual stellar mass surface density profiles of the \fable\ BCGs while the dark blue line shows the stacked profile.
    The vertical dashed line indicates the gravitational softening length of the simulation.
  }
  \label{fig:BCG_prof_Stott}
\end{figure*}

\subsubsection{Redshift evolution}

In Fig.~\ref{fig:BCG_prof_Stott} we plot radial profiles of stellar mass surface density at redshifts $z=1$ and $z=0.2$ in comparison to the stacked BCG profiles of \cite{Stott2011}. The \cite{Stott2011} sample includes a high-redshift sample of five BCGs at $0.8 < z < 1.3$ and a low-redshift sample of 19 clusters at $0.15 < z < 0.3$ with deep \textit{Hubble Space Telescope} Advanced Camera for Surveys (\textit{HST}/ACS) imaging data.

\cite{Stott2011} obtain a robust measurement of the typical stellar distribution of BCGs at each epoch by stacking the 1D surface brightness profiles in each sample and fitting the result with a S\'ersic profile.
The profiles are corrected to a common redshift ($z=1$ and $z=0.23$ for the high- and low-redshift samples, respectively) using \textit{k} and evolution corrections based on a \cite{Bruzual2003} simple stellar population (SSP) model with solar metallicity, formation redshift $z_{\mathrm{f}} = 3$ and a \cite{Chabrier2003} initial mass function.
We make use of the Python program EzGal \citep{Mancone2012} to generate \cite{Bruzual2003} SSP models with the same parameters in order to convert the best-fitting surface brightness profiles into stellar mass surface density profiles, which are shown as thick dashed lines in Fig.~\ref{fig:BCG_prof_Stott}.
The derived solar mass-to-light ratios are 0.88 and 2.71 for the high- and low-redshift samples, respectively.

Weak-lensing mass estimates for clusters in the high-redshift sample range from $M_{200} = 2.9 \times 10^{14} M_{\odot}$ to $M_{200} = 2.3 \times 10^{15} M_{\odot}$ \citep{Sereno2015c}.
Six \fable\ clusters have $M_{200} \geq 2.9 \times 10^{14} M_{\odot}$ at $z=1$ and we use these as our high-redshift comparison sample, although we caution that the median halo mass of the sample ($3.5 \times 10^{14} M_{\odot}$) is somewhat lower than that of the observed sample ($6.3 \times 10^{14} M_{\odot}$).
The individual profiles are shown as faint blue lines in Fig.~\ref{fig:BCG_prof_Stott} while the thick blue line shows the stacked profile in analogy with \cite{Stott2011}.
To construct our low-redshift comparison sample we use the fact that the low-redshift sample of \cite{Stott2011} has an average X-ray temperature similar to that of the high-redshift sample. Since the halo mass corresponding to a given X-ray temperature varies with redshift approximately like the self-similar expectation (see Paper II), we scale the lower mass threshold of the high-redshift sample, $M_{200} \geq 2.9 \times 10^{14} M_{\odot}$, by the factor $E(z=1.0) / E(z=0.2)$ to obtain an equivalent mass threshold of $M_{200} \geq 4.7 \times 10^{14} \, M_{\odot}$ for our low-redshift ($z=0.2$) sample. This yields a sample of nine \fable\ clusters with a median mass of $M_{200} = 6.9 \times 10^{14} \, M_{\odot}$.

There is a clear offset in normalisation between the simulated and observed stellar mass surface density profiles at both redshifts. The integrated stellar mass within 100~pkpc is $\approx 2-3$ times higher for the simulated stacked profiles, which is consistent with the overestimate in the BCG stellar masses shown in Fig.~\ref{fig:BCG_mass_vs_mass} given the uncertainty in the mass-to-light ratio.
The shapes of the simulated profiles are however consistent with the observations, particularly in the outer regions ($\gtrsim 10$~pkpc) where the vast majority of the stellar mass is situated.
The simulated and observed stacked profiles start to deviate at large radii ($\gtrsim 200$~pkpc), although we note that \cite{Stott2011} fit the stacked profiles to a surface brightness limit that corresponds to approximately $100$~pkpc and hence the shape of the profile is not well constrained outside this radius.
The simulated stacked profiles show an excess of stellar mass at small radii ($\lesssim 10$~pkpc) compared to the observed profiles. The flattening of the simulated profiles close to the gravitational softening length ($\approx 4$~pkpc; vertical dashed line) suggests that this may be a resolution effect. One interpretation is that the stellar mass that should reside at $\lesssim 4$~pkpc has been smoothed out to larger radii of $\sim 4-8$~pkpc, leading to a higher density in this region. Given that our simulated BCGs have somewhat high SFRs, it may be the case that too many stars form near the centre of the BCG and that their distribution is smoothed out to $\sim 10$~pkpc due to the gravitational softening.

In analogy with \cite{Stott2011} we fit a S\'ersic model to our stacked profiles at $z=1$ and $z=0.2$ and compare the best-fitting values for the effective radius, $r_e$, which is defined as the radius containing half of the light (or in this case the mass) of the galaxy. For consistency with the observations we fit the model using a least-squares fitting routine (in log-space) out to a radius of $100$ pkpc, which is approximately the radius at which the observed profiles drop below the surface brightness limit used in the fit. We take the uncertainty in the stacked profile to be the standard error of the mean at each radius.

Since $r_e$ is coupled to the S\'ersic index $n$ \citep{Graham1996}, we initially choose a fixed value of $n=4$ (a de Vaucouleur profile) and compare to the \cite{Stott2011} results where $n$ was also fixed to this value.
The best-fitting effective radii are $r_e = 18.5 \pm 1.7$ pkpc and $r_e = 23.6 \pm 2.1$ pkpc at $z=1$ and $z=0.2$, respectively. This corresponds to a size increase of $28 \pm 3$ per cent from high to low redshift, similar to the $35 \pm 3$ per cent increase found by \cite{Stott2011} who measure $r_e = 32.1 \pm 2.5$ pkpc and $r_e = 43.2 \pm 1.0$ pkpc for their high- and low-redshift samples, respectively.
When excluding the region inside the gravitational softening length, the best-fitting effective radii at $z=1$ and $z=0.2$ increase to $22.2 \pm 2.9$~pkpc and $26.0 \pm 3.5$~pkpc, respectively.
Note that in this case the size increase from high to low redshift ($\sim 17$ per cent) is even smaller.
Performing the fit to larger radii generally biases the best-fitting effective radius towards larger values. This is because the simulated profiles tend to fall off with radius more slowly than assumed in the de Vaucouleur model at $\gtrsim 100$~pkpc. For example, when fitting in the radial range $0-500$ pkpc the best-fitting radii are $r_e = 47.1 \pm 7.0$~pkpc and $r_e = 51.2 \pm 9.2$~pkpc at $z=1$ and $z=0.2$, respectively.
In this case the size growth is only $9 \pm 2$ per cent.
Overall these results suggest that a de Vaucouleur profile fit ($n=4$) to the simulated profiles, as used in many observational studies, leads to an inference of weak size evolution since $z = 1$, consistent with \cite{Stott2011} and in contrast to studies such as \cite{Bernardi2009} and \cite{Ascaso2011}.

However, the inferred size evolution is strongly dependent on the choice of model.
For example, allowing the S\'ersic index to vary (and fitting within $0-100$ pkpc) we measure $r_e = 21.4 \pm 4.5$ pkpc with $n = 7.0 \pm 1.5$ for the $z = 1$ stack and $r_e = 37.9 \pm 10.0$ pkpc with $n = 5.7 \pm 1.1$ for the $z = 0.2$ stack. This corresponds to a size increase of $77 \pm 26$ per cent from high to low redshift, significantly larger than the $\sim 28$ per cent increase found for the de Vaucouleur profile, albeit with large uncertainties.
\cite{Stott2011} also fit their stacked profiles with a free S\'ersic model, measuring $r_e = 47.6 \pm 13.7$ pkpc with $n = 5.4 \pm 0.9$ at $z \sim 1$ and $r_e = 57.9 \pm 4.5$ pkpc with $n = 4.8 \pm 0.2$ at $z \sim 0.2$.
Their inferred size growth therefore drops slightly ($\sim 22$ per cent) compared with the $n=4$ case ($\sim 35$ per cent).
The simulated profiles are more sensitive to changes in the S\'ersic index because they are, in general, poorly fit by a S\'ersic model for reasonable values of $n \lesssim 20$.
In particular, outside the fitting radius the simulated profiles show a more gradual change in slope than the free S\'ersic fits and are fairly close to a power law.
As a result, the best-fitting S\'ersic index diverges as we increase the maximum radius used in the fit.
The same effect has been found in a number of observational studies (e.g. \citealt{Graham1996, Gonzalez2005, Zibetti2005, Seigar2007, Stott2011}). These studies find that the outer envelopes of observed BCGs (which likely contain a significant proportion of ICL) are often poorly fit by a S\'ersic profile, biasing the fit to large $n$.
Since the size of this bias depends on the outer radius of the fit, the inferred sizes can be very sensitive to the depth of the data and the background subtraction procedure (e.g. \citealt{Bernardi2010}). These two factors may explain why a S\'ersic model is a good fit to the \cite{Stott2011} profiles but not to the simulations.

The true half-mass radii measured directly from the simulation depend on the outer radius of integration, although in most cases they imply a significant size evolution with redshift. For example, for an outer radius of $500$ pkpc the half-mass radii are $r_e = 39.0$ pkpc at $z=1$ and $r_e = 78.3$ pkpc at $z=0.2$, corresponding to a size increase of $\sim 100$ per cent. Alternatively, at an integration radius of $100$ pkpc we find $r_e = 11.2$ pkpc and $r_e = 19.5$ pkpc, which correspond to a size increase of $\sim 74$ per cent. This implies significant BCG size growth, with the caveat that we are implicitly assuming that the BCG abruptly ends (and the ICL begins) at a radius of $100$ pkpc when in reality there is no such clear distinction.
This size growth is consistent with \cite{Ascaso2011} who measure a size increase of $106 \pm 63$ per cent between $z \sim 0.5$ and $z \sim 0$. Indeed, \cite{Ascaso2011} fit a S\'ersic model combined with an exponential component out to $\sim 100$ pkpc, which is a better fit to their observed profiles and likely avoids some of the biases associated with the single S\'ersic fit described above.

The size growth inferred from the true half-mass radii increases as we enlarge the integration radius. This indicates that the build-up of stellar mass in BCGs between $z=1$ and $z=0.2$ occurs more rapidly at large radii.
Indeed, from Fig.~\ref{fig:BCG_prof_Stott} it is clear that the slope of the simulated profiles at $\gtrsim 100$ pkpc is noticeably shallower in the low redshift sample (the surface mass density scales with radius like $\sim r^{-1.7}$ at $z=0.2$ as opposed to $\sim r^{-2.0}$ at $z=1$).
This implies that, at $z \lesssim 1$, stellar mass growth occurs predominantly in the ICL at large radii.
This interpretation is consistent with the relatively mild growth of the BCG stellar mass at $z < 1$ discussed in Section~\ref{subsec:BCG_mass_evol}, although we caution that the build-up of ICL in our simulations may be overestimated due to enhanced tidal stripping of satellite galaxies as discussed in Section~\ref{subsec:BCG_mass_evol}.

Overall, our findings confirm those of previous observational studies (e.g. \citealt{Bernardi2010, Bernardi2013, Bai2014}) which show that the comparison between best-fitting model parameters is ambiguous due to the coupling between parameters (e.g. effective radius and S\'ersic index; see e.g. \citealt{Graham1996}) and systematics in the sky background measurement (and therefore the outer radius to which the BCG profile can be reliably fit).
Some of the biases associated with the former can likely be alleviated by a more suitable choice of model, for example the triple S\'ersic profile used in \cite{Kravtsov2018}, which we showed in the previous section provides a reasonable description of the BCG profiles out to large radii. However, in that comparison we also found that the profiles with the smallest fitting radii significantly underestimate the stellar mass surface density at large radii, similar to the single S\'ersic fits described above. This can considerably bias the inferred scale radius (and potentially the total stellar mass measurement), with important consequences for studies of BCG growth.
Our results suggest that careful background subtraction, sufficiently deep data (and/or stacking), and parametric models that accurately describe the outer profiles of BCGs and the surrounding ICL are all required if further progress in the study of BCG growth is to be made.
In future, treating the simulations in the same manner as real data using mock observations will help shed further light on these issues.

\section{Conclusions}\label{sec:conclusion}
In this work we have explored the baryon content of galaxy clusters and groups in the \fable\ suite of cosmological hydrodynamical simulations.
We have quantified the halo mass and redshift dependence of the total gas and stellar mass within $r_{500}$ (Sections~\ref{subsec:ICM_mass}, \ref{subsec:stellar_mass} and \ref{subsec:ICM_stellar_mass_z}) and provided a census of the stellar mass content in central galaxies, intracluster light and satellites across a wide halo mass range (Section~\ref{subsec:stellar_mass_components}).
We have further studied the stellar mass content of the BCGs that form at the centre of these massive haloes, including their correlation with host cluster mass (Section~\ref{subsec:BCG_mass_vs_mass}), their stellar mass evolution (Section~\ref{subsec:BCG_mass_evol}), the importance of in-situ star formation (Section~\ref{subsec:BCG_SFR}) and their stellar mass profiles at $z \sim 0$ and $z \sim 1$ (Section~\ref{subsec:BCG_mass_profiles}).
The main conclusions of this work are summarised in the following points.

\begin{itemize}

\item The total gas mass and total stellar mass of \fable\ groups and clusters as measured within $r_{500}$ are in excellent agreement with observational constraints based on X-ray hydrostatic mass estimates at $z \approx 0$ across two decades in halo mass ($M_{500} \approx 10^{13}$--$10^{15} M_{\odot}$; Fig.~\ref{fig:baryon_mass_vs_halo_mass}). Conversely, comparison with weak lensing-calibrated constraints implies that the simulated systems are somewhat too gas-rich.
Taken at face value, the level of baryon depletion implied by these weak lensing-based constraints presents a challenge to current models of cluster formation.

\item For a sample of \fable\ clusters with $M_{500} > 3 \times 10^{14} \, M_{\odot}$ we show that the total gas mass and total stellar mass within $r_{500}$ at fixed halo mass are approximately independent of redshift at $z \lesssim 1$ (Fig.~\ref{fig:baryon_mass_vs_z}), in agreement with recent constraints using SZ-selected cluster samples \citep{Chiu2018}. For lower mass samples the simulations predict significant redshift evolution in these quantities (see Section~\ref{subsec:ICM_stellar_mass_z}).
This has important implications for the growth of massive galaxy clusters, which accumulate lower mass objects from these evolving populations and yet show little redshift evolution themselves.

\item The fraction of the total stellar mass in satellite galaxies is significantly smaller in \fable\ clusters ($\sim 40$ per cent) than observed ($\sim 70$ per cent; \citealt{Kravtsov2018}). Conversely, the ICL mass fraction is significantly larger than observed (Fig.~\ref{fig:stellar_mass_components_z0}). These discrepancies can be explained if satellite galaxies in \fable\ are stripped of their stars (and perhaps disrupted completely) at an accelerated rate. This could be because the \fable\ model, like Illustris, produces galaxies that are too large at fixed stellar mass (see Appendix~\ref{A:size-mass}) and are thus more susceptible to tidal stripping. This implies a need to calibrate next-generation models to reproduce the observed galaxy size--mass relation to ensure that the impact of cluster-specific processes such as tidal stripping can be accurately predicted.

\item The predicted relation between BCG stellar mass and host cluster mass has a similar slope and intrinsic scatter to observational constraints at $z \approx 0$, but a somewhat higher normalisation (by $\sim 0.2-0.3$ dex; Fig.~\ref{fig:BCG_mass_vs_mass}). This offset is similar to, or less than, that of the recent \textsc{c-eagle} and IllustrisTNG simulations. There is tentative evidence that the offset with respect to observations persists up to $z \sim 1$ (Fig.~\ref{fig:BCG_mass_vs_z}).
The source of this discrepancy is currently unclear, with possible causes including an overabundance of massive galaxies ($\gtrsim 10^{11} M_{\odot}$) in the field environment, excess in-situ star formation within the BCGs, or enhanced tidal stripping of stars from satellite galaxies.

\item We also examine the evolution of stellar mass in the main progenitors of \fable\ BCGs, finding good agreement with several observational inferences at $z \lesssim 1$ \citep{Lidman2012, Lin2013, Bellstedt2016}. In particular, simulated BCGs show moderate stellar mass growth in the central $30$~pkpc at $0.3 < z < 1$ (a factor of $\sim 1.5$) which effectively halts at $z \lesssim 0.3$. Growth continues, however, in the outskirts of the BCG, signalling the late development of the ICL.
Indeed, the level of mass growth depends sensitively on the mass definition, which may explain some of the variation among observational constraints.

\item A comparison of the star formation rates of \fable\ BCGs to low-redshift ($z \sim 0.2$) observations reveals that the simulated BCGs tend to be either moderately to highly star-forming ($\gtrsim 1 M_{\odot} \, \mathrm{yr}^{-1}$), or not forming stars at all (Fig.~\ref{fig:BCG_SFR_vs_stellar_mass}). This apparent dichotomy may be the result of strong but intermittent radio-mode AGN feedback that is able to completely shut down star formation initially, but which acts too infrequently to maintain an overall modest level of star formation in BCGs as observed \citep{Fraser-McKelvie2014}.
Even so, the redshift evolution of the median SFR and sSFR is in excellent agreement with observational constraints out to at least $z = 1$ (Fig.~\ref{fig:BCG_SFR_vs_z}).
In-situ star formation contributes significantly to the stellar mass growth of \fable\ BCGs, providing more than half of the final assembled mass in the median. The bulk of the in-situ mass gain occurs at $z \gtrsim 1$ where SFRs are highest, although the contribution at lower redshifts remains significant (Fig.~\ref{fig:BCG_mass_growth_insitu}).

\item Stellar mass surface density profiles centred on \fable\ BCGs are in excellent agreement with $z \approx 0$ observations from \cite{Kravtsov2018}, which are based on triple-S\'ersic fits to the observed light profile out to $\sim 100$ pkpc (Fig.~\ref{fig:BCG_prof_Kravtsov}). The extrapolation of these fits to larger radii agrees well with the simulated profiles out to several hundred pkpc in the majority of cases, although the simulated profiles suggest that the stellar mass density can be underestimated if the outer radius of the fit is not sufficiently large.

\item We also study the evolution of the simulated profiles with comparison to the stacking analysis of BCG light profiles performed by \cite{Stott2011} for clusters at $z \sim 1$ and $z \sim 0.2$ (Fig.~\ref{fig:BCG_prof_Stott}). We find that a single S\'ersic model tends to underestimate the stellar mass density in the simulations beyond the surface brightness limit of the observations ($\gtrsim 100$ pkpc) as, on these scales, the simulated profiles follow a close to power law shape out to several hundred pkpc.
The slope at large radii is slightly shallower at $z = 0.2$ compared with $z=1$ such that the inferred size growth is highly sensitive to the outer radius of the fit. This highlights the need for sufficiently deep imaging and careful background subtraction in addition to an appropriate choice for the assumed light profile.
\end{itemize}

Given the lack of a well-defined transition between the BCG and surrounding ICL (see e.g. the stellar mass profiles in Fig.~\ref{fig:BCG_prof_Kravtsov}) and the sensitivity of inferred BCG mass growth to the definition of galaxy stellar mass (see Section~\ref{subsec:BCG_mass_evol}), it seems clear that a like-for-like comparison of simulations to data requires both mock observations and consistent mass definitions (such as apertures of fixed physical size).
Analyses such as these will enable reliable comparison, and perhaps calibration, of simulations to observations that will help to pinpoint and address the remaining discrepancies of our model.
For example, it is likely that a more sophisticated AGN feedback scheme, such as AGN jet modelling (e.g. \citealt{Bourne2017, Weinberger2017, Bourne2019}), and/or additional physical processes such as cosmic rays \citep{Sijacki2008, Pakmor2016a, Simpson2016, Pfrommer2017a, Ruszkowski2017} or anisotropic thermal conduction \citep{Kannan2016, Kannan2017, Yang2016b, Barnes2018}, are required to reliably suppress star formation in massive galaxies and to deplete cluster baryon fractions significantly below the universal average without over-heating or excessively expelling gas from the cluster core regions.
This is a salient concern for all existing hydrodynamical simulations, none of which are able to exactly reproduce the total baryon content of galaxy groups and clusters including the observed partitioning of baryons among their various gaseous and stellar phases.
Nonetheless, hydrodynamical simulations will doubtless continue to play an important role in our understanding of the formation and evolution of galaxies, groups and clusters.
This is especially true of the low-mass cluster/galaxy group regime probed in this study, which will become accessible in the near future with surveys such as SPT-3G \citep{Benson2014}, Advanced ACTpol \citep{Henderson2016} and eROSITA \citep{Merloni2012, Pillepich2018} with follow-up observations from X-ray and optical/infra-red instruments such as \textit{Chandra}, XMM-Newton, \textit{Spitzer}, DES and the James Webb Space Telescope \citep{Gardner2006}.

\section*{Acknowledgements}
The authors would like to thank Andrey Kravtsov for useful discussions and suggestions and for providing their data.
We are grateful to Volker Springel for making the {\sc arepo} moving-mesh code available to us and to the Illustris collaboration for their development of the Illustris galaxy formation model, which provided an excellent starting point for the development of the \fable\ model.
NAH is supported by the Science and Technology Facilities Council (STFC).
EP acknowledges support by the Kavli Foundation.
DS acknowledges support by the STFC and the ERC Starting Grant 638707 ``Black holes and their host galaxies: co-evolution across cosmic time''.
This work made use of:
the Cambridge Service for Data Driven Discovery (CSD3), part of which is operated by the University of Cambridge Research Computing Services on behalf of the STFC DiRAC HPC Facility (\href{www.dirac.ac.uk}{www.dirac.ac.uk}). The DiRAC component of CSD3 was funded by BEIS capital funding via STFC capital grants ST/P002307/1 and ST/R002452/1 and STFC operations grant ST/R00689X/1; 
The DiRAC@Durham facility managed by the Institute for Computational Cosmology on behalf of DiRAC. The equipment was funded by BEIS capital funding via STFC capital grants ST/P002293/1 and ST/R002371/1, Durham University and STFC operations grant ST/R000832/1. DiRAC is part of the National e-Infrastructure.

\section*{Data availability}
The data underlying this article will be shared on request to the corresponding author.

\bibliographystyle{mnras}
\bibliography{Fable_stars}

\clearpage
\appendix
\section{Galaxy size-mass relation}\label{A:size-mass}

\begin{figure}
  \includegraphics[width=0.975\columnwidth]{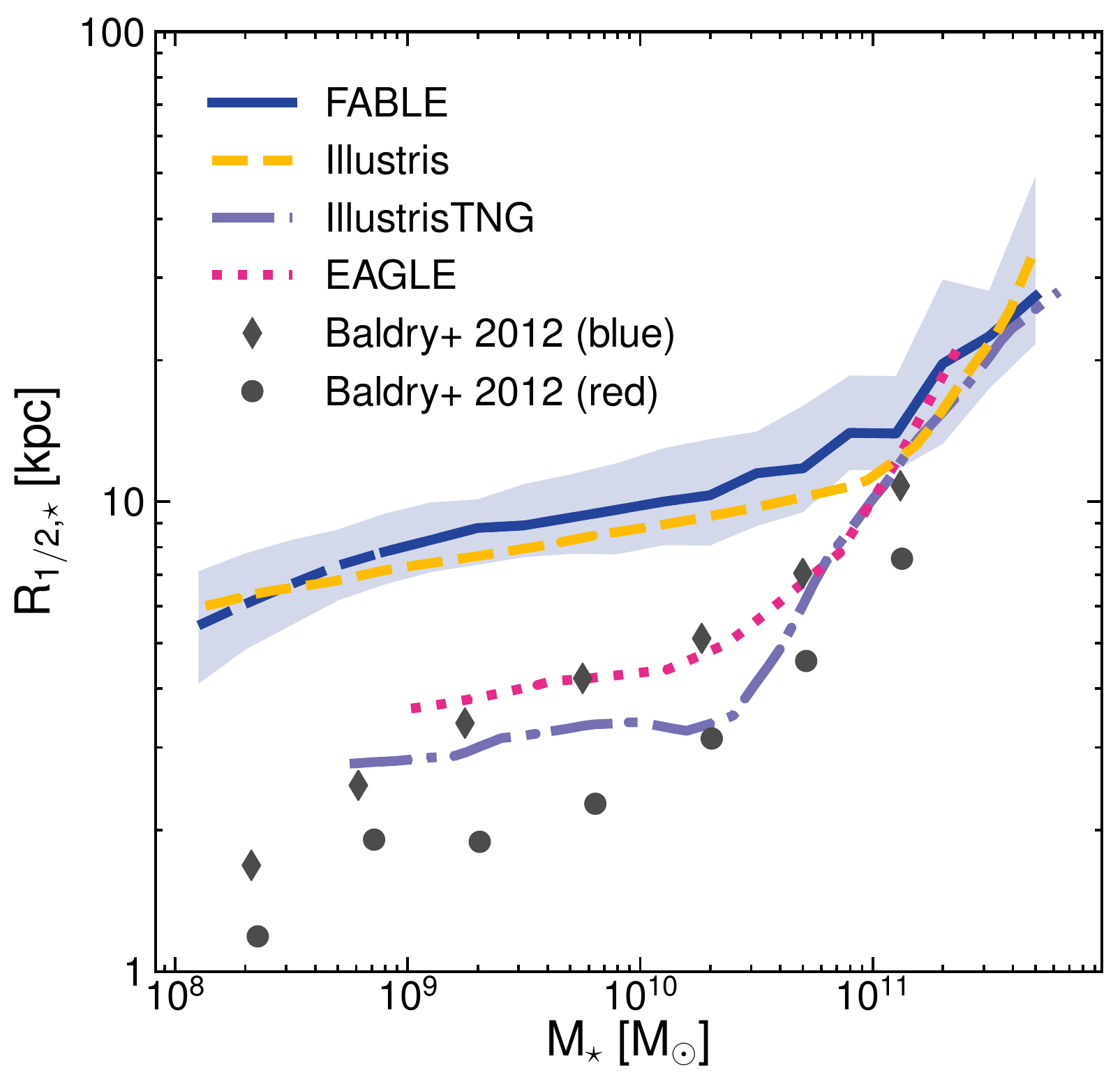}
  \caption{The median relationship between galaxy size and stellar mass for galaxies in the \textsc{fable}, Illustris, IllustrisTNG and \textsc{eagle} simulations (lines) compared to data from \protect\cite{Baldry2012} (symbols).
For \fable\ the line becomes dashed for galaxies that are resolved by fewer than 100 star particles on average. The shaded region encloses the $16$th to $84$th percentiles.
For the simulations the galaxy size is taken to be the three-dimensional stellar half-mass radius. For \fable, Illustris and IllustrisTNG the stellar mass is that bound to the galaxy within twice the stellar half-mass radius. For the \textsc{eagle} relation the stellar mass is the mass of stars within a spherical aperture of 30 pkpc. The \protect\cite{Baldry2012} data are based on half-light radii and best estimates of the total stellar mass of the galaxy.}
  \label{fig:size-mass}
\end{figure}

Figure~\ref{fig:size-mass} shows the median galaxy size--stellar mass relations at $z=0$ in \fable, Illustris, IllustrisTNG \citep{Genel2018} and \textsc{eagle} \citep{Furlong2017} (lines).
Symbols show the median relations between half-light radius and total stellar mass for blue and red galaxies in the Galaxy And Mass Assembly (GAMA) survey \citep{Baldry2012}.
We construct the \fable\ relation using all galaxies in our $40 \, h^{-1}$ comoving Mpc on-a-side volume for fair comparison with the other simulation results, which are also based on uniformly-sampled cosmological volumes. We note, however, that also including galaxies in our suite of zoom-in simulations has little effect on the median relation.

\fable\ galaxies with $M_{\star} < 10^{11} M_{\odot}$ are larger than
observed ones by roughly a factor of two. A similar offset is found in the original Illustris simulation (orange dashed line), as confirmed by previous studies \citep{Snyder2015, Bottrell2017, Furlong2017}.
The discrepancy is much improved in the IllustrisTNG model, which was calibrated to approximately reproduce, among other galaxy scaling relations, the observed galaxy size--mass relation at $z=0$ \citep{Pillepich2018a}.
This change is owed to a combination of several modifications to the galactic winds model, although an investigation of its precise origin has been postponed to future work \citep{Pillepich2018a}.

The \textsc{eagle} relation is also in reasonable agreement with the data \citep{Furlong2017}. Like IllustrisTNG, the calibration of the \textsc{eagle} model involved broad comparisons with observed present-day galaxy sizes \citep{Crain2015}.
This was not the case for the \fable\ model, which was calibrated to the present-day galaxy stellar mass function and the gas mass fractions of galaxy groups only (see Paper~I). Whereas the changes made to galactic winds and AGN feedback in \fable\ relative to Illustris lead to significantly better agreement with the observed galaxy stellar mass function, their affect on galaxy sizes appears to be minimal. As we discuss in the main body of the paper, the large sizes of \fable\ galaxies may have important implications for the evolution of the stellar mass in clusters due to, for example, enhanced tidal stripping and dwarf galaxy disruption.
As such, the approach taken by \textsc{eagle} and IllustrisTNG to calibrate their models to observed galaxy sizes may prove a necessary step in the development of new galaxy cluster simulations.

\bsp	
\label{lastpage}
\end{document}